\documentclass[aps,prd,preprint,tightenlines,preprintnumbers,showpacs,superscriptaddress,nofootinbib,eqsecnum,floatfix]{revtex4-1}

\usepackage{bm}
\usepackage{graphicx}
\usepackage{amsmath, amssymb}
\usepackage{hyperref}
\usepackage{dcolumn}
\usepackage{array}
\hypersetup{
    colorlinks=true,     
    linkcolor=blue,      
    citecolor=blue,      
    filecolor=blue,      
    urlcolor=blue        
}
\usepackage{rotating}
\usepackage{mathtools}
\usepackage{multirow}
\usepackage{slashed} 

\newcommand{\bi}{\begin{itemize}}
\newcommand{\ei}{\end{itemize}}

\newcommand{\be}{\begin{equation}}
\newcommand{\ee}{\end{equation}}

\newcommand{\bea}{\begin{eqnarray}}
\newcommand{\eea}{\end{eqnarray}}

\newcolumntype{P}[1]{>{\centering\arraybackslash}p{#1}} 

\newcommand{\chisq}{\ensuremath{\chi^2/\text{dof}}}

\newcommand{\vev}[1]{\ensuremath{\left\langle #1 \right\rangle} }

\newcommand{\Dslash}{\ensuremath{D\kern-0.6em/\kern0.15em}}
\newcommand{\tr}{\ensuremath{\mathop{\text{tr}}}}     
\newcommand{\Tr}{\ensuremath{\mathop{\text{Tr}}}}     

\newcommand{\cpt}{$\chi$PT}

\newcommand{\PiLR}{\ensuremath{\Pi_\text{LR}}} 

\newcommand{\spose}[1]{\hbox to 0pt{#1\hss}}

\newcommand{\inapprox}{\mathrel{\spose{\lower 3pt\hbox{$\mathchar"218$}}
 \raise 2.0pt\hbox{$\mathchar"232$}}}

\def\vev#1{\langle #1\rangle}
\def\SU{{\rm SU}}
\def\SO{{\rm SO}}
\def\U1{{\rm U}(1)}
\def\Eq#1{Eq.~(\ref{#1})}
\def\Eqs#1{Eqs.~(\ref{#1})}
\def\wcpt{W$\chi$PT}

\def\tk{\tilde{k}}
\def\ha{\hat{a}}

\def\hF{\hat{F}}
\def\hm{\hat{m}}
\def\hM{\hat{M}}

\def\ringB{\mathring{B}}
\def\ringF{\mathring{F}}
\def\ringm{\mathring{m}}
\def\ringW{\mathring{W}}

\begin{document}


\title{Spectroscopy of SU$(4)$ composite Higgs theory with two distinct fermion representations}



\author{Venkitesh Ayyar}
\affiliation{Department of Physics, University of Colorado, Boulder, Colorado 80309, USA}

\author{Thomas DeGrand}
\affiliation{Department of Physics, University of Colorado, Boulder, Colorado 80309, USA}

\author{Maarten Golterman}
\affiliation{Department of Physics and Astronomy, San Francisco State University,\\
San Francisco, CA 94132, USA}

\author{Daniel C.~Hackett}
\affiliation{Department of Physics, University of Colorado, Boulder, Colorado 80309, USA}

\author{William~I.~Jay}
\affiliation{Department of Physics, University of Colorado, Boulder, Colorado 80309, USA}

\author{Ethan T.~Neil}\email{ethan.neil@colorado.edu}
\affiliation{Department of Physics, University of Colorado, Boulder, Colorado 80309, USA}
\affiliation{RIKEN-BNL Research Center, Brookhaven National Laboratory, \\ Upton, New York 11973, USA}

\author{Yigal~Shamir}
\affiliation{Raymond and Beverly Sackler School of Physics and Astronomy,
Tel~Aviv University, 69978 Tel~Aviv, Israel}

\author{Benjamin Svetitsky}
\affiliation{Raymond and Beverly Sackler School of Physics and Astronomy,
Tel~Aviv University, 69978 Tel~Aviv, Israel}

\date{\today} 

\begin{abstract}
We have simulated the SU(4) lattice gauge theory coupled to dynamical fermions in the fundamental and two-index antisymmetric (sextet) representations simultaneously. 
Such theories arise naturally in the context of composite Higgs models that include a partially composite top quark.
We describe the low-lying meson spectrum of the theory and fit the pseudoscalar masses and decay constants to chiral perturbation theory.
We infer as well the mass and decay constant of the Goldstone boson corresponding to the non-anomalous U(1) symmetry of the model.
Our results are broadly consistent with large-$N_c$ scaling and vector-meson dominance.

\end{abstract}

\maketitle


\section{Introduction}
    \label{sec:intro}

Gauge theories coupled simultaneously to more than one fermion representation---``multirep'' theories---open a new dimension in the study of gauge dynamics.
Apart from the influence of each fermion species on the gauge field and vice versa, phase transitions and symmetry breaking in each species can affect the others dramatically.
Of course, QCD already contains light quarks, strange quarks, and heavy quarks, and the influence of each species on the others is an old and continuing object of QCD calculations.
The difference is that QCD's quarks are all equivalent, in that a tuning of the masses can change one into another.
Fermions in inequivalent representations, on the other hand, enter the dynamics with different strengths irrespective of their masses.

As usual, symmetries offer the clearest perspective on the physics of inequivalent fermions.
 Each species has its maximal flavor symmetry, while no symmetries mix the different species.
If all the fermions are made massless, the chiral symmetries of the species remain distinct.
One symmetry could break spontaneously while others do not.
This is a generalization of the old issue of scale separation, which was originally seen as a possible separation of a chiral scale from the confinement scale of the gauge theory \cite{Kogut:1982fn,Kogut:1984sb,Karsch:1998qj}.
It is possible that inequivalent representations, simultaneously coupled to the gauge field, define independent chiral scales.
This might find expression in the finite-temperature physics of the theory, in the form of distinct phase transitions for each fermion species as well as for the confinement physics of the gauge field.
Alternatively, one phase transition, possibly governed by the largest
quadratic Casimir of the fermion representations, might trigger all the others to occur at the same scale.

We present here the first results of our work on the SU(4) gauge theory with $N_f = 2$ Dirac fermions in each of two distinct representations, the fundamental $\textbf{4}$ and two-index antisymmetric $\textbf{6}$ (a real representation).
We have chosen this model because it is close to a model proposed by Ferretti for a hypercolor theory that yields a composite Higgs boson \cite{Georgi:1984af,Dugan:1984hq} and a partially composite top quark \cite{Kaplan:1991dc}.
Ferretti's model \cite{Ferretti:2014qta} contains 5 Majorana fermions in the sextet representation and 3 Diracs fermions in the fundamental;
simulating this fermion content requires the costly rational hybrid Monte Carlo (RHMC) algorithm, and so, instead, we study the theory with 4 Majoranas (equivalent to 2 Dirac fermions) in the sextet and 2 Diracs in the fundamental.
In Ferretti's model, the massless sextet Majorana fermions $\Psi$ condense to break their chiral symmetry according to $\textrm{SU}(5)\to\textrm{SO}(5)$, whereupon the Standard Model's Higgs multiplet appears as Nambu--Goldstone (NG) bosons; our symmetry breaking scheme is%
\footnote{This scheme is not directly useful
for model building since the $\textrm{SU}(4)/\textrm{SO}(4)$ coset does not accommodate the Higgs field.}
 $\textrm{SU}(4)\to\textrm{SO}(4)$.
The fundamental fermions  $\psi^{(\mathbf{4})}$ in Ferretti's model are brought in so that the theory will possess fermionic baryons constructed as $\psi^{(\mathbf{4})} \psi^{(\mathbf{4})} \Psi$ ``chimera'' bound states, to be used as top partners; they condense (again, in the chiral limit) according to $\textrm{SU}(3)_L\times\textrm{SU}(3)_R\to\textrm{SU}(3)_V$.
In our model, the corresponding symmetry-breaking scheme is $\textrm{SU}(2)_L\times\textrm{SU}(2)_R\to\textrm{SU}(2)_V$.
We believe that our model contains all the qualitative physics of Ferretti's model while offering a laboratory for developing quantitative techniques.

Multirep theories of physical significance are not easy to come by.
Apart from the phenomenological requirements, Ferretti's choice of model is constrained \cite{Ferretti:2013kya} by the simple fact that higher-representation matter fields push gauge theories into the conformal window unless the number of fermions is quite small.
It is essential that the gauge theory of hypercolor exhibit confinement and the concomitant breaking of global symmetries.

In this work we present results from the mesonic sector of the theory, leaving baryonic observables for another paper.
We have already explored the mesonic and baryonic spectrum of the SU(4) gauge theories with only fundamental \cite{DeGrand:2016pur} or only sextet fermions \cite{DeGrand:2015lna}.%
\footnote{A preliminary exploration of the chimera states---using configurations generated with only fundamental dynamical fermions---was presented in \cite{DeGrand:2016mxr}.}
Those results fit nicely into the body of work on QCD and its generalizations to larger values of $N_c$.
The analysis there, similar to QCD studies, related the gauge coupling $\beta$ and hopping parameter $\kappa$ to a physical scale $r_1$ (the Sommer scale) and the quark mass $m_q$, and used the latter as an abscissa for plotting particle masses and decay constants.
Here, of course, the space of bare couplings consists of the gauge coupling $\beta$ and {\em two\/}  hopping parameters $\kappa_4$ and $\kappa_6$ for the two fermion species.
We translate these into the scale parameter $t_0$, derived from the Yang--Mills gradient flow, and the two quark masses $m_4$ and $m_6$.

Our main tool for understanding the meson spectrum is a recent generalization of chiral perturbation theory (\cpt) to the low-energy sector of a two-representation theory \cite{DeGrand:2016pgq}.
This form of \cpt\ provides formulas for masses, decay constants, and chiral condensates at next-to-leading order, with $m_4$ and $m_6$ as independent variables.
These formulas contain an important qualitatively new piece of physics compared to QCD---communication between the different species.
They describe, for instance, the dependence of the masses of the NG bosons of all the broken chiral symmetries on both fermion masses.

Another new feature of the two-representation theory is the existence
of a non-anomalous singlet axial current, and a corresponding singlet NG
boson that must be included in the low-energy chiral theory.
This particle is denoted $\zeta$ in Ref.~\cite{DeGrand:2016pgq} and is of significant phenomenological interest for composite Higgs models \cite{Ferretti:2013kya,Ferretti:2016upr,Belyaev:2016ftv}.
In this work we do not probe this singlet pseudoscalar state
directly.
Nevertheless we extract information about it indirectly,
via its virtual contributions to the properties of the flavored NG bosons associated with chiral symmetry breaking of the individual representations.
In particular, its decay constant in the chiral limit is a parameter in the chiral Lagrangian and thus appears as a fit parameter, allowing us to infer its mass using the leading-order formula.

Besides the pseudoscalar channel, we calculate masses and matrix elements of the lightest vector bosons.
The vector is the lightest narrow resonance in QCD, and its properties are closely related to those of the pseudoscalars within the framework of vector meson dominance (VMD).
We explore the evidence for VMD in our theory and its consequences for the decay width of the vector.
This is of particular phenomenological interest, since in composite Higgs models, the vector resonance is often one of the first signatures expected in collider searches.

The paper is organized as follows.
In Sec.~\ref{sec:lattice} we describe the lattice theory,
the observables we use, and ensembles we generated.
In Sec.~\ref{sec:chiral} we describe our application of \cpt,
including the discretization effects of Wilson fermions,
and our scale setting method which is based on $t_0$.
In Sec.~\ref{sec:pseudoscalars} we present our results for the pseudoscalar spectrum and decay constants, including the flavor singlet $\zeta$.
We present the vector particles in Sec.~\ref{sec:vectors} and use VMD to estimate decay widths.
In Sec.~\ref{sec:conclusions} we discuss our results from the point of view of large-$N_c$ predictions, and present our overall conclusions.

The tables containing the various measured quantities have been collected together in  Appendix~\ref{app:data_tables}.
In Appendix~\ref{app:tech-latt} we explain technical aspects of our analysis of lattice data.
In Appendix~\ref{app:tech-other} we review the definition of the U(1) axial current and of the mass parameter in Wilson \cpt.
Finally, Appendix~\ref{app:renormalization} contains a calculation of perturbative $Z$-factors for the nHYP lattice action with dislocation-suppression.

\section{The Lattice Theory}
    \label{sec:lattice}
\subsection{Symmetries \label{sec:symmetries}}

The chiral symmetry of the fundamental fermions and its expected breaking are the same as in two-flavor QCD.  The specifics of chiral symmetry breaking for the sextet representation are less well-known, so we will discuss them briefly; a more detailed explanation is given in \cite{DeGrand:2015lna,DeGrand:2016pgq}.

The sextet representation of SU(4) is a real representation.
Our model has two Dirac fermions charged under this representation,
$\psi_i^{(\mathbf{6})}, \bar\psi_i^{(\mathbf{6})}$, $i=1,2,$
which are equivalent to four Majorana fermions $\Psi_I$, $I=1,\ldots,4$.
The global symmetry of the continuum theory is thus also SU(4).
Using the language of Majorana fermions,
the bilinear condensate $\vev{\overline\Psi_I \Psi_J}$ is symmetric in its Majorana-flavor indices.
Hence, after spontaneous symmetry breaking one expects the unbroken symmetry to be SO(4) \cite{Peskin:1980gc}.
One consequence of the enlarged symmetry is that $\bar{\psi}^{(\mathbf{6})} \psi^{(\mathbf{6})}$ mesons and $\psi^{(\mathbf{6})} \psi^{(\mathbf{6})}$ diquarks (both gauge-singlet objects) are members of a degenerate multiplet of the unbroken group.

As usual, the chiral symmetries of the theory are explicitly broken by the Wilson term in the lattice action.
The lattice theory thus has the same flavor symmetry as expected in the continuum theory after spontaneous symmetry breaking:
$\SU(2)_V \times \U1_B$ for the fundamental representation and $\SO(4)$ for the sextet.
Our use of Wilson fermions thus assumes that the spontaneous breaking of chiral symmetries is as would be forced by a bilinear condensate, and all measured correlation functions reflect this.

A special feature of the two-representation theory is the existence of a conserved $\U1$ axial current.
While the individual U(1) currents $J_{5\mu}^{(\textbf{4})}$ and $J_{5\mu}^{(\textbf{6})}$ are anomalous, one can form a linear combination $J_{5\mu}$ of these currents that decouples from $F\tilde{F}$.
Condensation of either fermion species then spontaneously breaks the non-anomalous axial symmetry, giving rise to a singlet NG boson that we denote $\zeta$.
We review the normalization of the U(1) current in Appendix~\ref{app:conserved_axial_current}.

\subsection{Lattice action and parameters \label{sec:action}}

Our lattice action contains gauge-field terms and two fermion actions, one for each representation:
\be
S=S_{\textrm{gauge}}+S_F^{({\bf 4})}+S_F^{({\bf 6})}.
\ee
Each fermion action is a Wilson--clover action built of gauge links constructed by nHYP smearing \cite{Hasenfratz:2001hp,Hasenfratz:2007rf}.
In $S_F^{({\bf 6})}$ the smeared links are promoted to the sextet representation \cite{DeGrand:2015lna}.
There are two hopping parameters, $\kappa_4$ and $\kappa_6$.
We set both clover coefficients equal to unity, $c_\text{SW} = 1$, a choice known to work well with nHYP smearing in QCD~\cite{Bernard:1999kc} and with fermions in higher representations \cite{Shamir:2010cq}.

The gauge-field action takes the form
\be
S_{\textrm{gauge}} = \beta S_\text{plaq} + \gamma S_\text{NDS}.
\ee
The first term is the usual plaquette action, while the second is an nHYP dislocation-suppression (NDS) term \cite{DeGrand:2014rwa}, constructed from the nHYP-smeared links.
The NDS term is designed to avoid singularities in the nHYP smearing.
For the present study, we hold the ratio $\gamma/\beta$ fixed at 1/125 and use $\beta$ as a free bare parameter.

Concurrent with the work described here, we are also studying the finite-temperature phase structure of the theory \cite{Ayyar:2017uqh,Ayyar:2018ppa}.  Comparison of the sextet-only limit of this theory to earlier published results \cite{DeGrand:2015lna} shows that the use of the NDS action removes the previously-observed bulk transition from the interesting region of parameter space (see also Ref.~\cite{Hasenfratz:2015ssa}).  In the multirep theory, we see no evidence for a bulk transition anywhere near the range of bare parameters at which we run, indicating that all of our ensembles correspond to the confined continuum phase with broken chiral symmetry.

We extract  masses and decay constants in the usual way from two-point correlation functions.
We denote pseudoscalar masses and decay constants in the representation $r$ by $M_{Pr}$ and $F_{Pr}$, respectively.
The corresponding quantities in the vector channel are denoted by $M_{Vr}$ and $F_{Vr}$.

We define the fermion masses $m_4$ and $m_6$ by imposing
the axial Ward identity (AWI),
\be \label{eq:pcac_continuum}
\partial_\mu \langle 0 | A_{\mu a}^{(r)}(x) \mathcal{O}_r(0) | 0 \rangle
= 2 m_r \langle 0 | P_a^{(r)}(x) \mathcal{O}_r(0) | 0 \rangle \ ,
\ee
where $x\ne 0$, and $a$ is an isospin index.
We use the local unimproved axial current $A_{\mu a}^{(r)}$
and pseudoscalar density $P_a^{(r)}$ in each representation $r$.  For the determination of the AWI mass, we do not renormalize these currents because the mass itself is not a physical observables; based on our perturbative renormalization of these currents described in Appendix~\ref{app:renormalization} (used for calculation of decay constants), the effect of including the renormalization would be small anyway, amounting to a few-percent shift of the masses.  For $\mathcal{O}_r$ we take a pseudoscalar source.
When the distinction between representations is irrelevant,
we will refer to the fermion mass defined by \Eq{eq:pcac_continuum}
as $m_\text{AWI}$.
Further information about our conventions and methods for spectroscopy is given in Appendix~\ref{app:tech-latt}.

\subsection{Scale setting \label{sec:scale}}

We set the scale in our simulations using the flow scale, $t_0$, introduced by L{\"u}scher~\cite{Luscher:2010iy}.
The flow scale is defined by the implicit equation
\be
t^2 \vev{E(t)}|_{t_0} = C,
\ee
where $E(t) = \frac{1}{4} G_{\mu\nu}^aG_{\mu\nu}^a(t)$ is constructed from the clover form of the field strength $G_{\mu\nu}^a$ at flow time $t$.
Here $C$ is a dimensionless number, conventionally \cite{Luscher:2010iy} taken to be 0.3 in QCD.
With this choice, $\sqrt{t_0}$ corresponds to a length scale of $0.14$~fm ({\em i.e.,} an energy scale of 1.4\ GeV) in QCD simulations~\cite{Bazavov:2015yea,DeGrand:2017gbi}.

For an arbitrary gauge theory, any value for $C$ is \emph{a priori} as good as any other.
However, for comparison to existing studies with different gauge groups, it is useful to let $C$ vary with $N_c$.
Arguments from large-$N_c$ QCD, supported by lattice data~\cite{Ce:2016awn,DeGrand:2017gbi}, suggest that $t_0 \sim N_c$ at leading order.
For the SU(4) simulations of this work we therefore use
\be \label{eq:def_t0}
t^2 \vev{E(t)}|_{t_0} = 0.3 \times \frac{4}{3} =  0.4\,.
\ee

Lattice calculations give masses as dimensionless numbers $Ma$ and gradient-flow scales as $t_0/a^2$.
Dimensionless products like $\hat M\equiv M\sqrt{t_0}$ eliminate the lattice spacing $a$, and our tables and figures will display such quantities.
To aid the intuition, one can mentally convert $M\sqrt{t_0}$ to $M/(1.4\ \textrm{GeV})$.

We return to the subject of scale-setting and connect it to \cpt\  in Section~\ref{sec:chiral} below.

\subsection{Ensembles \label{sec:ensembles}}

The ensembles used in this study are listed in Tables~\ref{table:bare_params_16x18}--\ref{table:bare_params_24x48} in Appendix~\ref{app:data_tables}\@.
They fall into three groups.
The short runs with the smallest lattices, of size $V = n_s^3 \times n_t = 16^3 \times 18$, were used for general orientation in the three-dimensional coupling space $(\beta,\kappa_4,\kappa_6)$.
The most important observables for this step were $\sqrt{t_0}$, the scale defined by gradient flow (see Tables~\ref{table:spec_t0_mq_16x18}--\ref{table:spec_t0_mq_24x48}), and the masses $M_{Pr}$ of the pseudoscalars constructed respectively from fermions in the $r={\bf 4}$ and~{\bf6} representations
(see Tables~\ref{table:spec_pseudoscalar_table_16x18}--\ref{table:spec_pseudoscalar_table_24x48}).

The goal of this orientation was to find couplings that give $t_0/a^2=O(1)$ along with pseudoscalar masses that are reasonably light, for subsequent comparison to \cpt.
It turned out that these short runs yielded results that are in themselves usable for the chiral fits to be presented below, and hence we include them in our analysis.

As can be seen in the tables, some ensembles differ in small changes to their $\kappa_r$ values.
Our orientation runs found that $t_0/a^2$ and $aM_{P}$ are often sensitive to these small changes.

We demanded that our ensembles satisfy the criterion $M_{Pr}L>4$ for both representations, where $L=n_sa$ is the spatial size of the lattice.
This is the familiar rule of thumb from QCD, based on the fact that leading-order finite-volume corrections are proportional to $e^{-M_\pi L}$; a more detailed study of finite-volume effects in our data is given in Appendix~\ref{ssec:finite_vol}.
We considered cutting data above a maximum value of $t_0/a^2$ beyond which finite-volume effects severely contaminate determination of the flow scale; such a cut was found to be unnecessary following the cuts on $M_PL$.
We did eliminate ensembles with $t_0/a^2<0.94$ because in these cases the flow did not enter a linear regime. These correspond to a large lattice spacing---in QCD language, $1/a<1.3$~GeV.

Having found interesting regions for study, we continued with high-statistics runs on lattices with $V=16^3 \times 32$.
Finally, we have four extended runs on lattices with $V=24^3\times48$.
These runs were done at large $t_0/a^2$ and small $\hM_P$,
so that the constraint $M_PL>4$ demanded an increase in $L/a$.

\begin{figure}[htb]
\includegraphics[width=0.6\textwidth]{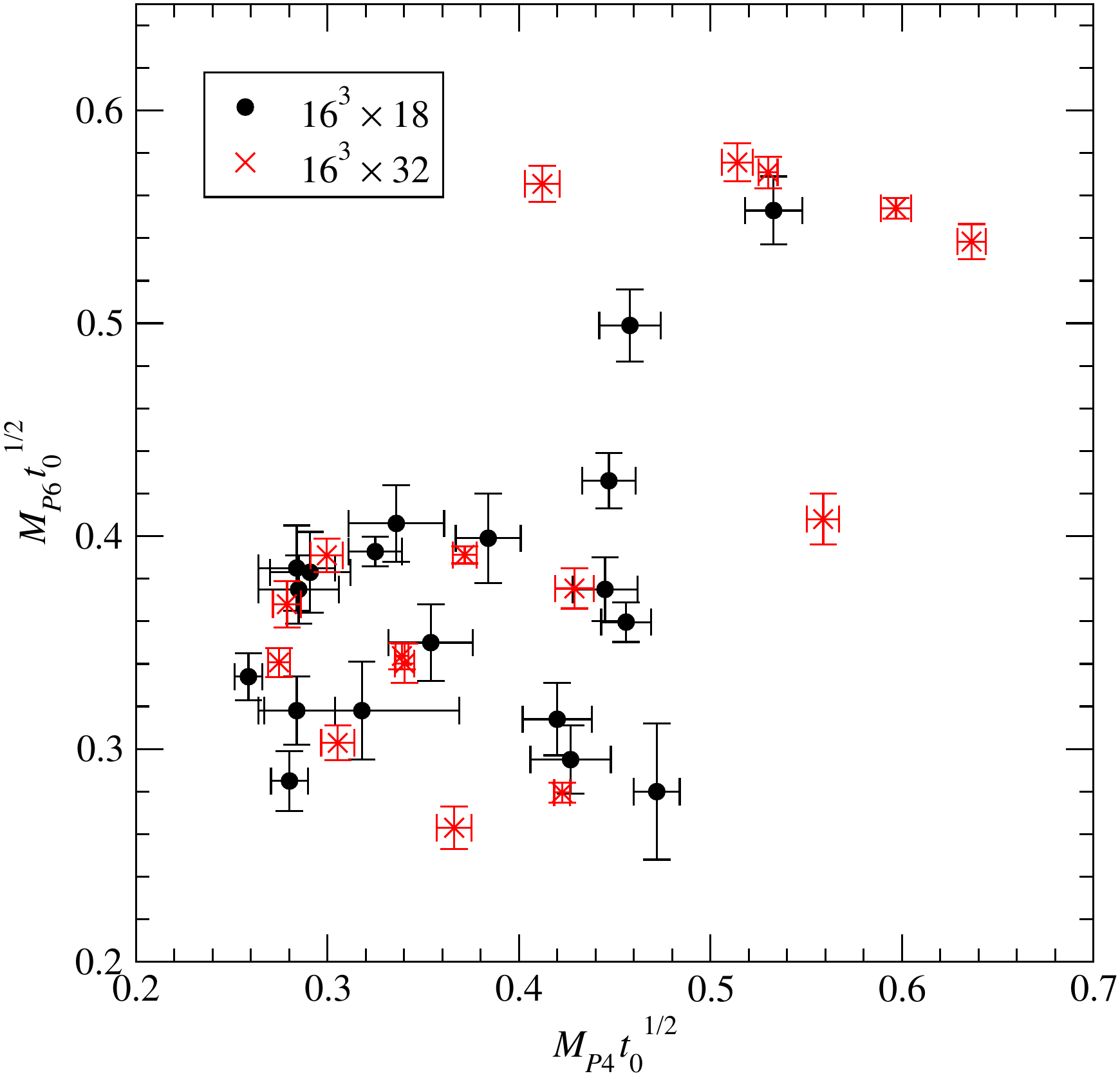}
\caption{Map of our ensembles in the plane of pseudoscalar masses $M_{Pr}$.
These are coarse lattices, with $0.94<\sqrt{t_0}/a<1.41$.
We define arbitrarily $\sqrt{t_0}=(1.4\ \textrm{GeV})^{-1}$ for comparison with QCD.
For most of these ensembles $1/a\simeq1.45$~GeV by this measure.
\label{fig:MP_map_lo_t0}}
\end{figure}

\begin{figure}[htb]
\includegraphics[width=0.65\textwidth]{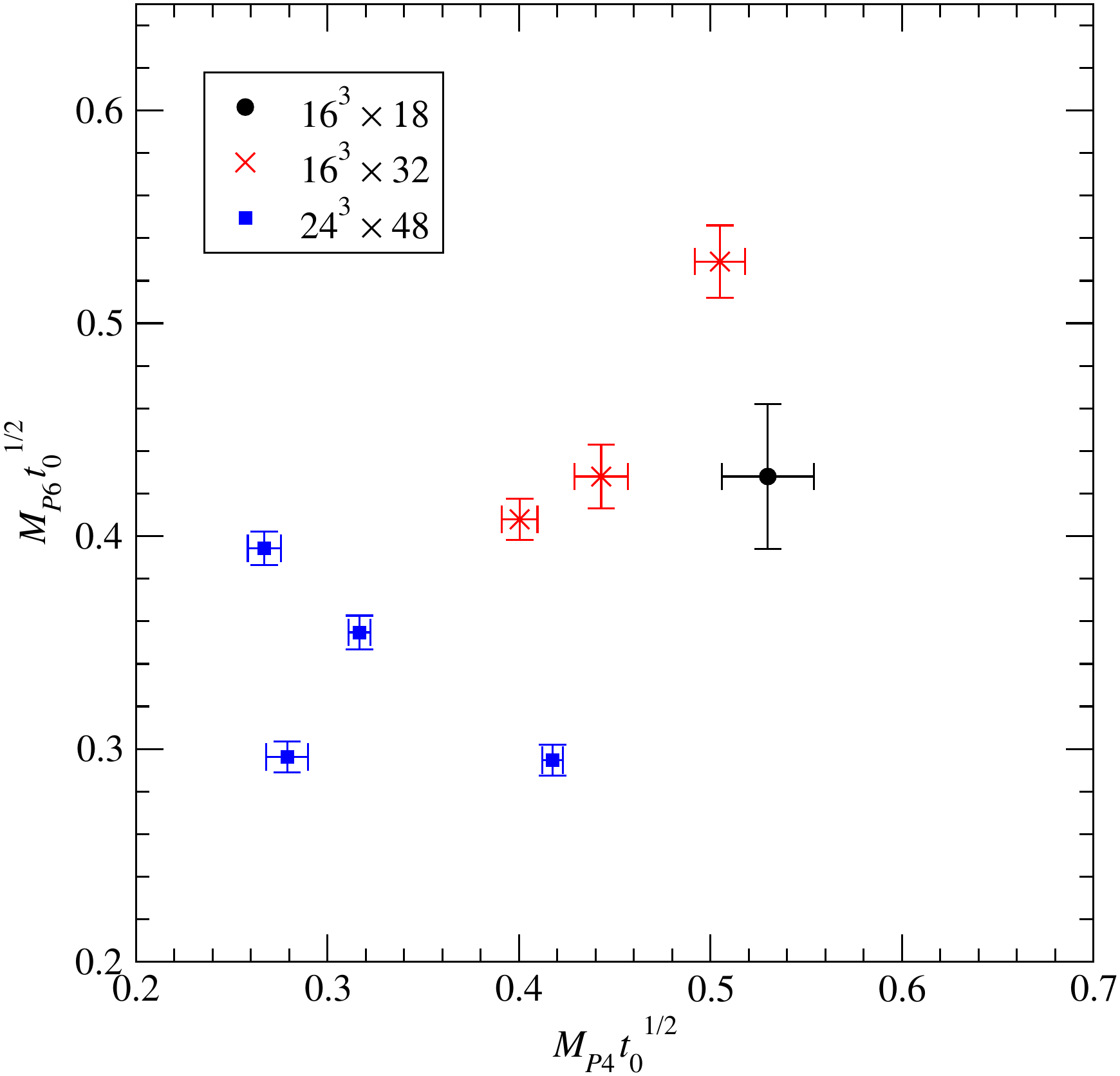}
\caption{Same as Fig.~\ref{fig:MP_map_lo_t0}, but here we plot ensembles on fine lattices,
$\sqrt{t_0}/a>1.41$.
If we fix $\sqrt{t_0}=(1.4\ \textrm{GeV})^{-1}$ then this means $1/a>2$~GeV.  The blue squares are all at $1/a\simeq 2.1$~GeV.
\label{fig:MP_map_hi_t0}}
\end{figure}

The pseudoscalar masses for all the ensembles are given in Tables~\ref{table:spec_pseudoscalar_table_16x18}--\ref{table:spec_pseudoscalar_table_24x48}.
To show our coverage of $M_P$ values, we map them in the $\left(M_{P4},M_{P6}\right)$ plane in Figs.~\ref{fig:MP_map_lo_t0} and~\ref{fig:MP_map_hi_t0}.
The first shows the pseudoscalar masses obtained for  $0.94<\sqrt{t_0}/a<1.41$, which translates to a cutoff of $1.3\ \textrm{GeV}<1/a<2\ \textrm{GeV}$ in QCD language (most are in the neighborhood of $\sqrt{t_0}/a=1.05$, or $1/a=1.45$~GeV).
The second plot represents ensembles in the range $1.41<\sqrt{t_0}/a<1.64$, or $2\ \textrm{GeV}<1/a<2.3\ \textrm{GeV}$.


\section{Chiral Perturbation Theory}
	\label{sec:chiral}

The standard framework for analyzing the light pseudoscalar sector is \cpt.
The generalization of \cpt\ to a theory with fermions in two different
representations was developed in Ref.~\cite{DeGrand:2016pgq}, and
the next-to-leading-order (NLO) results of this work provide the basis
for our fits for the pseudoscalar masses and decay constants.
We will also need Wilson chiral perturbation theory (\wcpt),
the extension of chiral perturbation theory to include
the discretization errors of Wilson fermions
\cite{Sharpe:1998xm,Bar:2003mh,Rupak:2002sm,Aoki:2003yv,Sharpe:2006pu,Golterman:2009kw}.

\subsection{\label{ssec:t0chi} Using a yardstick}

We need a yardstick with which to measure dimensionful quantities
as the fermion masses are varied.  In this paper, we use $\sqrt{t_0}$
for the characteristic length scale of every ensemble.
To measure an observable in units of $t_0$ simply means to multiply it
by the power of $t_0$ that renders it dimensionless.
Since $t_0$ itself admits a chiral expansion \cite{Bar:2013ora},
the resulting dimensionless quantity admits a chiral expansion
whenever the original dimensionful observable does.

To see how this works, consider a gauge theory
with mass-degenerate fermions of mass $m$, all in the same representation.
In continuum \cpt, the NLO expression for the decay constant is
\be
\label{FpiQCD}
  F^\text{NLO} = F \left( 1
  + c\, \frac{2Bm}{8\pi^2 F^2} \log\left( 2Bm/\mu^2 \right)
  + L\, \frac{2Bm}{F^2} \right) \ .
\ee
$B$ and $F$ are the familiar parameters of the LO lagrangian.
[Our normalization convention for the pseudoscalar decay constant
is larger by $\sqrt{2}$ than that commonly used
in the \cpt\ literature.
See \Eq{eq:mpi-def} below.]
We recall that the LO pseudoscalar mass is
\be
\label{MLO}
  (M^2)^\text{LO}=2Bm \ .
\ee
The remaining parameters in \Eq{FpiQCD} are $\mu$, the renormalization scale,
and $L$, which is a (dimensionless) linear combination of the NLO
low-energy constants (LECs), whose value depends on the choice of $\mu$.
The coefficient $c$ of the logarithmic term is a calculable number
that depends only on the fermion representation
and on the number of flavors \cite{Bijnens:2009qm}.

The NLO result for $t_0$ is
\be
\label{t0NLO}
  t_0^\text{NLO} = t_{0,{\rm ch}} \left( 1 + \tk_1 \frac{2Bm}{F^2} \right) \ ,
\ee
where $t_{0,{\rm ch}}$ is the value of $t_0$ in the chiral limit,
and $\tk_1$ is a new LEC.  Notice that this expression depends analytically
on the fermion mass $m$.
As was shown in Ref.~\cite{Bar:2013ora}, logarithmic corrections to $t_0$
occur for the first time at the next-to-next-to-leading order (N$^2$LO).

Combining \Eqs{FpiQCD} and~(\ref{t0NLO}) we obtain the NLO result
for the dimensionless product $\hat F\equiv F\sqrt{t_0}$,
\be
\label{Ft0}
  \hat F^\text{NLO} = F \sqrt{t_{0,{\rm ch}}} \left(
  1 + c\, \frac{2Bm}{8\pi^2 F^2} \log\left( 2Bm t_0 \right)
  + (L+\tk_1/2) \frac{2Bm}{F^2} \right) \ .
\ee
Here we have chosen the renormalization scale to be $\mu=t_{0,{\rm ch}}^{-1/2}$.
LECs are independent of the fermion mass, and to preserve this feature
we rescale them with $t_{0,{\rm ch}}$, for example defining $\ringF=F\sqrt{t_{0,{\rm ch}}}$.
Equation~(\ref{Ft0}) can then be written as
\be
\label{Fhat}
  \hF^\text{NLO} = \ringF \left(
  1 + c\, \frac{2\ringB\ringm}{8\pi^2 \ringF^2} \log\big( 2\ringB\ringm \big)
  + (L+\tk_1/2) \frac{2\ringB\ringm}{\ringF^2} \right) \ .
\ee
The expansion parameter is now $\ringm$, which is
the fermion mass $m$ measured in units of $t_{0,{\rm ch}}$.

Equation~(\ref{Fhat}) is inconvenient because $\ringm$ is not known
for a given ensemble until $t_{0,{\rm ch}}$ is known.
Finding $t_{0,{\rm ch}}$ (in units of $t_0$ of the given ensemble)
requires a complicated fitting procedure that we wish to avoid.
Instead, we opt for rescaling all observables of a given ensemble,
including the fermion mass, with $t_0$ of the same ensemble.
Introducing $\hat m\equiv m\sqrt{t_0}$ we now use \Eq{t0NLO} to
relate the rescaled masses,
\be
\label{mhring}
  \ringm
  = \hm \left( 1 - \tk_1 \frac{Bm}{F^2} \right) \ ,
\ee
which allows us rewrite \Eq{Fhat} as
\be
\label{Fmhat}
  \hF^\text{NLO} = \ringF \left(
  1 + c\, \frac{2\ringB\hm}{8\pi^2 \ringF^2} \log\big( 2\ringB\hm \big)
  + (L+\tk_1/2) \frac{2\ringB\hm}{\ringF^2} \right) \ .
\ee
The transition from $\ringm$ to $\hm$ left no trace,
because the difference is a higher-order correction.
More generally, at NLO the transition from $\ringm$ to $\hm$
can always be absorbed into a redefinition of the LECs.
(A case where the redefinition is non-trivial is the NLO result for
the pseudoscalar mass.)

An appealing feature of \Eq{Fmhat} is that it looks the same as \Eq{FpiQCD}.
In particular, the coefficient of the logarithmic term is unchanged.
The only minor change is that the coefficient of the NLO analytic term
is now $L+\tk_1/2$ instead of $L$.  (At N$^2$LO things would
become technically more complicated, because N$^2$LO logarithmic corrections
for $t_0$ would have to be incorporated as well.)
It can be checked that this nice feature generalizes to an arbitrary fermion
content.  In the NLO fit formulae
that we present below, all the logarithmic terms
will thus have the same coefficients as in the usual continuum NLO
expressions \cite{Bijnens:2009qm,DeGrand:2016pgq}.

\subsection{\label{ssec:wchpt} Wilson chiral perturbation theory}

The extension of continuum chiral perturbation theory to include
the discretization errors of Wilson fermions goes under the name
of Wilson chiral perturbation theory, or \wcpt.
In the light pseudoscalar sector,
\wcpt\ allows us to extrapolate both to the chiral limit, $m\to 0$,
and to the continuum limit, $a\to 0$.
\wcpt\ comes in two variants, depending on the choice of
a power counting scheme.  In this paper we follow the ``generic small mass,''
or GSM, power counting, defined by
\be
\label{GSM}
  p^2 \sim m \sim a \ ,
\ee
where $p$ is an external momentum, $m$ is the fermion mass, and $a$ is the lattice spacing, all measured in terms of a typical hadronic scale.
The alternative power counting scheme, known as the ``large cutoff effects,''
or LCE, power counting, is defined by
\be
\label{LCE}
  p^2 \sim m \sim a^2 \ .
\ee
The GSM scheme is appropriate when the fermion mass is not too small,
and $O(a^2)$ effects may be considered as subleading corrections.
(We must, of course, remain within the chiral regime, meaning that
$\hm=m\sqrt{t_0}$ is small.)  In particular, our determination of the
critical surface $\kappa_r^c$, where the mass of fermions in
representation $r$ vanishes, is done via extrapolation
from the GSM regime.
As a result, we do not probe the possible existence of an Aoki phase.
For more details, see Appendices~\ref{ssec:kappa_crit}
and~\ref{ssec:appendix_wcpt}.

The fermion mass appearing in the LO lagrangian of \wcpt\ is the so-called
\emph{shifted mass}, defined by
\be
\label{mshift}
  m_\text{shifted} = m_\text{ctm} + aW_0/B ,
\ee
where $m_\text{ctm}$ is the fermion mass of continuum \cpt,
and $W_0$ is a new LEC from \wcpt.  The difference between the shifted
and continuum masses vanishes in the continuum limit.
For this lattice study, we need to know how the shifted mass $m_\text{shifted}$
compares to the fermion mass $m_\text{AWI}$ measured in our simulations
via the axial Ward identity \Eq{eq:pcac_continuum}.
As was shown in Ref.~\cite{Aoki:2009ri},
$m_\text{shifted} = m_\text{AWI}$, up to corrections that are
higher order in either of the above power counting schemes.
In view of the important role that this result plays in our analysis,
we briefly summarize the derivation of Ref.~\cite{Aoki:2009ri}
in Appendix~\ref{ssec:appendix_wcpt}.  For our chiral fits we thus define
\be
\label{mhAWI}
  \hm = m_\text{AWI} \sqrt{t_0} \ .
\ee

The last ingredient we need for our fits is the lattice spacing.
Since we are measuring all dimensionful quantities in units of $t_0$,
it is natural to adopt a \emph{mass-dependent} prescription,
and to measure also the lattice spacing in units of $t_0$.
We thus introduce
\be
\label{ahat}
  \ha \equiv a/\sqrt{t_0} \ .
\ee
The Wilson discretization effects of any hatted (dimensionless) observable
will be accounted for by an expansion in $\ha$.

In QCD, it is common to choose a \emph{mass-independent} scale-setting
prescription, whereby the lattice spacing is a function
of the bare coupling $\beta$, but is independent of the bare fermion masses
(see for example
Refs.~\cite{Bazavov:2009bb,Bruno:2016plf}).
In brief, for every constant-$\beta$ plane, this procedure requires
finding the point where certain dimensionless quantities
(such as $M_\pi/F_\pi$ and $M_K/F_K$) attain their real-world values.
The value in lattice units of some dimensionful observable
at the reference point is then used to determine the lattice spacing
in physical units.

Here we have opted for mass-dependent scale setting
because of several important differences.
First, the BSM context does not provide us with any experimental results
that could be used to define a reference point.
This problem might be circumvented by invoking the chiral limit
as a reference point on each constant-$\beta$ plane.  This, however,
has the undesirable feature that the scale setting procedure
would necessarily involve an extrapolation.

Second, in our model, as in many other models that have been studied
in the BSM context, we observe a rapid change of $t_0/a^2$
with the fermion mass, especially when the latter becomes light.
Moreover, this phenomenon is quite general, and is seen for virtually any
quantity that might be used to set the scale; its proper interpretation
is thus that the lattice spacing itself is changing rapidly.
The underlying reason is that, in comparison with QCD, BSM theories
tend to have a large number of fermionic degrees of freedom,
which have a strong screening effect on the bare coupling.
When we consider the dependence of a hatted quantity, such as $\hM_P$,
on the hatted mass parameter, $\hm$, we expect to see some deviations from
the continuum values, but such scaling violations should be small
when the bare coupling is small enough.
By contrast, as explained above, the lattice spacing $\ha$ itself can vary rapidly
with the fermion mass(es).  By using the mass-dependent scale-setting
prescription of \Eq{ahat} we can incorporate this effect into our analysis.
As we will see, the remaining scaling violations in the hatted quantities are small and amenable to \wcpt.

\subsection{\label{ssec:chifit} Summary of \cpt\ formulae}

Our central fits below will include terms through NLO in the GSM power counting.
These formulae depend exclusively on the dimensionless quantities
we have introduced in the previous subsections.
The NLO expressions for the pseudoscalar masses of the two representations are
\begin{align}
\label{eq:chipt_mp4sq}
(\hM_{P4}^2)^\text{NLO} &= 2 \hm_4 \ringB_4
	\left(
		1 +
		L^M_{44} \hm_4 +
		L^M_{46} \hm_6 +
		\frac{1}{2} \Delta_4 -
		\frac{4}{5} \Delta_\zeta
	\right) \\
	& \quad +
	\ringW^M_{44} \ha \hm_4 +
	\ringW^M_{46} \ha \hm_6 +
	\ringW^M_4 \ha^2 \ ,
\allowdisplaybreaks\nonumber\\
\label{eq:chipt_mp6sq}
(\hM_{P6}^2)^\text{NLO} &= 2 \hm_6 \ringB_6
	\left(
		1 +
		L^M_{66} \hm_6 +
		L^M_{64} \hm_4 -
		\frac{1}{4} \Delta_6 -
		\frac{1}{5} \Delta_\zeta
	\right) \\
	& \quad +
	\ringW^M_{66} \ha \hm_6 +
	\ringW^M_{64} \ha \hm_4 +
	\ringW^M_6 \ha^2 \ ,
\nonumber
\end{align}
while the expressions for the decay constants are
\begin{align}
\label{eq:chipt_fp4}
(\ringF_{P4})^\text{NLO} &= \ringF_4
	\left(
		1 +
		L^F_{44} \hm_4 +
		L^F_{46} \hm_6 -
		\Delta_4
	\right)
	+
	\ringW^F_4 \ha \ ,
\\
\label{eq:chipt_fp6}
(\ringF_{P6})^\text{NLO} &= \ringF_6
	\left(
		1 +
		L^F_{66} \hm_6 +
		L^F_{64} \hm_4 -
		2 \Delta_6
	\right)
	+ \ringW^F_6 \ha \ .
\end{align}
The one-loop chiral logarithms enter as
\begin{align}
\label{eq:Delta}
    \Delta_4 &= \frac{2 \hm_4 \ringB_4}{8 \pi^2 \ringF_4^2}
                \log \left( 2 \hm_4 \ringB_4 \right) \ ,
\allowdisplaybreaks\\
    \Delta_6 &= \frac{2 \hm_6 \ringB_6}{8 \pi^2 \ringF_6^2}
                \log \left( 2 \hm_6 \ringB_6 \right) \ ,
\allowdisplaybreaks\nonumber\\
    \Delta_\zeta &= \frac{\hat{M}_\zeta^2}{8 \pi^2 \ringF_\zeta^2}
                \log \left( \hat{M}_\zeta^2 \right) \ ,
\nonumber
\end{align}
where the dimensionless mass-squared of the singlet NG boson
is {\em defined} by
\be \label{eq:mzeta}
  \hat{M}_\zeta^2 = \frac{8}{5} \left(\frac{ 2 \ringF_4^2 \hm_4 \ringB_4
  + \ringF_6^2 \hm_6 \ringB_6 }{\ringF_\zeta^2} \right) \ .
\ee
This corresponds to the LO result of Ref.~\cite{DeGrand:2016pgq},
rescaled by $t_0$.
Further technical details related to the $\zeta$ and our conventions for
the conserved axial current appear in
Appendix~\ref{app:conserved_axial_current}.

The most important parameters in the expressions above are the LO LECS
of the continuum two-representation theory (rescaled by $\sqrt{t_{0,{\rm ch}}}$):
$\ringB_4$, $\ringB_6$, $\ringF_4$, $\ringF_6$, and $\ringF_\zeta$.
The dimensionless parameters $L^M_{rs}$ and $L^F_{rs}$, $r=4,6,$
are linear combinations of the continuum NLO LECs and of similar NLO LECs
originating from the chiral expansion of the flow scale [cf.\ \Eq{t0NLO}].
The general form of the analytic NLO continuum terms was discussed
in \cite{DeGrand:2016pgq}.
Because we do not have enough independent quantities to distinguish
the individual NLO LECs, we instead consider $L^M_{rs}$ and $L^F_{rs}$
as the parameters for the fit.
Finally, the various $\ringW$ parameters account for the NLO analytic terms
of \wcpt\ in the GSM power-counting scheme.
Overall, these formulae contain 21 undetermined parameters,
which we will fit below using 172 correlated points of data: four data points
for each of our 43 ensembles.

We have not presented NLO fit formulae for the mass and decay constant
of the singlet NG boson $\zeta$.  We do not make use
of these formulae in this work because we have not calculated
fermion-disconnected diagrams, which is technically challenging,
and so we do not have direct access to the singlet sector.
Nevertheless, through their dependence on $\Delta_\zeta$,
virtual $\zeta$ loops contribute to the masses and decay constants
of the other NG bosons at NLO.  In the next section we will explore
what can be learned about the singlet sector from this effect.

Another interesting quantity is the chiral condensate in each representation.
At lowest order in \cpt\ (equivalently, in the corresponding chiral limit,
$\hm_r\to 0$), the fermion condensate per flavor is given by
\be \label{eq:condensates}
\hat{\Sigma}_r = -\ringB_r \ringF_r^2 \ .
\ee
Instead of measuring the condensates directly---a formidable task
with Wilson fermions---we will make use of \Eq{eq:condensates}
to extract their values in the (double) chiral limit
from our analysis of the pseudoscalar masses and decay constants.

\section{Pseudoscalar mesons}
	\label{sec:pseudoscalars}

\subsection{Masses and decay constants}

\begin{figure}[htb]
\includegraphics[width=0.8\columnwidth]{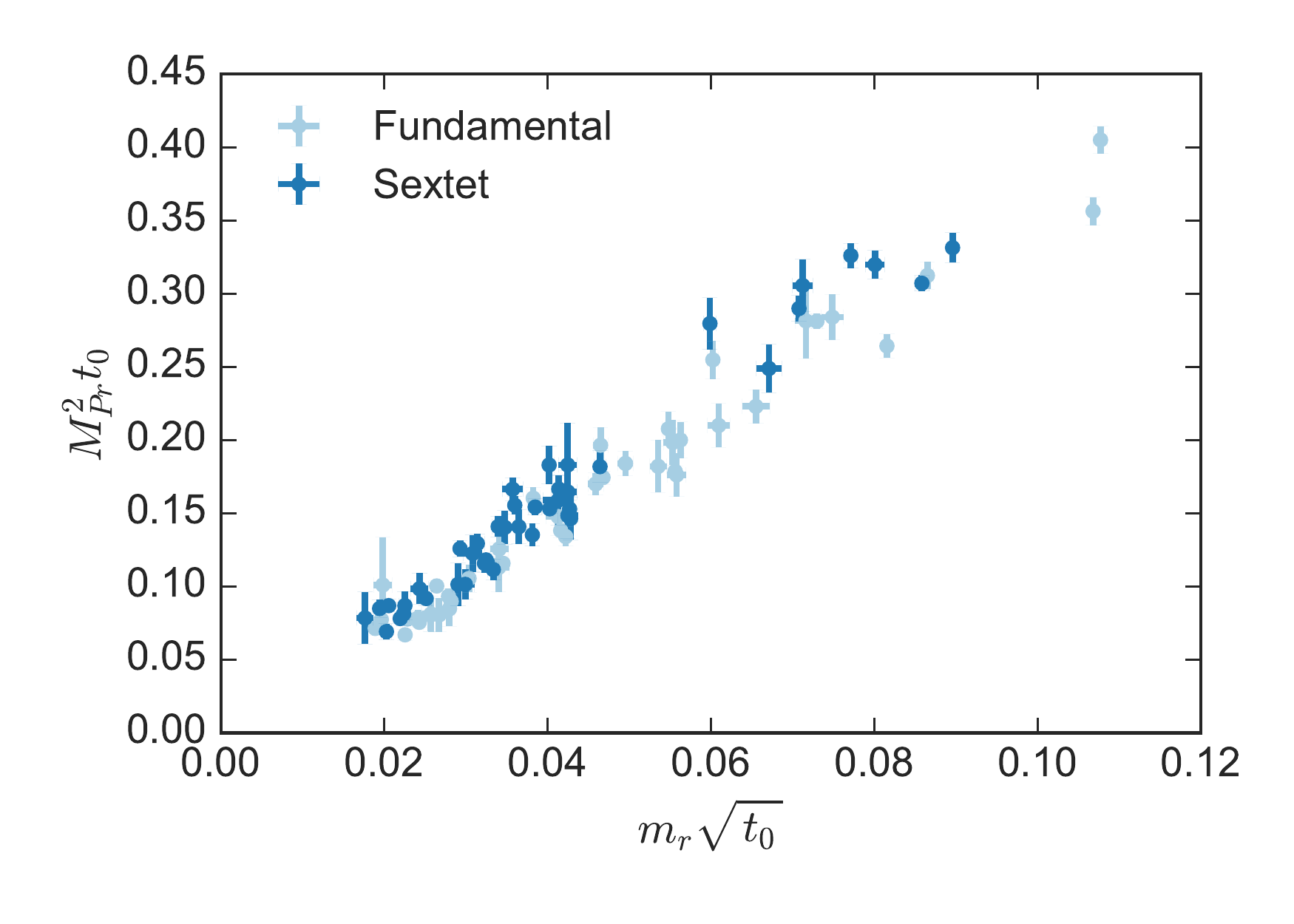}
\caption{Squared mass of the two pseudoscalar species, each plotted against the AWI mass of the corresponding fermion species, all in units of the flow scale $t_0$.}
\label{fig:gmor}
\end{figure}

We begin with the pseudoscalar mesons, which become NG bosons
in the chiral limit.  For a first look, we plot in Fig.~\ref{fig:gmor}
the squared masses $\hM_{Pr}^2$.
The sextet mass $\hM_{P6}$ is plotted against the AWI mass $\hm_6$
of the sextet fermion, ignoring the dependence on the fundamental fermion
mass $\hm_4$, and likewise for $\hM_{P4}$, plotted against $\hm_4$.
As expected from leading-order chiral perturbation theory, the overall behavior of each squared mass is approximately linear.
One supposes that the scatter around the straight lines is due to the hidden dependence on the other fermion mass, as well as corrections from NLO and from lattice artifacts.
We will examine this hypothesis shortly.

\begin{figure}[htb]
\includegraphics[width=0.8\columnwidth]{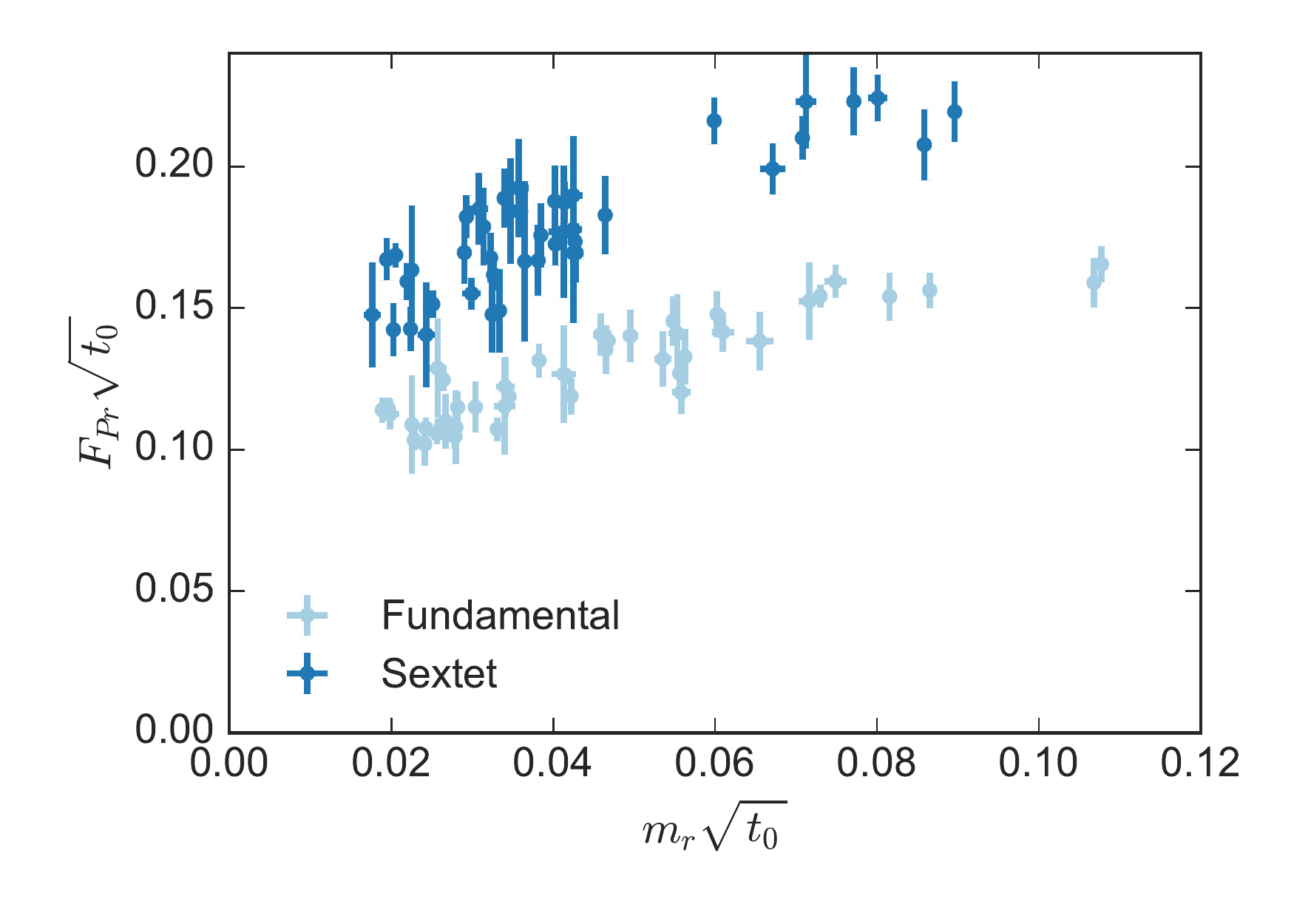}
\caption{Decay constant of each pseudoscalar species plotted against the mass of the corresponding fermion species, in units of the flow scale $t_0$.}
\label{fig:fpi}
\end{figure}

The pseudoscalar decay constants are defined by
\be \label{eq:mpi-def}
\big\langle 0 \big| A_{4a}^{(r)} \big| P_b^{(r)} \big\rangle
= \delta_{ab} M_{Pr} F_{Pr} \ ,
\ee
at zero spatial momentum, which is the convention that gives
$F_\pi \simeq 130$~MeV in QCD.
We calculate $F_{Pr}$ with the procedure described in
Appendix~\ref{ssec:decay_constants}, renormalizing according to the analogue of
\Eq{defFV}.
We plot the (rescaled) decay constants $\hF_{Pr}$ in Fig.~\ref{fig:fpi}.
The data show a steady rise with $\hm_r$.
The same qualitative behavior is seen in QCD, where the pion decay constant is an increasing function of the quark mass.

We have presented in Sec.~\ref{ssec:chifit} the predictions of \cpt\ in NLO for pseudoscalar observables.
We conduct a joint fit of the four observables $\hM_{Pr}^2$ and $\hF_{Pr}$ to the NLO formulae of Eqs.~(\ref{eq:chipt_mp4sq})--(\ref{eq:chipt_fp6}).
On each ensemble, we use single-elimination jackknife to construct the $6\times 6$ correlation matrix among pseudoscalar masses, decay constants, and AWI masses of the fermions.
These correlation matrices enter into the $\chi^2$ that is minimized for the fit.
We do not include correlations with the flow scale $t_0$, which has negligible error compared to the other quantities we extract.

The full NLO fit to 21 parameters
 and $172-21=151$ degrees of freedom gives $\chi^2$/DOF = 0.48.
Table~\ref{table:chipt_central_fit_table} contains the resulting values for the LECs and demonstrates the presence of important lattice artifacts in our data.
For the masses, the most significant terms are the
$\mathcal{O}\left( \hat{m}_{r} \hat{a} \right)$ artifacts, in the same
representation.  For the decay constants, the
$\mathcal{O}\left( \hat{a} \right)$ artifacts are also significant.
From an empirical perspective, these four NLO Wilson terms form a necessary minimal set of artifact terms for modeling the data.

\begin{table}[h]
\centering
\setlength{\tabcolsep}{12pt} 

\begin{tabular}{lld}
\toprule
LO:	&$\ringB_4 \rule{0ex}{3ex}$ & 2.4(2) \\
	&$\ringB_6           $ & 2.7(1) \\
	&$\ringF_4           $ & 0.114(7) \\
	&$\ringF_6           $ & 0.17(1) \\
	&$\ringF_\zeta       $ & 0.16(2) \\ [5pt]
continuum NLO:&$L^F_{44}		$& 3.4(5) \\
	&$L^F_{46}              $ & 1.4(6) \\
	&$L^F_{64}              $ & 0.3(4) \\
	&$L^F_{66}              $ & 3.9(5) \\
	&$L^M_{44}              $ & 0.1(7) \\
	&$L^M_{46}              $ & 3.(1) \\
	&$L^M_{64}              $ & 0.4(7) \\
	&$L^M_{66}              $ & 0.5(7) \\[5pt]
lattice NLO:&$\ringW_{4}^{F}  	   $ & -0.055(6) \\
	&$\ringW_{6}^{F}           $ & -0.08(1) \\
	&$\ringW_{4}^{M}           $ & 0.01(1) \\ 
	&$\ringW_{44}^{M}          $ & -1.9(3) \\ 
	&$\ringW_{46}^{M}          $ & -0.6(3) \\ 
	&$\ringW_{6}^{M}           $ & 0.001(9) \\ 
	&$\ringW_{64}^{M}          $ & 0.1(2) \\ 
	&$\ringW_{66}^{M}          $ & -2.5(4) \\ 
\hline\hline
\end{tabular}
\caption{Parameter values from a joint fit to the full NLO \cpt\ formulae.}
\label{table:chipt_central_fit_table}
\end{table}

\begin{figure}[htb]
	\includegraphics[width=0.85\columnwidth]{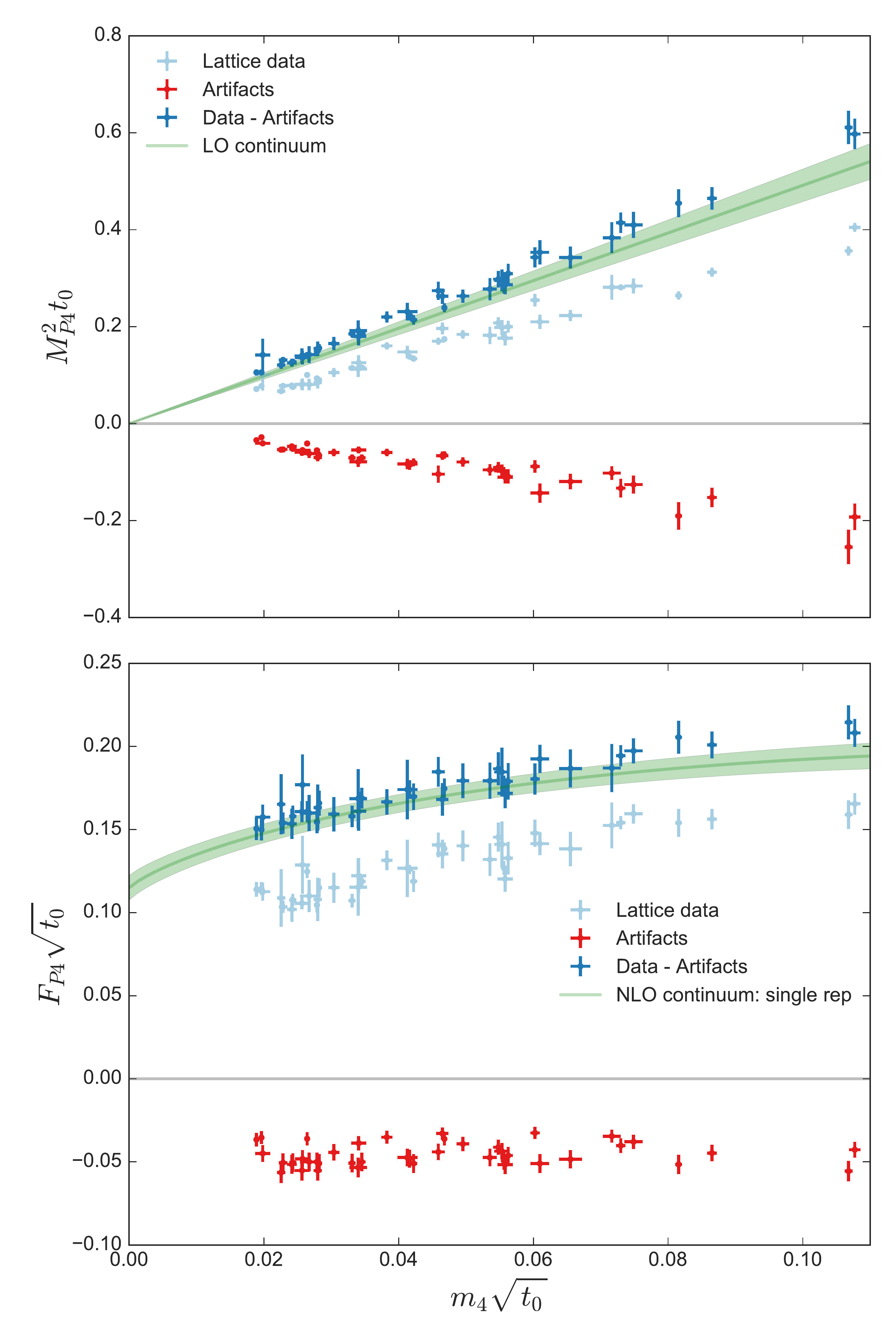}
	\caption{Breakdown of the contribution of lattice artifacts in the joint fit to \cpt\ for the fundamental masses and decay constants.\label{fig:breakdowns_chipt_fund}}
\end{figure}

\begin{figure}[htb]
	\includegraphics[width=0.85\columnwidth]{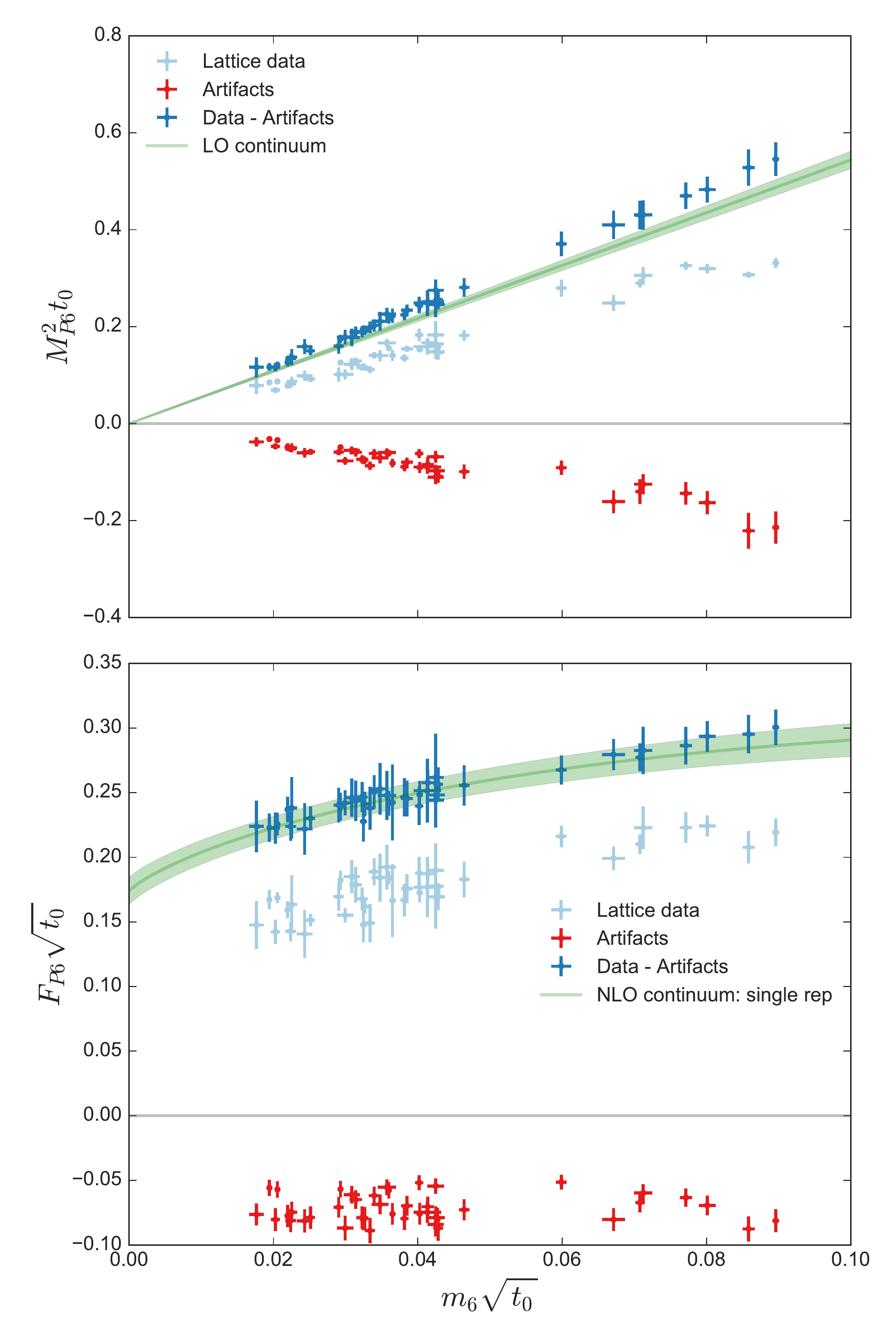}
	\caption{Breakdown of the contribution of lattice artifacts in the joint fit to \cpt\ for the sextet masses and decay constants.\label{fig:breakdowns_chipt_sextet}}
\end{figure}

Figures~\ref{fig:breakdowns_chipt_fund} and~\ref{fig:breakdowns_chipt_sextet}
illustrate the sizes of the Wilson artifacts (red) emerging from this fit.
In these figures, the ``corrected" data (dark blue) result from subtracting
the lattice artifacts from the data (light blue),
allowing us to extrapolate to the continuum limit, $\ha\to 0$.
The corrected data follow fairly well the tree-level formula for
the pseudoscalar masses and the continuum NLO result for the decay constants,
respectively, both indicated by green bands.
(The bands represent $1\sigma$ in the fit parameters.)
In order to display a smooth curve for the continuum NLO result for the
decay constants, we have included only the same-representation terms when
drawing the green band (indicated by ``continuum NLO SREP'' in the figure).
The remaining scatter and deviation in the subtracted data (dark blue)
is evidence of coupling between the representations.

Table~\ref{table:chipt_central_fit_table} demonstrates that all five
leading-order LECs are well-determined by the NLO fit.
We note that the singlet decay constant
$\ringF_\zeta$ is larger than $\ringF_4$ and similar in size to $\ringF_6$.
Because measurement of chiral logarithms is known to be a difficult task in QCD
studies, we return to the question of the stability of this result below.

Turning our attention to the NLO LECs, we examine the communication between
the representations.  The ratios $L^M_{46}/L^M_{44}$ and $L^F_{46}/L^F_{44}$
quantify the relative influence of the sextet fermions
on $\hM_{P4}^2$ and $\hF_{P4}$, respectively, in the continuum theory.
Similarly, the ratio $\ringW_{46}^{M} / \ringW_{44}^{M}$ measures
the relative influence of the sextet artifact term compared
to the fundamental artifact term in $\hat{M}_{P4}^2$.
Taking into account correlations, the following ratios are different from zero at the $2\sigma$ level,
\begin{align}
L^F_{46}/L^F_{44}           &= +0.4(2) \ ,
\\
\ringW_{46}^{M} / \ringW_{44}^{M}  &= +0.30(15) \ .
\end{align}
The converse influence of the fundamentals upon the sextets follows
from exchanging \hbox{$(4 \leftrightharpoons 6)$}.
The ratios $L^M_{64}/L^M_{66}$, $L^F_{64}/L^F_{66}$,
and $\ringW_{64}^{M}/\ringW_{66}^{M}$ are all consistent with zero.
Despite the large uncertainties, this suggests that the sextets influence
the fundamentals significantly, while the converse is not true.
The same qualitative conclusion is also evident, for instance,
in the NLO continuum behavior of the decay constants.
Figure~\ref{fig:breakdowns_chipt_sextet} shows that subtracted data (dark blue)
are, to good approximation, a smooth function of $\hat{m}_6$ only.
In contrast, the corresponding fundamental result in
Fig.~\ref{fig:breakdowns_chipt_fund} (also in dark blue) exhibits
a conspicuous jaggedness, indicating important dependence
on the sextet fermion mass.

\subsection{Stability of the NLO fit} \label{ssec:stability}

\begin{figure}[htb]
\includegraphics[width=\textwidth]{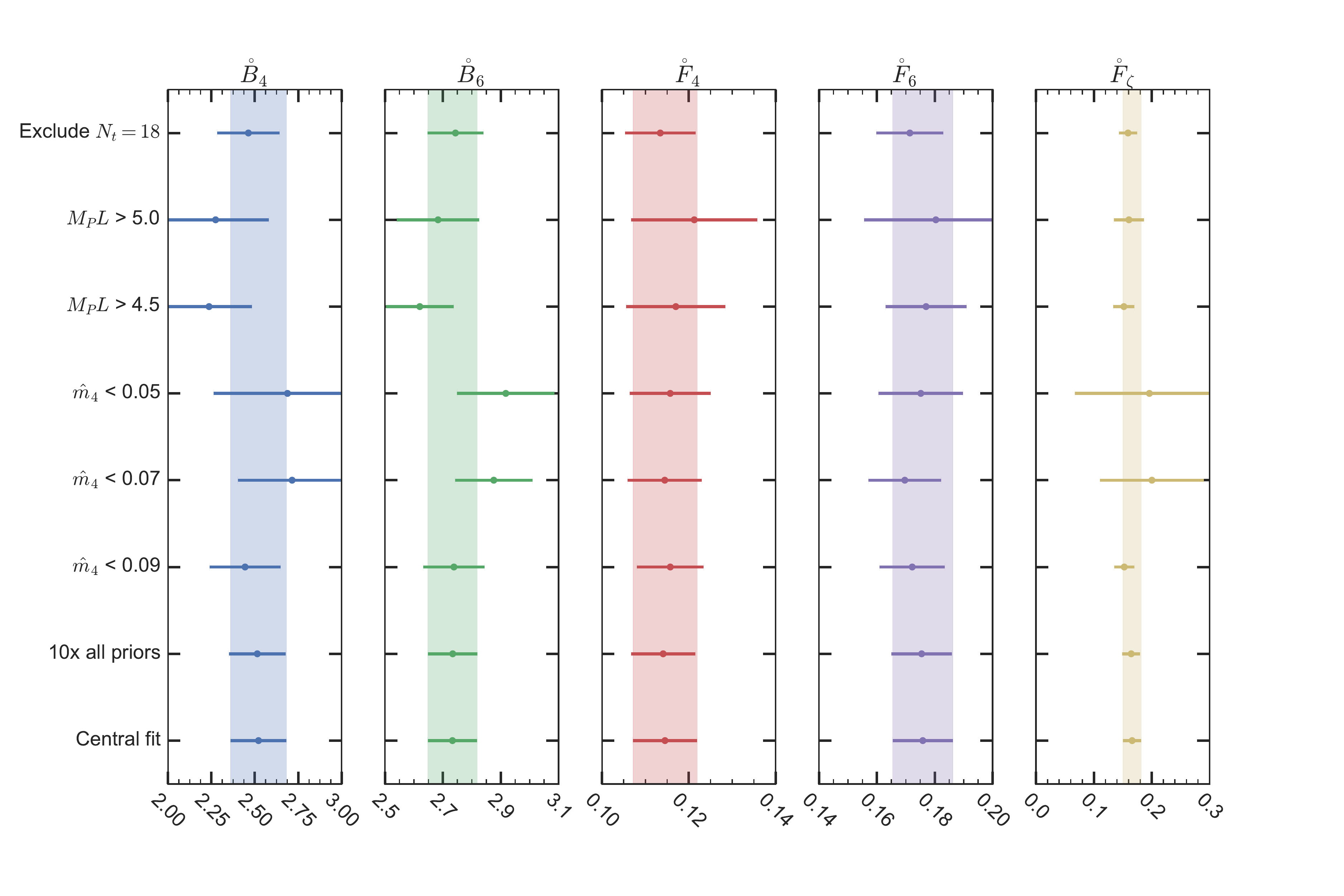}
\caption{
	Exploring the stability of leading-order LECs in chiral fits.
	We take the NLO result to define our central values,
	which appear at the bottom of each column.  The variations
	are described in the text.
\label{fig:chipt_fit_stability}}
\end{figure}

In this subsection we explore the stability of the NLO fit.
First, since we are using priors to ensure convergence
of the non-linear fitting procedure, it is important to verify
that our results were not biased by them.  To this end, we have redone
the fit using the results of the first fit as initial guess,
while multiplying the width of all priors by 10.
Figure~\ref{fig:chipt_fit_stability} shows the results of both fits
for the 5 LO LECs in the two lines at the bottom.  The results are indistinguishable, indicating that the LO LECs were not influenced by the priors.  (The same is also true for the NLO LECs.)

The chiral fit provides \emph{a-posteriori} justification for the use of
the GSM power-counting scheme, where $\mathcal{O}(a^2)$ terms are
not part of the LO lagrangian.
Both fermion masses in our ensembles lie roughly in the range
\be
0.02 \lesssim \hm_r \lesssim 0.10 \ .
\ee
The range of lattice spacings we explore is
\be
  0.4 \lesssim \ha^2 \lesssim 1.1 \ .
\ee
[Recall that our scale setting implies
$\ha^2=a^2/t_0$ by definition, see \Eq{ahat}.]
The $\mathcal{O}(m_r)$ contribution to the pseudoscalar masses
is $2 \ringB_r \hm_r$, while the $\mathcal{O}(a^2)$
contribution is $\ringW_{r}^{M} \ha^2$.
For our fermion masses and lattice spacings,
these contributions lie approximately within the following ranges
\begin{align}
2 \ringB_r \hm_r:&\ \  5 \times [ 0.02 , 0.1] \approx [0.1, 0.5] \ ,
\\
\ringW_r^{M} \ha^2:&\ \ 0.01 \times [0.4, 1.1] \approx [0.004, 0.01] \ .
\end{align}
We see that the $\mathcal{O}(m_r)$ terms are at least an order of magnitude
larger than the $\mathcal{O}(a^2)$ terms, showing that the GSM power counting
is the appropriate one (as long as this picture is not upset by large N$^2$LO corrections, see below).

Measurement of the LECs also provides information about the convergence
of the chiral expansion.
With our convention for the decay constant, the expansion parameters
of continuum \cpt\ are $\xi_r \equiv 2 \ringB_r \hat{m}_r /8 \pi^2 \ringF_r^2$\@.
With the central-fit values for the LECs, these fermion masses correspond
 to the following ranges for the expansion parameters,
\begin{align}
0.09 \lesssim  \xi_4 &\lesssim 0.50 \label{eq:cpt4_convergence} \ ,
\\
0.04 \lesssim  \xi_6 &\lesssim 0.20 \label{eq:cpt6_convergence} \ .
\end{align}
We see that the the maximum of the sextet expansion parameter $\xi_6$ is
smaller by a factor of 2.5 than the fundamental expansion parameter $\xi_4$.
The main reason is that $\ringF_6$ is significantly larger than $\ringF_4$,
as might be expected based on the relative dimension of the two representations (see Sec.~\ref{ssec:largeN}).

It is quite plausible that $\xi_6$ is sufficiently small that the expansion
in $m_6$ converges well over our entire ensemble set.  The same
may not be true for $\xi_4$, whose value can be as large as 0.5.
In the next three lines of Fig.~\ref{fig:chipt_fit_stability}
we study the influence on the LO LECs of dropping ensembles at the high end of the $\hm_4$ range:
$\hm_4>0.09$, $>0.07$, and finally $>0.05$.
We see that truncating our data set has only a modest effect on the $\ringF_r$ and $\ringB_r$ parameters.
On the other hand, since we only obtain $\ringF_\zeta$ through NLO logarithms,
it is not surprising that the increase in the error bar of $\ringF_\zeta$
is much more pronounced.  Indeed, when we restrict to $\hm_4<0.05$,
$\ringF_\zeta$ is only $2\sigma$ away from zero.

The next two lines in Fig.~\ref{fig:chipt_fit_stability} investigate the possible influence of finite-volume effects
on our central analysis; further discussion appears in Appendix~\ref{ssec:finite_vol}.  The minimum cutoff on $M_P L$ in data used in the central fit is varied from its initial value of 4.0 in our main analysis to 4.5 and 5.0, excluding more data that may be expected to have the largest finite-volume contamination.  Finally, in the top line we repeat our fit with all $V=16^3 \times 18$ ensembles excluded from the analysis, in order to test for systematic effects in our correlator analysis due to the smaller time extent.  No significant change to our results is seen in any case.

The main systematic uncertainty about this non-QCD system is the neglect
of N$^2$LO corrections.
We do not really know how high can we go in $\xi_4$
and $\xi_6$ if we want these corrections to remain below a certain level.
While our
stability tests give us some insight, we do not have enough data for a
quantitative study of N$^2$LO\@.
Nevertheless, we take the smallness of $\xi_6$ and our stability tests
on $\hm_4$ as evidence that the data are in the regime where NLO ChPT applies, even if we do not have enough information to quantify the corresponding systematic error.

\subsection{The singlet Goldstone boson $\zeta$ \label{ssec:zeta}}

\begin{figure}[htb]
\includegraphics[width=0.8\textwidth]{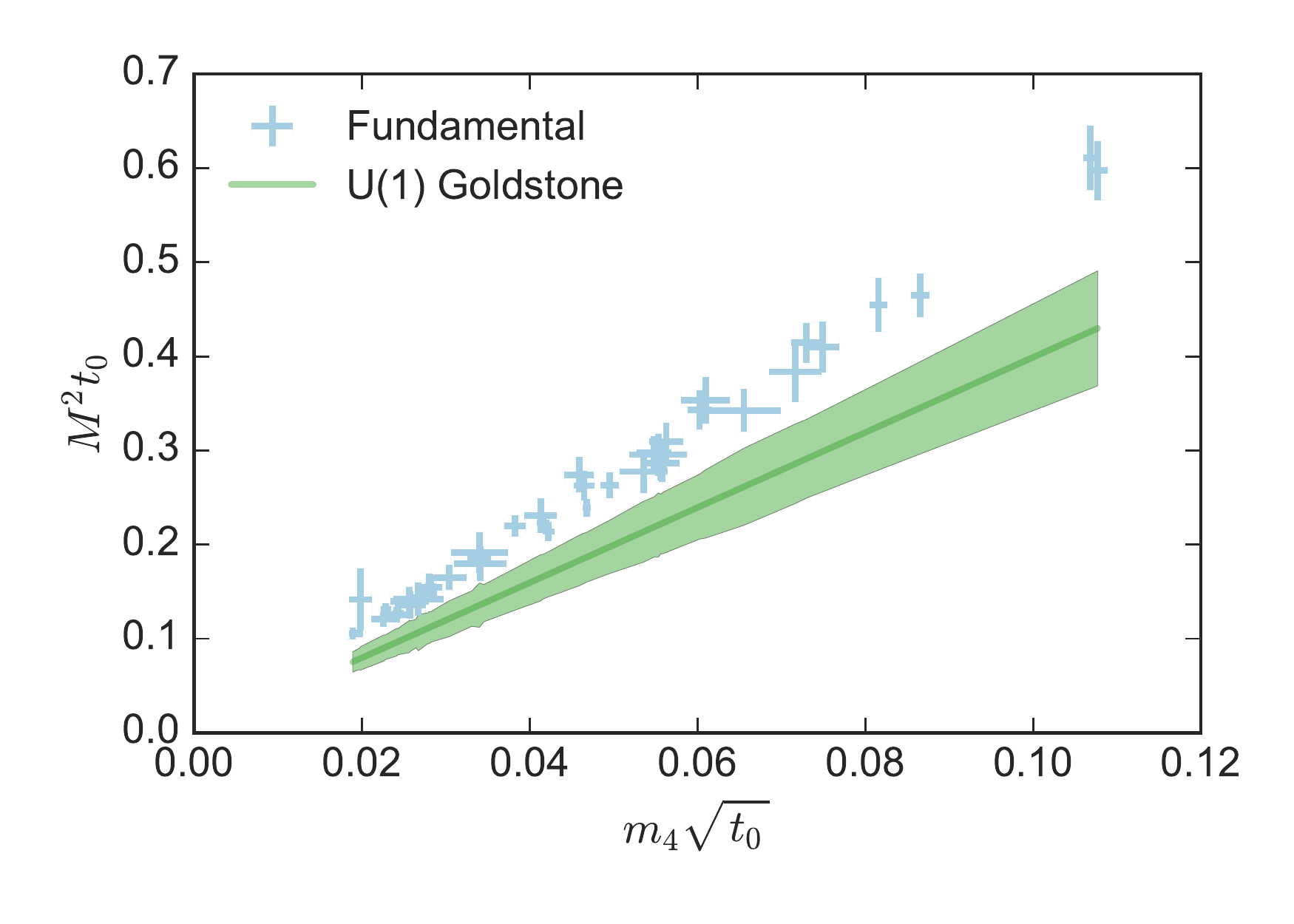}
\caption{
Mass squared $\hM_\zeta^2$ of the non-anomalous NG boson
in the combined continuum ($\ha\to 0$) and chiral-sextet ($\hm_6\to 0$) limits,
as extracted using \Eq{eq:mzeta} and the central fit's parameters,
plotted against $\hm_4$.
The pseudoscalar mass $\hM_4^2$ in the fundamental sector
in the same limit is shown for comparison.
\label{fig:mzeta_ferretti_limit}}
\end{figure}

As explained in Sec.~\ref{ssec:chifit}, the chiral fit in the
fundamental and sextet sectors allows us to probe the $\zeta$ meson as well.
We examine its mass in the chiral-sextet limit, $\hm_6 \to 0$.
Figure~\ref{fig:mzeta_ferretti_limit} shows $\hat M_\zeta^2$,
constructed using \Eq{eq:mzeta} and the parameters of the central fit,
in the continuum ($\ha\to 0$) limit, as a function of the mass $\hm_4$ of the fundamental fermions.
The figure shows that the singlet boson is consistently lighter than the pseudoscalar  of the fundamental sector in this limit.

We can make a conservative prediction regarding the $\zeta$ mass as follows.
As we have just explained, we do not know how large $\hm_4$ can be while keeping the N$^2$LO corrections below, say, 10\% or 20\%.
Lowering the maximal value of $\hm_4$ raises the uncertainty in $\ringF_\zeta$, as seen in Fig.~\ref{fig:chipt_fit_stability}.
Still, even if we lower the maximal value of $\hm_4$ so as to,
say, double the uncertainty of $\ringF_\zeta$,
we would still find that $M_\zeta^2<M_4^2$ at the 1$\sigma$ level.

The chiral-sextet limit is interesting for
composite-Higgs models.
In many models, including those proposed by
Ferretti and Karateev \cite{Ferretti:2013kya}, the symmetries of the Standard Model
are embedded into the unbroken global symmetries, so that neither the fundamental nor the sextet fermions
are required to be strictly massless.
Nonetheless, successful models
are likely to have very light sextet fermions, because a large sextet mass
would prevent the Higgs field from condensing even after the generation
of a potential from the coupling of the
Higgs to Standard Model fields.

\section{Vector Mesons}
	\label{sec:vectors}

\subsection{Masses and decay constants}

We now turn to our results for vector masses and decay constants.
Vector-meson decay constants appear in the literature with a variety
of conventions.  We define $F_{Vr}$ to have units of energy,
\be \label{eq:mv-def}
  \big\langle 0 \big| V_{ia}^{(r)} \big| V_{jb}^{(r)} \big\rangle
  = \delta_{ab}\delta_{ij} F_{Vr} M_{Vr}\ ,
\ee
where the vector meson is at rest.  The indices are $i,j=1,2,3,$ for the
spatial directions, and as usual, $a,b=1,2,3,$ for isospin.
This definition is frequently used in the phenomenology literature
on precision electroweak observables, for example Ref.~\cite{Peskin:1991sw}.

\begin{figure}[htb]
\includegraphics[width=0.8\columnwidth]{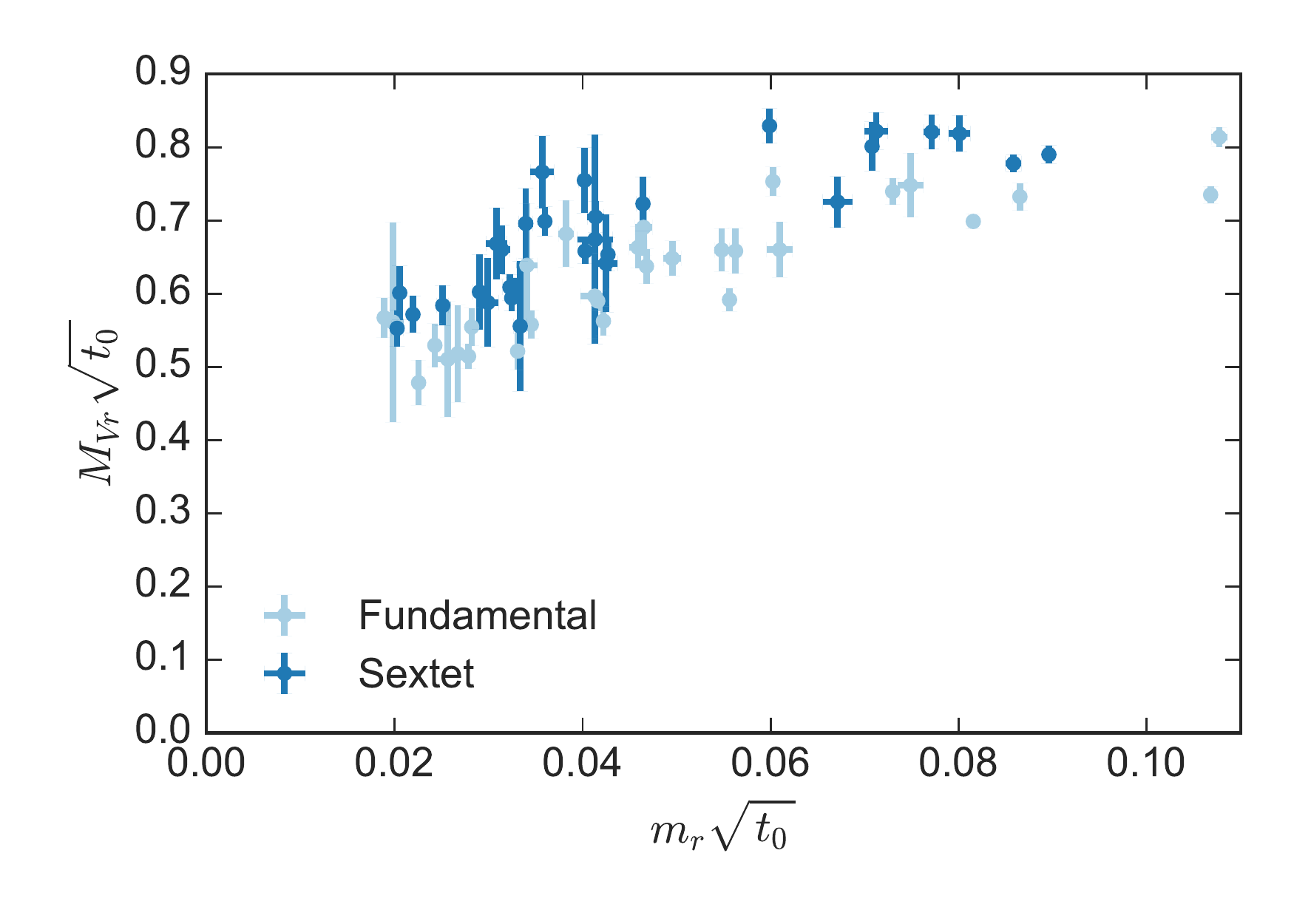}
\caption{Vector masses vs fermion masses in units of the flow scale $t_0$.\label{fig:mv}}
\end{figure}

\begin{figure}[htb]
\includegraphics[width=0.8\columnwidth]{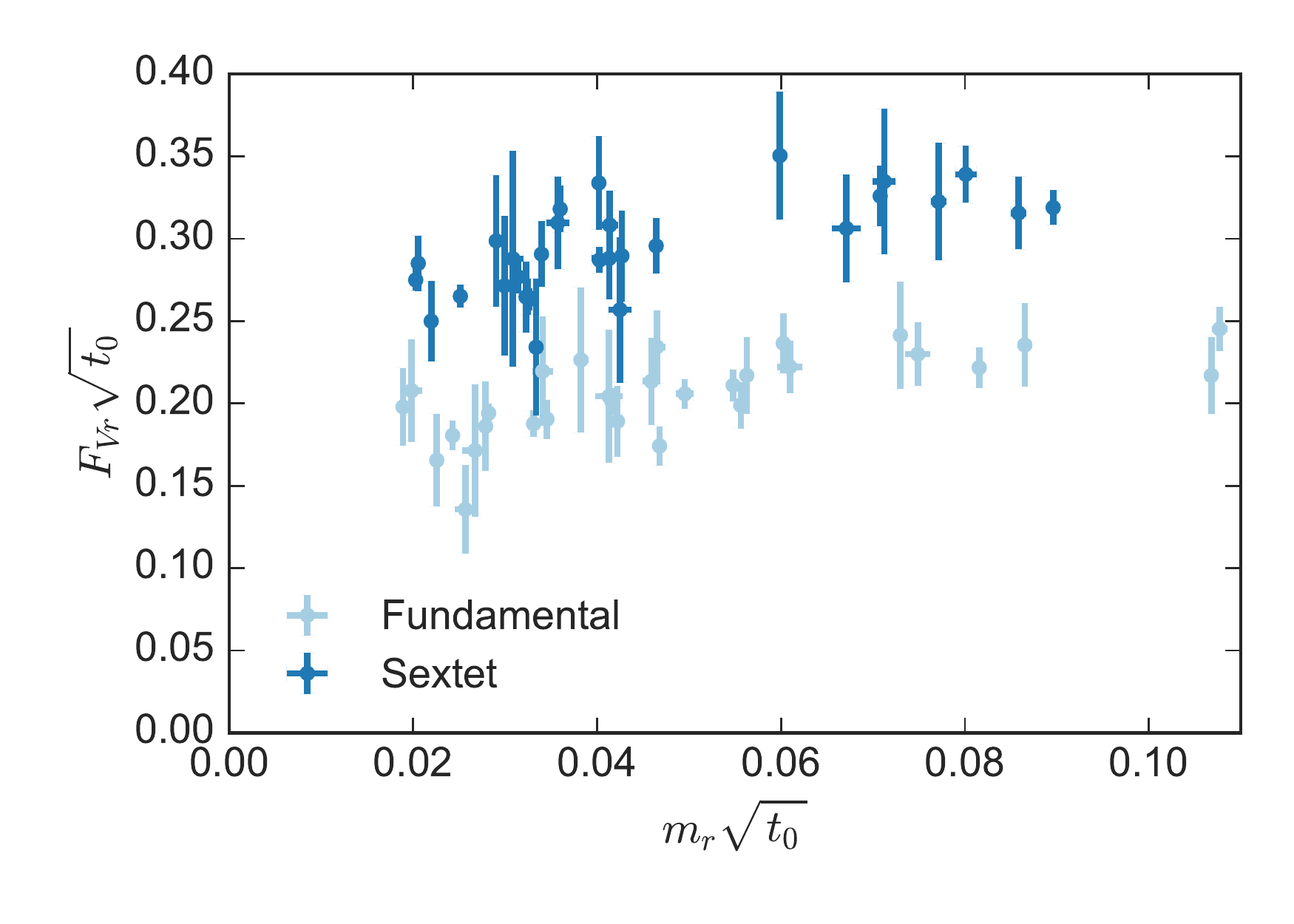}
\caption{Vector decay constants vs fermion masses in units of the flow scale $t_0$.\label{fig:fv}}
\end{figure}

Figures~\ref{fig:mv} and~\ref{fig:fv} show results for $\hM_{Vr}$
and $\hF_{Vr}$, respectively, each plotted against the fermion mass $\hm_r $
in the same representation.  As before, we measure all quantities
in units of $t_0$.
The data for these plots are listed in Tables~\ref{table:spec_vector_table_16x18}--\ref{table:spec_vector_table_24x48}.
Both quantities shows a modest, plausibly linear rise against the fermion mass, albeit with a large spread.

\begin{figure}[htb]
\includegraphics[width=0.8\columnwidth]{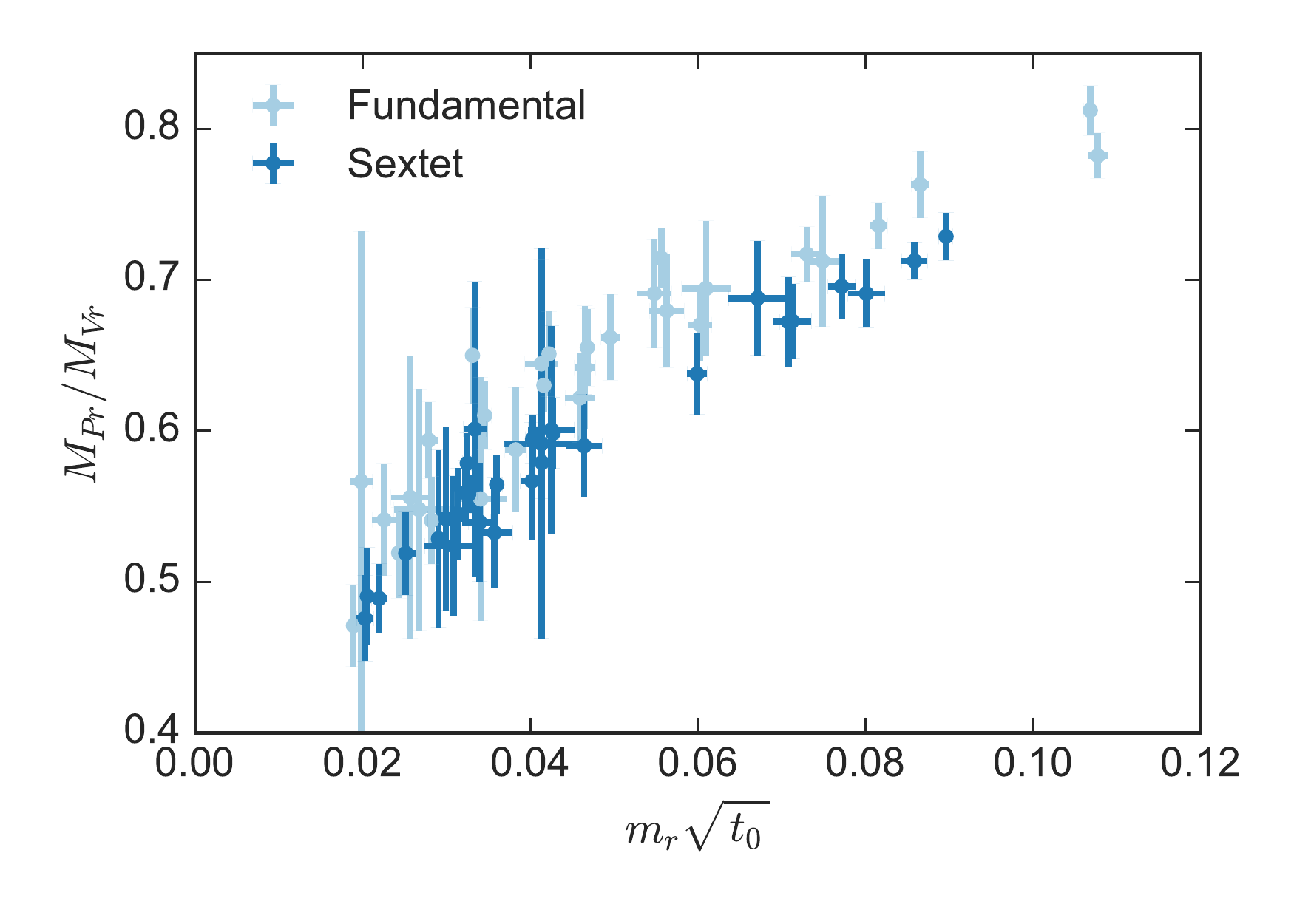}
\caption{The mass ratio $M_{Pr}/M_{Vr}$ in a fixed representation.}
\label{fig:mps_by_mv}
\end{figure}

Figure~\ref{fig:mps_by_mv} shows the ratio of the pseudoscalar and vector masses, $M_{Pr}/M_{Vr}$, again plotted against the fermion mass $\hm_r $
in the same representation.  This ratio is greater than
or equal to a half for all but the smallest masses. 
Because the decay $V \rightarrow PP$ is $p$-wave, the vector is stable if  $M_{P}/M_{V} > 0.5 \sqrt{1 - 4 k_{\rm min}^2/M_{V}^2}$, where $k_{\rm min} = 2\pi / L$ is the minimum nonzero momentum.  This condition is satisfied for both representations on all of our ensembles, so the vectors are indeed stable.

\begin{figure}[htb]
	\includegraphics[width=0.85\columnwidth]{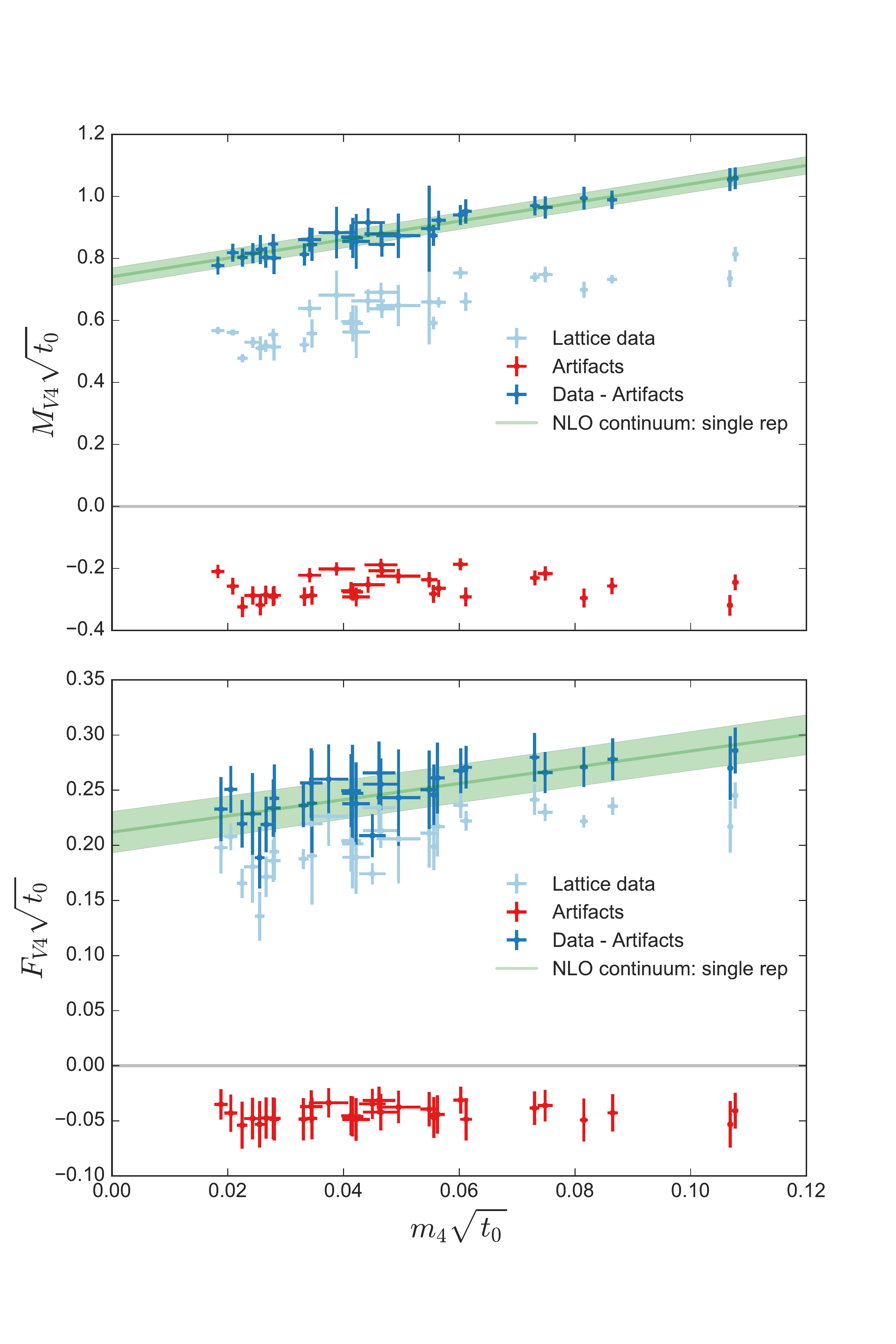}
	\caption{Breakdown of the contribution of lattice artifacts in the empirical models for the vector masses and decay constants in the fundamental representation.\label{fig:breakdowns_vector_fund}}
\end{figure}

\begin{figure}[htb]
	\includegraphics[width=0.85\columnwidth]{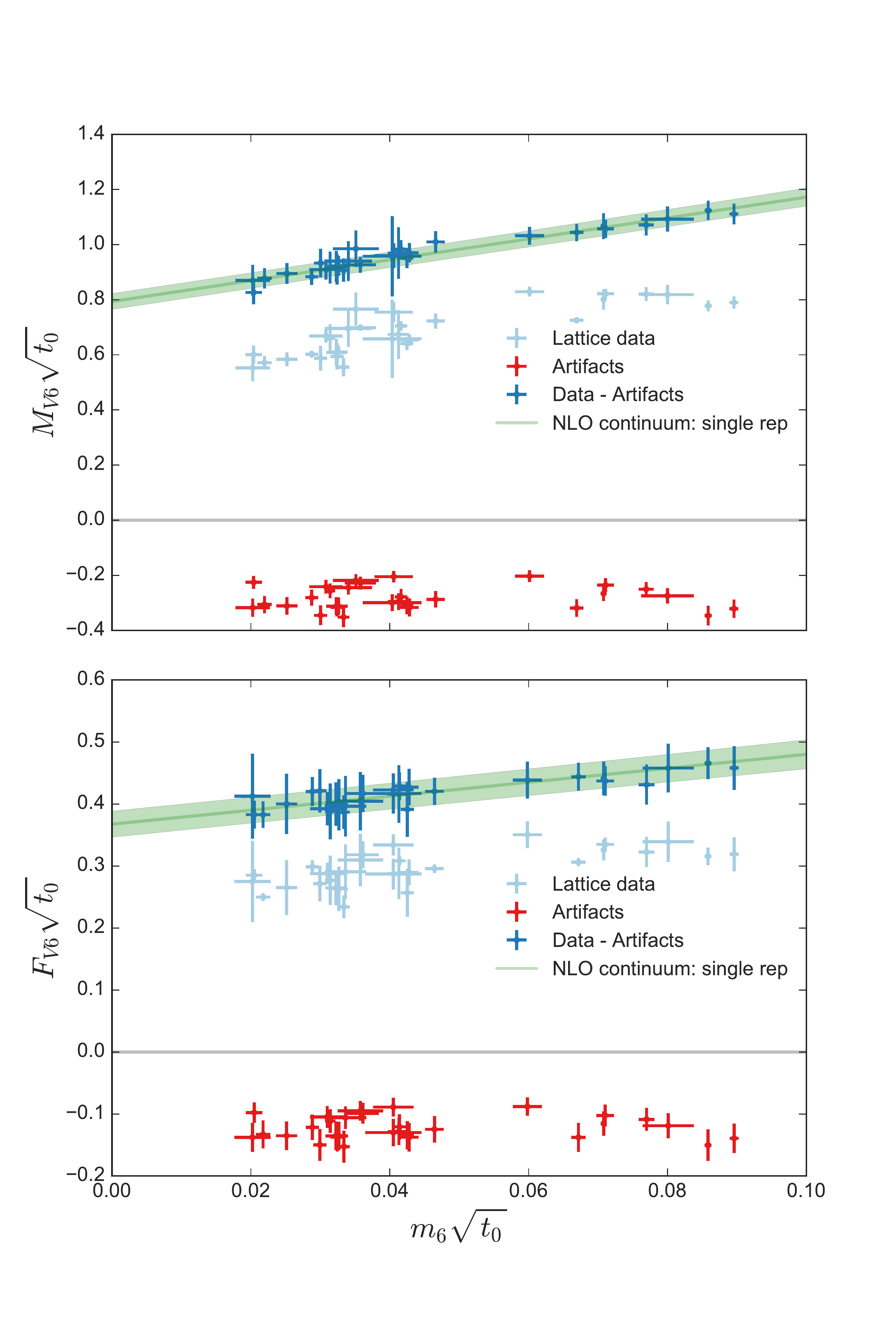}
	\caption{Breakdown of the contribution of lattice artifacts in the empirical models for the vector masses and decay constants in the sextet representation.\label{fig:breakdowns_vector_sextet}}
\end{figure}

We model $\hM_{Vr}$ and $\hF_{Vr}$ as linear functions of
the fermion mass in the same representation and of the lattice spacing,
for example,
\begin{align}
\hat{M}_{V4} = c_0 + c_1 \hat{m}_{4} + c_2 \hat{a} \ .
\end{align}
For this analysis, we restrict ourselves to the 30 ensembles for which we were able to measure the vector decay constants (see Tables~\ref{table:spec_vector_table_16x18}---\ref{table:spec_vector_table_24x48}).
	The individual correlated fits are successful, with typical $\chi^2$/DOF $\lesssim 1.0$ for $30 - 3 = 27$ degrees of freedom.
Figures~\ref{fig:breakdowns_vector_fund} and~\ref{fig:breakdowns_vector_sextet} illustrate the contribution of lattice artifacts to these fits in the same manner as for the pseudoscalars above; the green bands represent the linear continuum terms in each fit.

\subsection{Vector meson dominance and the KSRF relations}
\begin{figure}[htb]
\includegraphics[width=0.8\columnwidth]{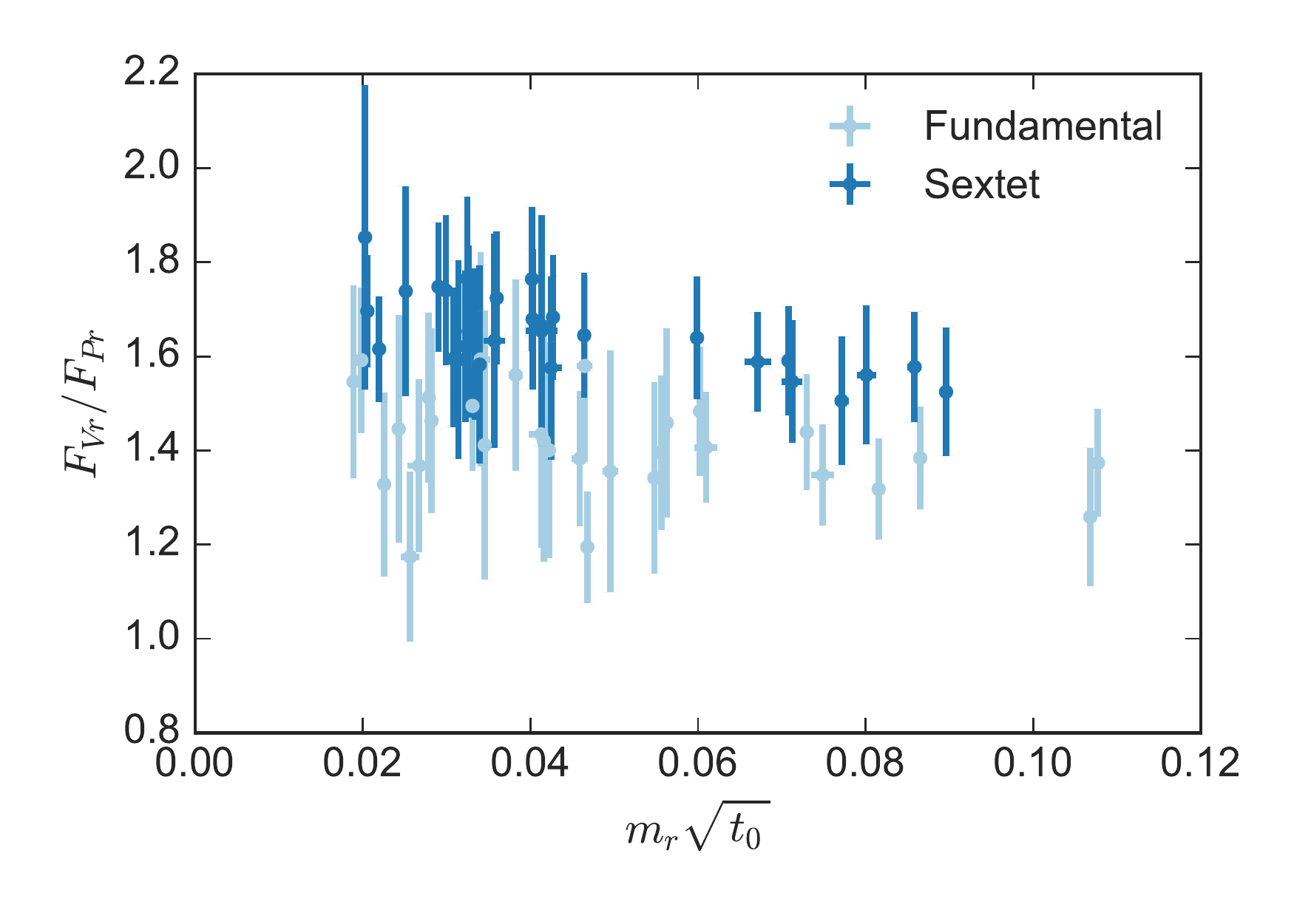}
\caption{
	The ratio of the vector and pseudoscalar decay constants in each representation.
        The KSRF prediction is a constant value of $\sqrt{2}$.
	\label{fig:fv_by_fp}
}
\end{figure}

The pseudoscalar and vector decay constants are related through the hypothesis
of vector meson dominance (VMD).
Kawarabayashi, Suzuki, Riazuddin, and Fayyazuddin (KSRF)
showed long ago~\cite{Kawarabayashi:1966kd, Riazuddin:1966sw, Johnson:1970xe}
that VMD leads to the prediction
\be \label{eq:KSRF}
F_V = \sqrt{2} F_P \ ,
\ee
independent of representation.
 Figure~\ref{fig:fv_by_fp} shows the ratio $F_{Vr}/F_{Pr}$ in each representation, after subtracting lattice artifacts.
The KSRF prediction is qualitatively successful.
(In QCD, the experimental value is roughly 1.66.)

\begin{figure}[htb]
\includegraphics[width=0.8\columnwidth]{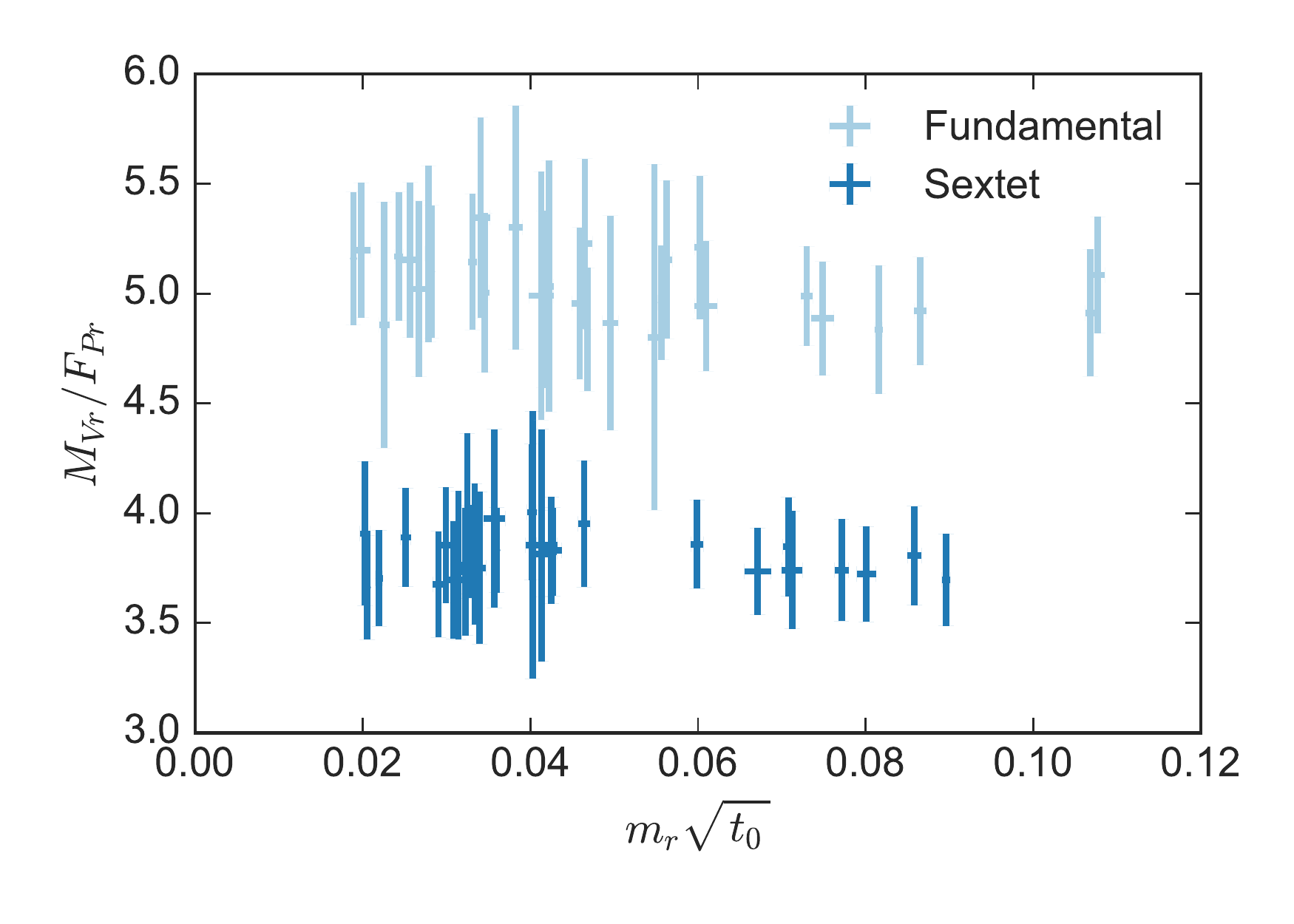}
\caption{
	The ratio of the vector mass and pseudoscalar decay constant in a fixed representation.
	KSRF identify this quantity with the coupling $g_{\textit{VPP}}$.
        In QCD, this ratio is roughly  5.9.
	\label{fig:mv_by_fp}
}
\end{figure}

\begin{figure}[htb]
\includegraphics[width=0.7\columnwidth]{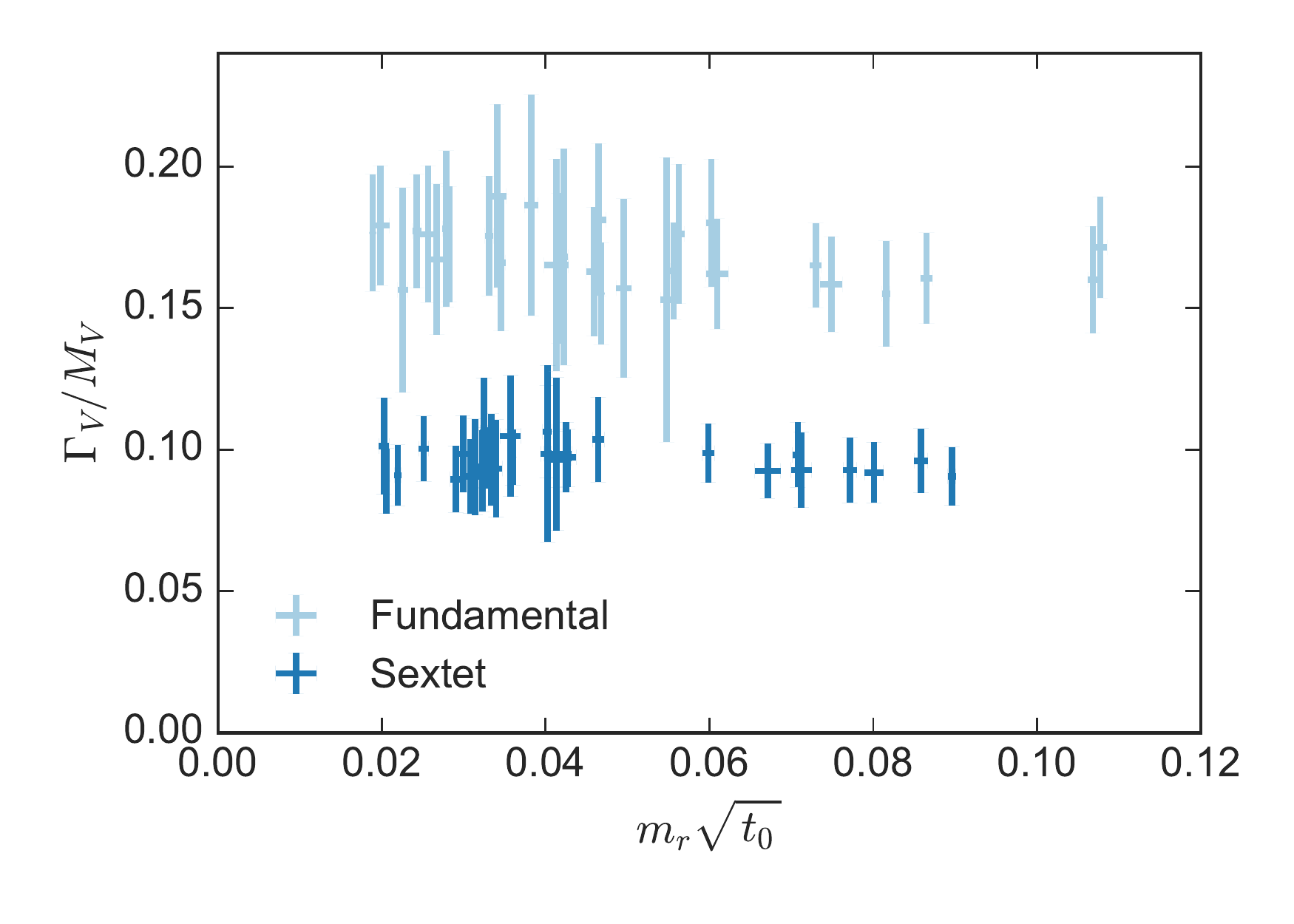}
\caption{
        Tree-level estimates for the width-to-mass ratio of the vector mesons according to KSRF.
        The KSRF estimate for this ratio is roughly 0.23 in QCD.
        \label{fig:gammav_by_mv}
        }
\end{figure}

Another result of KSRF is that the on-shell coupling constant $g_{\textit{VPP}}$
mediating the decay of a vector into two pseudoscalars is given by
\be
g_{VPP} = \frac{M_V}{F_P} \ .
\ee
We plot this ratio in Fig.~\ref{fig:mv_by_fp}.
As already noted, in our ensembles the vector meson is stable.
Nevertheless, we may use the KSRF result as a phenomenological estimate
for the behavior close to the chiral limit.
Using the tree-level formula for the $V \rightarrow PP$ decay width
in the limit where $M_{Pr}\ll M_{Vr}$,
\be \label{eq:vector_width1}
\Gamma_{V \rightarrow PP} \simeq \frac{g^2_{\textit{VPP}} M_V}{48 \pi} \ ,
\ee
we can estimate the the mass-to-width ratio for each vector resonance,
\be \label{eq:vector_width}
\frac{\Gamma_{V \rightarrow PP}}{M_V} \simeq \frac{M_V^2}{48\pi F_P^2} \ .
\ee
From Fig.~\ref{fig:mv_by_fp} we thus obtain Fig.~\ref{fig:gammav_by_mv}.
For the physical $\rho$ meson, this ratio has a value of roughly 0.23.
(The experimental value is~0.19.)

\section{Discussion}
    \label{sec:conclusions}
\subsection{\label{ssec:largeN}  Large-$N_c$ counting}

We want to put our results in context with comparisons to QCD\@.
Large-$N_c$ counting (for $N_c$ colors) is a way to do that.
Any quantity $Q$ is expected to scale across $N_c$ as
\be
Q(N_c)= N_c^p\left(Q_0 + \frac{Q_1}{N_c}+\frac{Q_2}{N_c^2}+\cdots \right),
\label{largeNc}
\ee
where $p$ is some characteristic exponent
determined by large-$N_c$ considerations,
and the $Q_i$ are a set of expansion coefficients.
Before we get to the (limited) comparisons between different $N_c$'s
we can make, our theory gives us a unique opportunity to compare
the expansion coefficients for different representations.
More specifically, if we neglect all the subleading corrections,
our data allow us to compare the leading expansion coefficient $Q_0$
between the fundamental and two-index antisymmetric representations,
for various obesrvables.

We start with meson masses, which are predicted to be independent of $N_c$
[$p=0$ in \Eq{largeNc}].  Figs.~\ref{fig:gmor} and~\ref{fig:mv}
reveal near-independence of representation of the pseudoscalar
and vector masses when plotted against the corresponding fermion mass.
This is further supported by the near equality of $\ringB_4$ and $\ringB_6$ in
Table~\ref{table:chipt_central_fit_table}.  We conclude that $Q_0$
is roughly independent of representation for the pseudoscalar and vector
meson masses.

\begin{figure}[htb]
\includegraphics[width=0.75\textwidth]{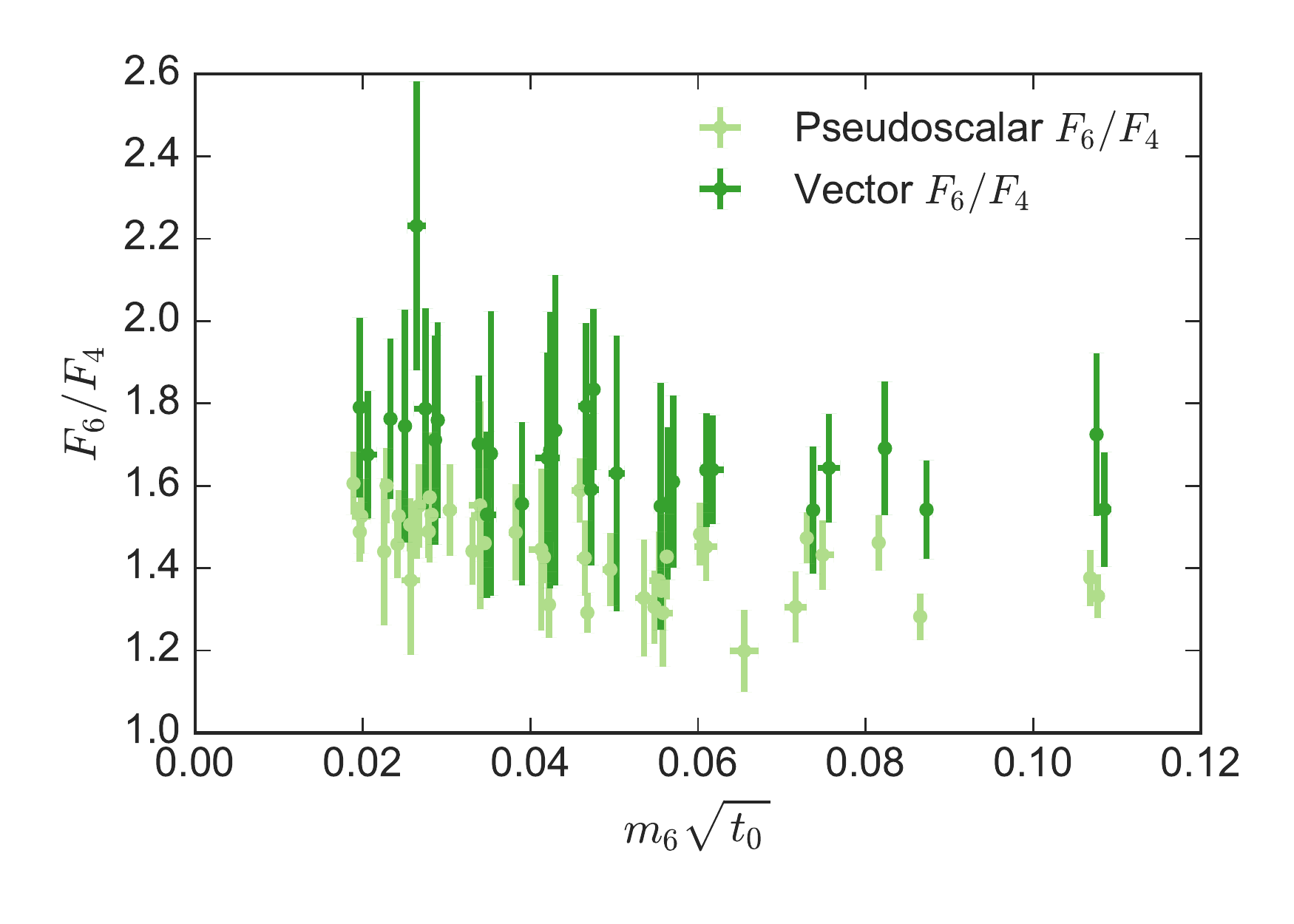}
\caption{
	The ratio of the decay constants in the sextet and fundamental representations in both the pseudoscalar (dark green) and vector (light green) channels.
 \label{fig:f6_by_f4}
}
\end{figure}

Decay constants scale as $\sqrt{N_c}$
for single-index fermions
and as $N_c$ for two-index fermions.
As for the leading expansion coefficient,
a possible guess might be that the product $N_c^p Q_0$ follows
the leading large-$N_c$ behavior of $(\dim\  r)^{1/2}$.
This would imply that there exists a constant $c$
such that $Q_0 \approx c$ for mesons made of fundamental-representation
fermions, while $Q_0 \approx c/\sqrt{2}$ for mesons made out of fermions
in the two-index antisymmetric representation.
For $N_c=4$ we thus expect $F_6/F_4\approx\sqrt{N_c/2}=\sqrt{2}$.
The ratio $F_6/F_4$ is plotted in Fig.~\ref{fig:f6_by_f4} for
the pseudoscalar and vector mesons, showing good (perhaps too good) agreement
with this expectation.
Another consequence is that the ratio $F_{Vr}/F_{Pr}$ is expected to be roughly independent of representation $r$, in agreement with Fig.~\ref{fig:fv_by_fp} and the KSRF relation.

In this context, we can also compare the fermion condensates in the two representations.
We can use the results of the chiral fit to calculate these (per flavor), each in its corresponding chiral limit.  Using \Eq{eq:condensates} we find
$\hat{\Sigma}_6/\hat{\Sigma}_4 = 2.4(3)$
for the ratio of condensates in the (double) chiral limit,
reflecting our empirical findings that $\ringB_6/\ringB_4\simeq 1$
and $\ringF_6/\ringF_4\simeq\sqrt{2}$.

Finally, we compare our results for the ratio $M_{Vr}/F_{Pr}$ to the SU(3) case.
We expect that this ratio for either SU(4) representation will be smaller
than its value in QCD, which is $770\ \textrm{MeV}/130\ \textrm{MeV}=5.9$.
This is borne out by Fig.~\ref{fig:mv_by_fp}.  At a more quantitative level,
large-$N_c$ scaling predicts this value to be
$5.9 \times \sqrt{3/4} \approx 5.1$ for the fundamental representation,
in good agreement with the data.  If we regard
the fundamental representation of SU(3) as the two-index antisymmetric one,
we obtain a large-$N_c$ prediction of $5.9 \times 3/4 \approx 4.4$
for the sextet.  While the average value of this ratio is smaller
for the sextet, one can say that this prediction is consistent with
the general trend of our data.  If we add to this the KSRF relation,
we find correspondingly smaller values for the width to mass ratio
of our vector mesons compared to the physical $\rho$ meson.

\subsection{\label{ssec:conclusions}  Conclusions}

In this paper we have described the low-lying mesonic spectrum of
SU(4) gauge theory coupled to dynamical fermions in the fundamental
and sextet representations.
These multirep simulations are the first of their kind.
Our choice of this theory was inspired by its close similarity
to a composite-Higgs model first studied by Ferretti \cite{Ferretti:2014qta}.

Our analysis focused on the masses of the pseudoscalar and vector states
and their associated decay constants.
Using the extension of \cpt\ that accounts for the
discretization errors of Wilson fermions, we carried out
a global fit of the pseudoscalar masses and decay constants of the
two representations, to NLO in the GSM power-counting scheme.
We found significant lattice artifacts, which we were able to subtract,
obtaining predictions for continuum-limit values.
Our chiral fit provides mild evidence for coupling between the two
fermion representations, a novel feature of multirep theories.

Through both the mass terms and the Wilson terms, our lattice setup
incorporates the expected symmetry breaking patterns:
$\textrm{SU}(2)_L\times\textrm{SU}(2)_R\to\textrm{SU}(2)_V$
in the fundamental sector, and
$\textrm{SU}(4)\to\textrm{SO}(4)$ in the sextet sector.  We did not carry out
a dedicated study of alternative symmetry breaking patterns.
Still, the success of the chiral fits provides some confirmation that
the above symmetry breaking patterns are the right ones.

The main theoretical uncertainty of our chiral fits concerns the size
of N$^2$LO effects.  Thanks to a large decay constant,
the chiral expansion converges quickly in the sextet sector, supporting
the hypothesis that N$^2$LO effects are small in this sector.  In the fundamental sector
the chiral expansion converges more slowly.
Hence, keeping N$^2$LO effects
below a certain comfortable level might require the exclusion of ensembles
where $\hm_4$ is on the high side.  More quantitative statements cannot be made
given our data.

The correlation functions that we calculated probe directly the
pseudoscalar states made purely of fundamental or purely of sextet fermions.
This is reflected in the stability of the LO parameters $\ringF_r$
and $\ringB_r$ if we constrain the maximal value of $\hm_4$ to successively
smaller values.  The last LO parameter, the decay constant $\ringF_\zeta$
of the axion-like singlet NG boson, is not well-determined because we have not calculated propagators in the $\zeta$ channel.
The $\zeta$ meson does contribute through virtual loops
to the correlation functions we have studied.  Accordingly,
our fits depend on $\ringF_\zeta$, but only through NLO logarithmic terms.
$\ringF_\zeta$ is more sensitive to the upper limit on $\hm_4$; as a result, so is our prediction for the mass of the $\zeta$ boson.
Nonetheless, we have argued that the $\zeta$
is lighter than the fundamental-sector NG bosons, $M_\zeta<M_4$, in the sextet-chiral limit
$\hm_6\to 0$, a limit which is interesting for the phenomenology of
Ferretti's model.
In a full composite-Higgs model, however,
  the masses of all pseudoscalar states can receive
  important corrections from the couplings to Standard Model fields.

In modeling our results for the vector mesons,
we found that the ratio of pseudoscalar to vector
decay constants agrees well with the KSRF result based on vector meson
dominance.  As discussed in Sec.~\ref{ssec:largeN},
comparing the KSRF prediction for the decay rate
of the vector meson in the chiral limit to the QCD case shows reasonable
agreement with large-$N_c$ counting.

Although our estimates for $\Gamma_V / M_V$ depend on the well-motivated but non-rigorous assumption of vector meson dominance, the resulting narrowness is almost certainly generic.
In large-$N_c$, the widths of mesons made of fundamental-representation fermions scales as $1/\sqrt{N_c}$ and thus they become narrower as $N_c$ increases.
Insofar as large-$N_c$ provides the cleanest explanation for the narrowness and existence of mesons in QCD, the vector mesons should become narrower in theories with more colors.
We proposed that in multirep theories, the generalization of $\sqrt{N_c}$ is $(\dim r)^{1/2}$, a hypothesis supported by our data.
This result is good news for phenomenologists looking to constrain models like the Ferretti model, since narrower states typically provide clearer signals in collider data.

As we have mentioned, we are also exploring the phase diagram of this multirep theory
 \cite{Ayyar:2017uqh}.
We have been looking for---and not finding---scale separation between the representations in the confinement and chiral transitions.
We are also studying the baryon spectrum, a particularly interesting sector of the theory given its connection to top-quark physics and partial compositeness in the Ferretti model.

Other interesting avenues for the future work in this model (or multirep composite Higgs theories more generally) include quantities related to the Higgs potential.
The contribution of the Standard Model's gauge fields to the Higgs potential, \PiLR, is conceptually identical to the physics of electromagnetic splittings between pions in QCD and has been the subject of a recent pilot study on the lattice~\cite{DeGrand:2016htl}.
The top-quark contribution to the Higgs potential is considerably more challenging \cite{Golterman:2015zwa,Golterman:2017vdj}.

\begin{acknowledgments}

Computations for this work were carried out with resources provided by the USQCD Collaboration, which is funded
by the Office of Science of the U.S.\ Department of Energy;  with the Summit supercomputer, a joint effort of the University of Colorado Boulder and Colorado State University, which is supported by the National Science Foundation (awards ACI-1532235 and ACI-1532236), the University of Colorado Boulder, and Colorado State University; by the Cy-Tera Project, which is co-funded by the European Regional Development Fund and the Republic of Cyprus through the Research Promotion Foundation; and the DECI resource ARCHER based in the United Kingdom at the University of Edinburgh with support from PRACE.

This work was supported in part by the U.S.\ Department of Energy under grants DE-SC0010005 (Colorado) and DE-SC0013682 (M.~G.~), and by the Israel Science Foundation under grant no.~449/13 (Tel Aviv).  Brookhaven National Laboratory is supported by the U. S. Department of Energy under contract DE-SC0012704.

\end{acknowledgments}

\appendix

\clearpage
\section{Data tables \label{app:data_tables}}

\begin{table}[h] 
\centering
\setlength{\tabcolsep}{12pt} 
\begin{ruledtabular}
\begin{tabular}{clllc}
Ensemble &   $\beta$ & $\kappa_4$ & $\kappa_6$ &   Configurations\\
\hline
1  &  7.2   &  0.13173 &  0.13423 & 67 \\
2  &  7.2   &  0.1318  &  0.1341  & 29 \\
3  &  7.2   &  0.132   &  0.134   & 42 \\
4  &  7.3   &  0.1314  &  0.1333  & 17 \\
5  &  7.3   &  0.1315  &  0.1333  & 17 \\
6  &  7.308 &  0.1304  &  0.1339  & 29 \\
7  &  7.31  &  0.1305  &  0.1339  & 17 \\
8  &  7.32  &  0.13    &  0.134   & 17 \\
9  &  7.33  &  0.1314  &  0.1332  & 17 \\
10 &  7.33  &  0.1314  &  0.1333  & 17 \\
11 &  7.33  &  0.1315  &  0.1335  & 17 \\
12 &  7.4   &  0.1307  &  0.133   & 17 \\
13 &  7.4   &  0.131   &  0.133   & 29 \\
14 &  7.5   &  0.13    &  0.132   & 17 \\
15 &  7.5   &  0.13    &  0.1325  & 29 \\
16 &  7.5   &  0.13    &  0.1327  & 29 \\
17 &  7.5   &  0.13    &  0.1328  & 29 \\
18 &  7.5   &  0.1305  &  0.1327  & 29 \\
19 &  7.75  &  0.129   &  0.131   & 29 \\
20 &  7.75  &  0.129   &  0.1315  & 29 \\
\end{tabular}
\end{ruledtabular}

\caption{
	List of ensembles with $V=16^3 \times 18$ generated for this study.
	Configurations are separated by 4 Monte Carlo trajectories.
}
\label{table:bare_params_16x18}
\end{table}

\begin{table}[h] 
\centering
\setlength{\tabcolsep}{12pt} 
\begin{ruledtabular}
\begin{tabular}{clllc}
Ensemble &   $\beta$ & $\kappa_4$ & $\kappa_6$ & Configurations\\
\hline
21 &  7.25  &  0.13095 &  0.13418 & 61 \\
22 &  7.25  &  0.13147 &  0.13395 & 71 \\
23 &  7.276 &  0.13157 &  0.13364 & 96 \\
24 &  7.3   &  0.13117 &  0.13363 & 61 \\
25 &  7.3   &  0.13118 &  0.13361 & 96 \\
26 &  7.3   &  0.13162 &  0.1334  & 71 \\
27 &  7.308 &  0.1304  &  0.13393 & 96 \\
28 &  7.33  &  0.1314  &  0.1332  & 96 \\
29 &  7.4   &  0.1307  &  0.133   & 96 \\
30 &  7.55  &  0.129   &  0.1325  & 84 \\
31 &  7.55  &  0.13    &  0.1325  & 84 \\
32 &  7.65  &  0.128   &  0.131   & 49 \\
33 &  7.65  &  0.129   &  0.1308  & 49 \\
34 &  7.65  &  0.13    &  0.131   & 84 \\
35 &  7.65  &  0.13    &  0.132   & 84 \\
36 &  7.75  &  0.128   &  0.131   & 84 \\
37 &  7.75  &  0.129   &  0.1308  & 54 \\
38 &  7.75  &  0.1295  &  0.1315  & 34 \\
39 &  7.85  &  0.129   &  0.1308  & 44 \\
\end{tabular}
\end{ruledtabular}

\caption{
	List of ensembles with $V=16^3 \times 32$.
	Configurations are separated by 10 Monte Carlo trajectories.
}
\label{table:bare_params_16x32}
\end{table}

\begin{table}[h] 
\centering
\setlength{\tabcolsep}{12pt} 
\begin{ruledtabular}
\begin{tabular}{clllc}
Ensemble &   $\beta$ & $\kappa_4$ & $\kappa_6$ &  Configurations\\
\hline
40 &  7.51 &  0.1307 &  0.1328  & 133 \\
41 &  7.55 &  0.13   &  0.1327  & 80 \\
42 &  7.55 &  0.1305 &  0.1325  & 91 \\
43 &  7.55 &  0.1307 &  0.13234 & 80 \\
\end{tabular}
\end{ruledtabular}

\caption{
	List of ensembles with $V=24^3 \times 48$.
	Configurations are separated by 10 Monte Carlo trajectories.
}
\label{table:bare_params_24x48}
\end{table}


\begin{table}[h]
\centering
\setlength{\tabcolsep}{12pt} 
\begin{ruledtabular}
\begin{tabular}{clll}
Ensemble &   $t_0/a^2$	& $\hm_4$ & $\hm_6$ \\
\hline
1  &  1.07(2) &  0.024(1) &  0.022(1) \\
2  &  0.92(3) &  0.026(2) &  0.030(2) \\
3  &  0.89(2) &  0.023(1) &  0.033(1) \\
4  &  0.99(3) &  0.034(3) &  0.043(4) \\
5  &  0.93(2) &  0.028(2) &  0.043(2) \\
6  &  1.07(3) &  0.056(2) &  0.024(2) \\
7  &  1.26(2) &  0.054(3) &  0.023(3) \\
8  &  1.20(3) &  0.066(4) &  0.018(2) \\
9  &  1.15(3) &  0.027(3) &  0.043(3) \\
10 &  1.22(1) &  0.026(2) &  0.037(1) \\
11 &  1.40(2) &  0.020(1) &  0.029(1) \\
12 &  1.26(2) &  0.041(2) &  0.041(4) \\
13 &  1.45(2) &  0.030(2) &  0.039(2) \\
14 &  1.09(3) &  0.061(3) &  0.067(4) \\
15 &  1.33(2) &  0.056(2) &  0.046(2) \\
16 &  1.49(4) &  0.055(3) &  0.035(3) \\
17 &  1.67(2) &  0.055(2) &  0.031(1) \\
18 &  1.89(3) &  0.034(3) &  0.031(3) \\
19 &  1.99(6) &  0.075(2) &  0.071(2) \\
20 &  2.38(6) &  0.072(3) &  0.043(2) \\
\end{tabular}
\end{ruledtabular}

\caption{
	Measured gradient flow scale $t_0$ and fermion masses $\hat m_r=m_r\sqrt{t_0}$ in the ensembles with volume $V=16^3 \times 18$.
}
\label{table:spec_t0_mq_16x18}
\end{table}

\begin{table}[h]
\centering
\setlength{\tabcolsep}{12pt} 
\begin{ruledtabular}
\begin{tabular}{clll}
Ensemble &   $t_0/a^2$	& $\hm_4$ & $\hm_6$ \\
\hline
21 &   1.093(9) &   0.0422(7) &  0.0203(10) \\
22 &   1.135(9) &  0.0279(11) &  0.0251(12) \\
23 &  1.128(24) &   0.0243(7) &   0.0326(7) \\
24 &  1.132(12) &   0.0345(8) &  0.0323(14) \\
25 &  1.100(10) &   0.0331(5) &   0.0325(5) \\
26 &   1.111(9) &   0.0228(6) &   0.0381(8) \\
27 &  1.174(10) &   0.0556(7) &   0.0220(9) \\
28 &  1.095(12) &   0.0282(7) &   0.0427(7) \\
29 &  1.226(10) &   0.0416(8) &   0.0403(8) \\
30 &  1.418(12) &  0.0865(11) &  0.0414(15) \\
31 &  1.845(18) &  0.0495(11) &  0.0340(13) \\
32 &   0.916(5) &   0.1068(8) &  0.0858(15) \\
33 &   1.067(5) &  0.0816(10) &   0.0896(8) \\
34 &  1.463(15) &  0.0459(18) &  0.0801(22) \\
35 &  2.294(22) &  0.0382(13) &  0.0357(21) \\
36 &  1.556(12) &  0.1077(12) &  0.0708(10) \\
37 &  1.754(15) &  0.0730(19) &  0.0771(16) \\
38 &  2.621(20) &  0.0465(13) &  0.0402(14) \\
39 &  2.670(22) &  0.0602(14) &  0.0599(12) \\
\end{tabular}
\end{ruledtabular}

\caption{
Same as Table~\ref{table:spec_t0_mq_16x18}, but in the ensembles with 
	 volume $V=16^3 \times 32$.
}
\label{table:spec_t0_mq_16x32}
\end{table}

\begin{table}[h]
\centering
\setlength{\tabcolsep}{12pt} 
\begin{ruledtabular}
\begin{tabular}{clll}
Ensemble &   $t_0/a^2$	& $\hm_4$ & $\hm_6$ \\
\hline
40 &  2.260(16) &   0.0196(4) &   0.0194(9) \\
41 &  2.166(11) &   0.0468(5) &   0.0205(4) \\
42 &  2.182(12) &   0.0264(5) &   0.0293(6) \\
43 &   2.118(6) &   0.0189(5) &   0.0360(7) \\
\end{tabular}
\end{ruledtabular}

\caption{
	Same as Table~\ref{table:spec_t0_mq_16x18}, but in the ensembles with 
	 volume $V=24^3 \times 48$.
}
\label{table:spec_t0_mq_24x48}
\end{table}


\begin{table}[h]\centering
\setlength{\tabcolsep}{12pt} 
\begin{ruledtabular}
\begin{tabular}{cllll}
Ensemble 	&	$\hM_{P4}$	&	$\hM_{P6}$	&	$\hF_{P4}$	& 	$\hF_{P6}$\\
\hline
1  &  0.28(1) &  0.29(1) &  0.102(8)  &  0.143(8)  \\
2  &  0.28(2) &  0.32(2) &  0.106(4)  &  0.155(6)  \\
3  &  0.26(1) &  0.33(1) &  0.109(17) &  0.149(15) \\
4  &  0.34(3) &  0.41(2) &  0.115(17) &  0.178(33) \\
5  &  0.29(2) &  0.38(2) &  0.108(13) &  0.170(11) \\
6  &  0.42(2) &  0.31(2) &  0.120(8)  &  0.141(19) \\
7  &  0.43(2) &  0.30(2) &  0.132(10) &  0.163(23) \\
8  &  0.47(1) &  0.28(3) &  0.138(10) &  0.148(19) \\
9  &  0.28(2) &  0.39(2) &  0.110(10) &  0.169(10) \\
10 &  0.29(2) &  0.38(2) &  0.129(17) &  0.166(28) \\
11 &  0.32(5) &  0.32(2) &  0.113(5)  &  0.170(11) \\
12 &  0.38(2) &  0.40(2) &  0.127(17) &  0.177(23) \\
13 &  0.33(1) &  0.39(1) &  0.115(9)  &  0.176(11) \\
14 &  0.46(2) &  0.50(2) &  0.142(7)  &  0.199(9)  \\
15 &  0.45(1) &  0.43(1) &  0.133(10) &  0.183(14) \\
16 &  0.45(2) &  0.38(2) &  0.141(14) &  0.184(19) \\
17 &  0.46(1) &  0.36(1) &  0.145(9)  &  0.179(14) \\
18 &  0.35(2) &  0.35(2) &  0.122(11) &  0.185(13) \\
19 &  0.53(2) &  0.55(2) &  0.159(6)  &  0.223(17) \\
20 &  0.53(2) &  0.43(3) &  0.153(14) &  0.190(20) \\
\end{tabular}
\end{ruledtabular}

\caption{
Measured pseudoscalar masses $\hat M_{Pr}=M_{Pr}\sqrt{t_0}$ and decay constants $\hat F_{Pr}=F_{Pr}\sqrt{t_0}$ in the ensembles with volume $V=16^3 \times18$.}
\label{table:spec_pseudoscalar_table_16x18}
\end{table}

\begin{table}[h]\centering
\setlength{\tabcolsep}{12pt} 
\begin{ruledtabular}
\begin{tabular}{cllll}
Ensemble 	&	$\hM_{P4}$	&	$\hM_{P6}$	&	$\hF_{P4}$	& 	$\hF_{P6}$\\
\hline
21 &   0.366(9) &  0.263(10) &   0.119(6) &   0.142(9) \\
22 &   0.305(9) &   0.303(8) &   0.105(4) &   0.151(5) \\
23 &   0.275(6) &   0.341(7) &   0.108(4) &   0.162(5) \\
24 &   0.340(5) &   0.340(9) &   0.119(4) &   0.168(9) \\
25 &   0.339(3) &   0.344(6) &   0.107(4) &  0.148(13) \\
26 &   0.279(7) &  0.368(11) &   0.103(4) &  0.167(13) \\
27 &   0.423(4) &   0.279(5) &   0.127(4) &   0.159(7) \\
28 &   0.300(8) &   0.391(8) &   0.115(6) &   0.173(6) \\
29 &   0.372(6) &   0.391(4) &   0.126(3) &   0.173(8) \\
30 &   0.559(8) &  0.408(12) &   0.156(6) &   0.187(7) \\
31 &  0.429(10) &   0.375(9) &   0.140(9) &  0.189(10) \\
32 &   0.597(8) &   0.554(5) &   0.159(9) &  0.208(13) \\
33 &   0.514(8) &   0.576(9) &   0.154(8) &  0.219(11) \\
34 &   0.412(9) &   0.565(9) &   0.141(7) &   0.224(8) \\
35 &   0.400(9) &  0.408(10) &   0.132(6) &  0.192(17) \\
36 &   0.636(7) &   0.538(8) &   0.166(6) &   0.210(8) \\
37 &   0.530(5) &   0.571(7) &   0.154(4) &  0.223(12) \\
38 &  0.443(14) &  0.428(15) &   0.135(9) &  0.188(13) \\
39 &  0.505(13) &  0.529(17) &   0.148(8) &   0.216(8) \\
\end{tabular}
\end{ruledtabular}

\caption{
Same as Table~\ref{table:spec_pseudoscalar_table_16x18}, but in the ensembles with volume $V=16^3 \times 32$.}
\label{table:spec_pseudoscalar_table_16x32}
\end{table}

\begin{table}[h]\centering
\setlength{\tabcolsep}{12pt} 
\begin{ruledtabular}
\begin{tabular}{cllll}
Ensemble 	&	$\hM_{P4}$	&	$\hM_{P6}$	&	$\hF_{P4}$	& 	$\hF_{P6}$\\
\hline
40 &   0.278(4) &  0.291(10) &   0.114(4) &   0.167(7) \\
41 &   0.418(5) &   0.295(7) &   0.139(4) &   0.169(4) \\
42 &   0.317(6) &   0.355(8) &   0.125(4) &   0.182(8) \\
43 &   0.267(9) &   0.394(8) &   0.114(4) &   0.184(5) \\
\end{tabular}
\end{ruledtabular}

\caption{
	Same as Table~\ref{table:spec_pseudoscalar_table_16x18}, but in the ensembles with volume $V=24^3 \times48$.}
\label{table:spec_pseudoscalar_table_24x48}
\end{table}


\begin{table}[h] 
\centering
\setlength{\tabcolsep}{12pt} 
\begin{ruledtabular}
\begin{tabular}{cllll}
Ensemble 	&	$\hM_{V4}$	&	$\hM_{V6}$	&	$\hF_{V4}$	& 	$\hF_{V6}$\\
\hline
1  &   0.50(3) &   0.57(4) &       -- &       -- \\
2  &   0.51(8) &   0.59(6) &  0.14(3) &  0.27(4) \\
3  &   0.48(3) &   0.56(9) &  0.17(3) &  0.23(4) \\
4  &   0.56(6) &   0.67(5) &       -- &       -- \\
5  &   0.54(7) &   0.61(4) &       -- &       -- \\
6  &   0.62(2) &   0.62(6) &       -- &       -- \\
7  &   0.61(4) &   0.60(8) &       -- &       -- \\
8  &   0.62(4) &   0.61(16)&       -- &       -- \\
9  &   0.52(7) &   0.64(7) &  0.17(4) &  0.26(4) \\
10 &   0.56(8) &   0.63(5) &       -- &       -- \\
11 &   0.56(14)&   0.60(5) &  0.21(3) &  0.30(4) \\
12 &   0.60(6) &   0.67(14)&  0.20(4) &  0.29(3) \\
13 &   0.60(8) &   0.69(4) &       -- &       -- \\
14 &   0.66(4) &   0.73(4) &  0.22(2) &  0.31(3) \\
15 &   0.66(3) &   0.72(4) &  0.22(2) &  0.30(2) \\
16 &   0.66(4) &   0.68(9) &       -- &       -- \\
17 &   0.66(3) &   0.66(3) &  0.21(1) &  0.28(1) \\
18 &   0.64(8) &   0.67(5) &  0.22(3) &  0.29(7) \\
19 &   0.75(4) &   0.82(3) &  0.23(2) &  0.34(4) \\
20 &   0.74(4) &   0.77(10)&       -- &       -- \\
\end{tabular}
\end{ruledtabular}

\caption{Measured vector masses $\hat M_{Vr}=M_{Vr}\sqrt{t_0}$ and decay constants $\hat F_{Vr}=F_{Vr}\sqrt{t_0}$ in the ensembles with volume $V=16^3 \times 18$.
Some ensembles did not yield reliable measurements of $F_{Vr}$ because of insufficient statistics.
The figures and tables omit data from such ensembles.
}
\label{table:spec_vector_table_16x18}
\end{table}

\begin{table}[h] 
\centering
\setlength{\tabcolsep}{12pt} 
\begin{ruledtabular}
\begin{tabular}{clllllc}
Ensemble 	&	$\hM_{V4}$	&	$\hM_{V6}$	&	$\hF_{V4}$	& 	$\hF_{V6}$\\
\hline
21 &   0.56(2) &   0.55(3) &  0.19(2) &  0.275(1) \\
22 &   0.51(2) &   0.58(3) &  0.17(3) &  0.265(1) \\
23 &   0.53(3) &   0.61(1) &  0.18(1) &  0.263(1) \\
24 &   0.56(2) &   0.61(2) &  0.19(1) &  0.265(2) \\
25 &   0.52(3) &   0.59(2) &  0.19(1) &  0.265(1) \\
26 &   0.50(3) &   0.62(2) &       -- &        -- \\
27 &   0.59(2) &   0.57(3) &  0.20(1) &  0.250(2) \\
28 &   0.55(3) &   0.65(2) &  0.19(1) &  0.290(3) \\
29 &   0.59(1) &   0.66(2) &  0.20(1) &  0.287(1) \\
30 &   0.73(2) &   0.71(2) &  0.24(3) &  0.308(2) \\
31 &   0.65(2) &   0.70(5) &  0.21(1) &  0.291(2) \\
32 &   0.74(1) &   0.78(1) &  0.24(2) &  0.316(2) \\
33 &   0.70(1) &   0.79(1) &  0.22(1) &  0.319(1) \\
34 &   0.66(3) &   0.82(3) &  0.22(3) &  0.339(2) \\
35 &   0.68(5) &   0.77(5) &  0.20(5) &  0.310(3) \\
36 &   0.81(1) &   0.80(3) &  0.25(1) &  0.326(2) \\
37 &   0.74(2) &   0.82(2) &  0.23(4) &  0.322(4) \\
38 &   0.69(4) &   0.76(5) &  0.24(2) &  0.334(3) \\
39 &   0.75(2) &   0.83(2) &  0.24(2) &  0.350(4) \\
\end{tabular}
\end{ruledtabular}

\caption{Same as Table~\ref{table:spec_vector_table_16x18}, but in the ensembles with volume $V=16^3 \times 32$.
}
\label{table:spec_vector_table_16x32}
\end{table}

\begin{table}[h] 
\centering
\setlength{\tabcolsep}{12pt} 
\begin{ruledtabular}
\begin{tabular}{clllllc}
Ensemble 	&	$\hM_{V4}$	&	$\hM_{V6}$	&	$\hF_{V4}$	& 	$\hF_{V6}$\\
\hline
40 &   0.57(6) &   0.61(2) &       -- &       -- \\
41 &   0.64(2) &   0.60(4) &  0.17(1) &  0.29(2) \\
42 &   0.59(3) &   0.66(5) &       -- &       -- \\
43 &   0.57(3) &   0.70(2) &  0.20(2) &  0.32(1) \\
\end{tabular}
\end{ruledtabular}

\caption{Same as Table~\ref{table:spec_vector_table_16x18}, but in the ensembles with volume $V=24^3 \times 48$.
}
\label{table:spec_vector_table_24x48}
\end{table}

\clearpage
\section{Technical matters---lattice\label{app:tech-latt}}
\subsection{Correlator fitting} \label{ssec:correlator_fitting}

In calculating correlation functions,
we use a smeared source operator on the $t=0$ time slice while using both point and smeared operators at the sink, projected onto zero spatial momentum.
Smearing is done after fixing to the Coulomb gauge, always with smearing radius $r_0 = 6a$.
On the large lattices we use antiperiodic boundary conditions in time for the fermion propagators.
For the $V= 16^3 \times 18$ ensembles, on the other hand, we superimpose propagators computed with periodic and anti-periodic boundary conditions, which effectively doubles the temporal size of the lattice---a technique sometimes called the ``P+A trick''
(see Ref.~\cite{DeGrand:2007tx} and references therein).

After restricting each correlator to a range $[t_{\rm min}, t_{\rm max}]$ we find acceptable fits with single-exponential models, that is, without inclusion of excited states.
In each representation, we extract the fermion mass $m_r$
from the axial Ward identity (\ref{eq:pcac_continuum}) via joint fits
to the $\vev{AP}$ and $\vev{PP}$ correlators with a point sink.

We use the publicly available Python packages \texttt{lsqfit} \cite{Lepage:lsqfit} and \texttt{gvar} \cite{Lepage:gvar} for nonlinear fitting and classical error propagation.  When computing ratios of quantities derived from different fits, we use single-elimination jackknife to propagate errors including correlations.

\begin{figure}[htb]
\includegraphics[width=0.8\textwidth]{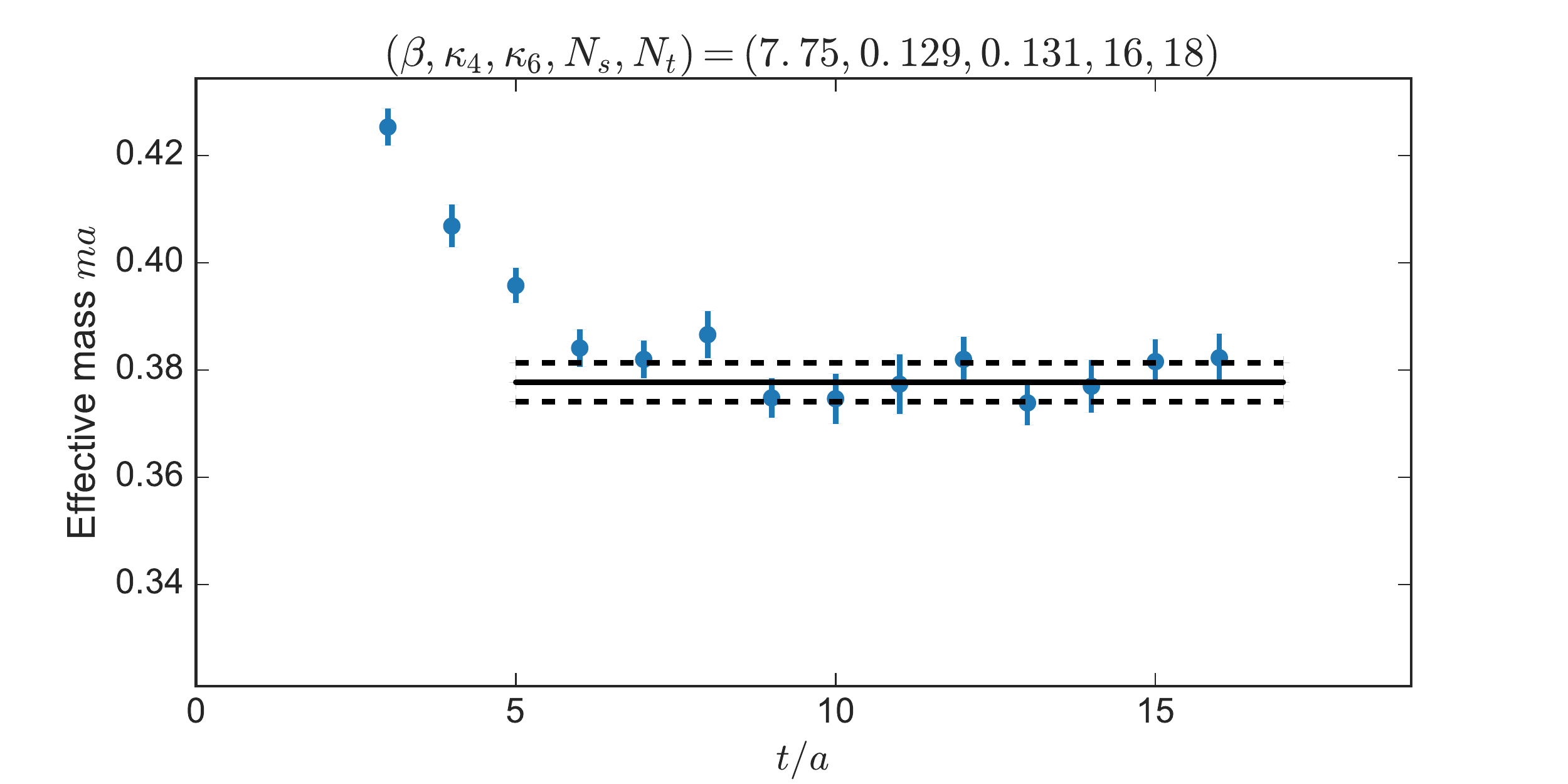}
\includegraphics[width=0.8\textwidth]{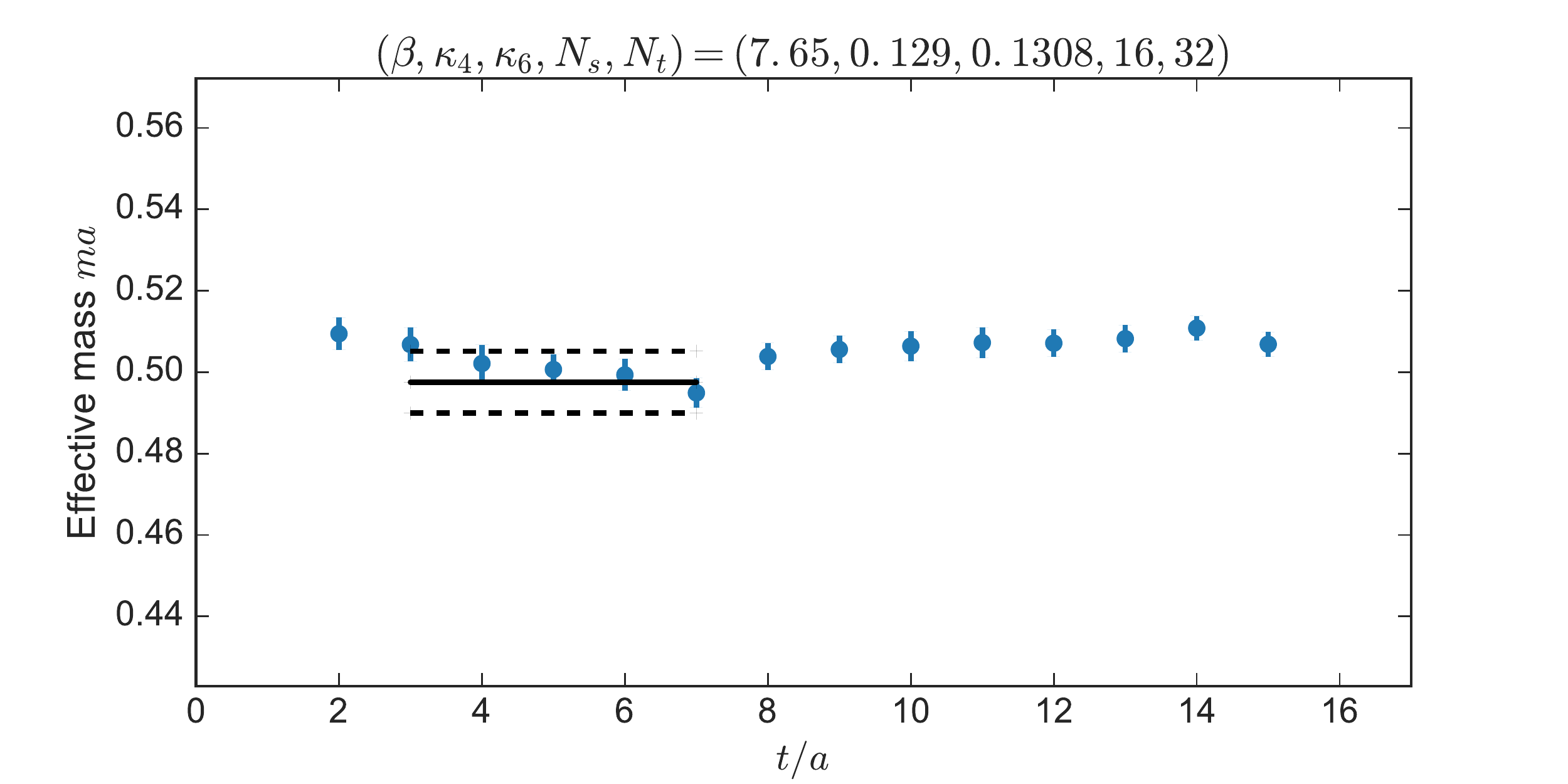}
\caption{
	Representative plot showing the effective mass $ma$ extracted from
	a smeared-source, point-sink pseudoscalar correlator on a typical
	$16^3 \times 18$ ensemble (top) and $16^3 \times 32$ ensemble (bottom)
	used in this study.  The black lines indicate the mass and error (including range-fit 
	systematic uncertainty) extracted from a full nonlinear fit of the correlator. The horizontal 
	width of the black lines indicates the starting and ending times $(t_i,t_f)$ used in the best range fit as
	determined by the MILC criterion.  Note that in the lower panel, the small difference between the central value
	from the best fit and the effective mass at larger $t/a$ is covered by the statistical + systematic error band, showing
	that our range-fit uncertainty is working as expected.
	\label{fig:eff_mass_compare}
	}
\end{figure}

For each correlator, our fitting procedure is as follows.
First, we vary the initial and final times $[t_{\rm min}, t_{\rm max}]$ used in the fits, amounting to a grid search over possible range fits.  The best fit is chosen automatically using a criterion from the QCD literature with a preference for small $\chisq$, large fit ranges, and well-determined fit parameters~\cite{Bitar:1990cb}.
We maximize
\be
Q\equiv \frac{p \times N_{\rm dof} }{\sum_n \sigma_{p_n}^2},
\label{eq:criterion}
\ee
where $p$ is the unconstrained $p$-value, $N_{\rm dof}$ denotes the number of degrees of freedom in the fit, and $\sigma_{p_n}$ denotes the statistical error in the $n$th fit parameter.
Although this criterion is ultimately arbitrary, it coincides with intuition about which fits ought to be considered good and removes  subjective bias.
We confirm that masses emerging from this procedure are consistent with expectations from effective mass plots; a representative comparison is shown in Fig.~\ref{fig:eff_mass_compare}.  We have also experimented with a ``two-state" double-exponential ansatz,
observing no significant changes in the ground-state masses within the combined statistical and systematic errors estimated using this procedure.

For an estimate of the systematic uncertainty associated with our fit-choice procedure, we compute the spread in the model parameters emerging from all nominally good fits satisfying $Q\ge 0.1$.
We then combine the statistical and systematic uncertainties conservatively using
\be
\sigma_\text{tot} = \sigma_\text{stat} + \sigma_\text{syst}.
\ee
The systematic error assigned by this procedure is often comparable to the statistical error, and is occasionally significantly larger.
The error estimates for the fermion masses $m_r$ in Tables~\ref{table:spec_t0_mq_16x18}--\ref{table:spec_vector_table_24x48} include this fit-range systematic.

\subsection{Decay constants and operator renormalization}\label{ssec:decay_constants}

The lattice operators appearing in the correlation functions are subject to finite renormalization in order to obtain the continuum-normalized operators that are required for determination of decay constants.  We carry out this procedure in lattice perturbation theory including tadpole improvement; the procedure is described in Appendix~\ref{app:renormalization}.  The explicit relationship between lattice and continuum operators is given by \Eq{eq:Qmua}.

Simultaneous fits to the smeared-source, point-sink ($s,p$) and smeared-source, smeared-sink ($s,s$) correlation functions allow us to extract the mass, decay constant, and smeared amplitude.
For example, the $\vev{VV}$ correlators in representation $r$ give us the vector decay constant $F_{Vr}$ defined in \Eq{eq:mv-def}.
The fit functions are
\begin{align}
C_{Vr}^{(s,p)}(t) &= \frac{A^s_{Vr} A^p_{Vr}}{2 M_{Vr}} \left( e^{-M_{Vr} t} + e^{-M_{Vr}(T-t)} \right), \\
C_{Vr}^{(s,s)}(t) &= \frac{A^s_{Vr}\,^2}{2 M_{Vr}} \left( e^{-M_{Vr} t} + e^{-M_{Vr}(T-t)} \right),
\end{align}
giving the vector mass $M_{Vr}$ and the point and smeared amplitudes $A^p_{Vr}$ and $A^s_{Vr}$, respectively.
In order to obtain decay constants with continuum normalizations, we apply the renormalization factors of \Eq{eq:Qmua}.
The result is
\be
\label{defFV}
F_{Vr} = Z_{Vr} \left(1 - \frac{3\kappa_r}{4\kappa_{r}^c} \right) \frac{A_{Vr}^p}{M_{Vr}}\ .
\ee

\subsection{Fermion mass determination and $\kappa_r^c$}\label{ssec:kappa_crit}

To determine the critical values $\kappa_{r}^c$, which enter into the field normalization for decay constants defined in Appendix~\ref{ssec:decay_constants}, we perform a global fit to the AWI fermion masses in units of the flow scale $t_0$ as given in Tables~\ref{table:spec_t0_mq_16x18}--\ref{table:spec_t0_mq_24x48}.  We use the model function
\be
\sqrt{t_0} m_{4} = c_0 + c_1 \beta + \kappa_4 (d_0 + d_1 \beta) + \kappa_6 (d_0' + d_1' \beta)
\label{eq:B7}
\ee
and similarly for $\sqrt{t_0} m_{6}$ (with a separate set of coefficients).
We find that these terms, which are a subset of all possible combinations of the bare parameters $\{\beta, \kappa_4, \kappa_6\}$ through quadratic order, are sufficient to provide reasonable fit quality.

\begin{figure}[ht]
\includegraphics[width=0.48\textwidth]{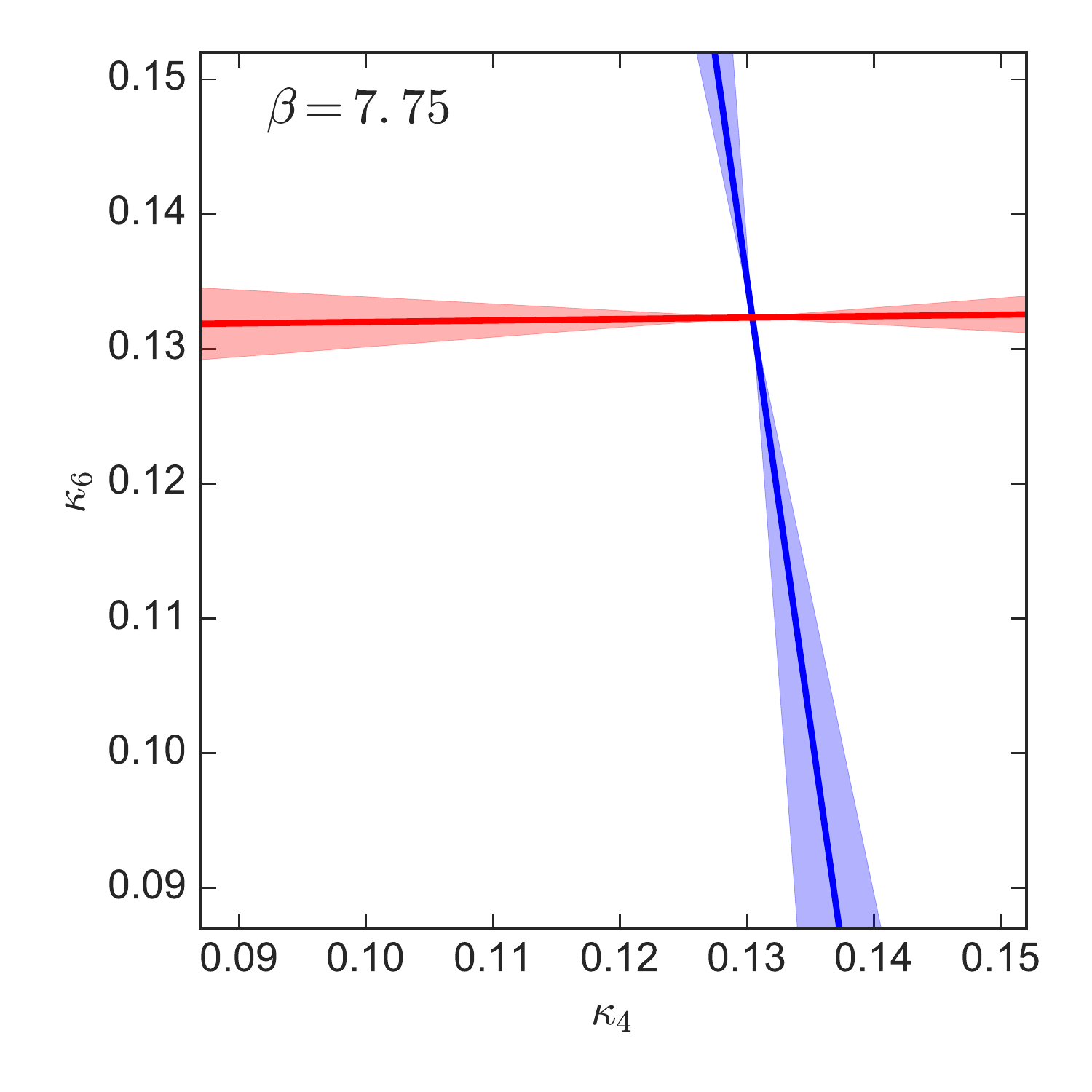}
\includegraphics[width=0.48\textwidth]{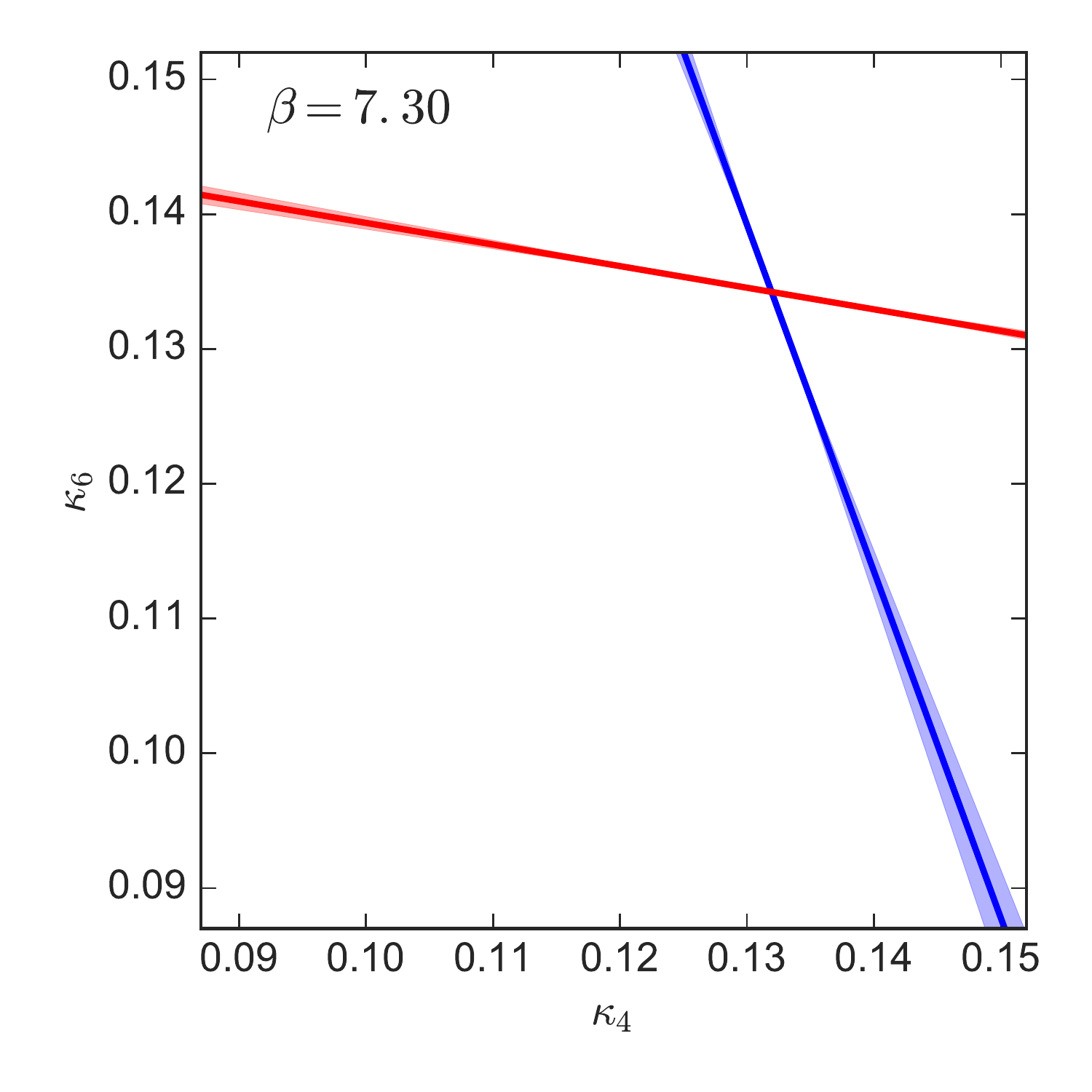}
\caption{$\kappa_c$ curves with uncertainties for both representations at two values of the bare coupling $\beta$, based on our fits to $\sqrt{t_0} m_{r}$ as described in the text.
The red (horizontal) curve is where $m_{6}$ vanishes, defining the function $\kappa_6^c(\beta,\kappa_4)$.
The blue (vertical) curve is where $m_{4}$ vanishes, defining $\kappa_4^c(\beta,\kappa_6)$.  \label{fig:kappa_c}}
\end{figure}

Since we are interested in the regions where $m_r\to 0$, we use only those ensembles that have $\sqrt{t_0} m_{r} < 0.08$, a value determined empirically by inspecting our data for deviations from the simple analytic behavior of \Eq{eq:B7}.  Our fits give $\chisq$ of 16/24 and 21/24 for fitting $\sqrt{t_0} m_{4}$ and $\sqrt{t_0} m_{6}$, respectively.  The resulting $\kappa_c$ curves at two $\beta$ values are shown in Fig.~\ref{fig:kappa_c}.

As noted above, because $\kappa_c$ is determined by extrapolation, we do not probe the existence of a possible Aoki phase.

\subsection{Study of finite-volume corrections}\label{ssec:finite_vol}

All of our ensembles satisfy the criterion $M_{Pr} L > 4$ for both pseudoscalar meson masses.  Using \Eq{eq:mzeta} with the results of our central fit, we further verify that $M_\zeta L > 4$ in all cases.  Although this cut is known to provide a useful threshold for the suppression of finite volume effects in QCD, since we are studying a new system a more cautious treatment is worthwhile.  Here we present three different analyses and arguments which lead us to estimate that finite-volume effects are no more than a few percent for our data, and are thus not resolved within the uncertainty of our results.  One particular source of finite-volume effects can be the freezing of the evolution of topological charge \cite{Brower:2003yx,Aoki:2007ka}; we measure the topological charge $Q$ on our ensembles using the Wilson flow to smear the gauge fields out to $t/a^2 = 5.0$ and find an acceptable distribution of $Q$ in all cases.

\begin{figure}[htb]
\includegraphics[width=0.8\textwidth]{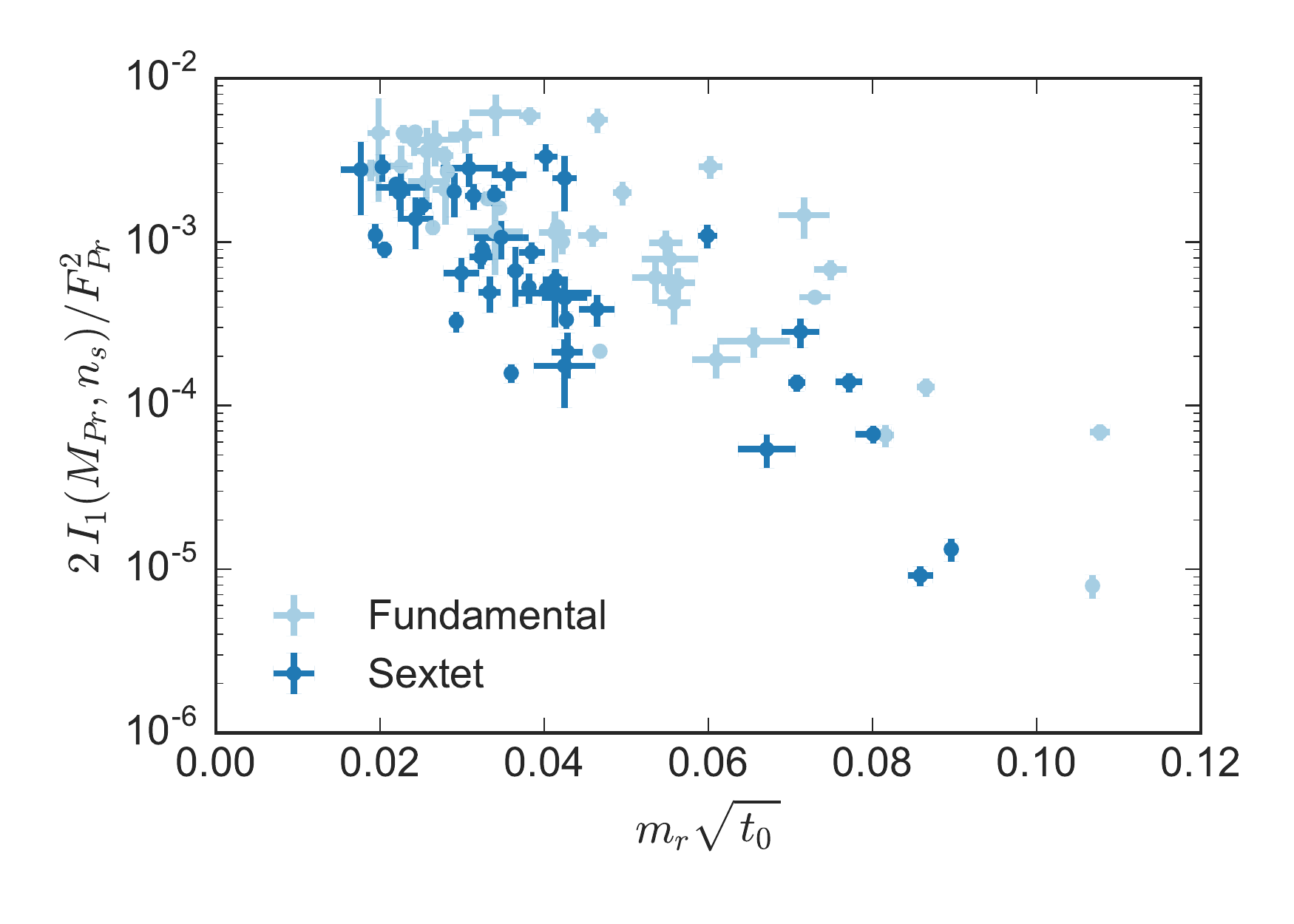}
\caption{
	Quantifying the size of leading-order finite volume corrections, \Eq{eq:I1}.
	\label{fig:finite_volume}
	}
\end{figure}

First, to obtain a theoretical estimate of the expected size of finite-volume effects, we consider the size of leading-order finite-volume correction to tadpole diagrams in chiral perturbation theory \cite{Gasser:1986vb,DeGrand:2016pur,Golterman:2009kw}.
The dimensionless figure of merit for this effect is $2 I_1(M_{Pr},L)/F_{Pr}^2,$ where
\be
 I_1(M,L) = 6 \left( \frac{M^2}{16 \pi^2} \right) \sqrt{ \frac{8 \pi}{(M L)^3} } \,e^{-M L}.
\label{eq:I1}
\ee
This quantity gauges the effect of mesons interacting with their finite-volume image points.
In Fig.~\ref{fig:finite_volume} we plot $2 I_1(M_{Pr},L)/F_{Pr}^2$ for all of our ensembles, with each representation $r$ plotted against the corresponding $\hm_r$.
We therefore expect that finite volume corrections do not exceed a few percent in the ensembles of this study.  This formula assumes the applicability of chiral perturbation theory, which requires that $F_{Pr} L \gtrsim 1$; over all of our ensembles we find that $F_{P4} L \gtrsim 1.3$ and $F_{P6} L \gtrsim 1.9$.

Second, we recalculate the observables $M_{Pr}$, $F_{Pr}$, and $t_0$ on ensembles with several spatial volumes at two sets of bare parameters, $(\beta, \kappa_4, \kappa_6) = (7.75, 0.128, 0.131)$ and $(7.75, 0.129, 0.1308)$.  These bare parameters match the production $V = 16^3 \times 32$ ensembles 36 and 37 as listed in Table~\ref{table:bare_params_16x32}.  Four ensembles hold $N_t = 32$ fixed and vary the spatial volume as $N_s = 12, 14, 16, 18$; the fifth ensemble at each point has $N_s = 24$ and $N_t = 48$.  

Results of this test are shown in Figs.~\ref{fig:fv_test_1}-\ref{fig:fv_test_3} below.  For both sets of bare parameters, all observables down to the smallest $N_s = 12$ are seen to be within $\pm 5\%$ of the central value obtained on $N_s = 24$, and within 2-3\% for $N_s = 16$ which is the smallest spatial volume used in our central analysis.

Finally, we have included explicit variation of the finite-volume cut on pseudoscalar meson masses (i.e. minimum cut on $M_{P_r} L$) in the stability analysis of our central chiral fit, as presented in Sec.~\ref{ssec:stability} and Fig.~\ref{fig:chipt_fit_stability}.  We also consider the effects of the finite temporal direction by cutting the $N_t = 18$ ensembles out of the analysis.  All of the fit results are seen to be stable at the one-sigma level as we vary the finite-volume cut.  We conclude that finite-volume effects are not significant in our results at the level of precision we obtain.

\begin{figure}[htb]
\includegraphics[width=0.8\textwidth]{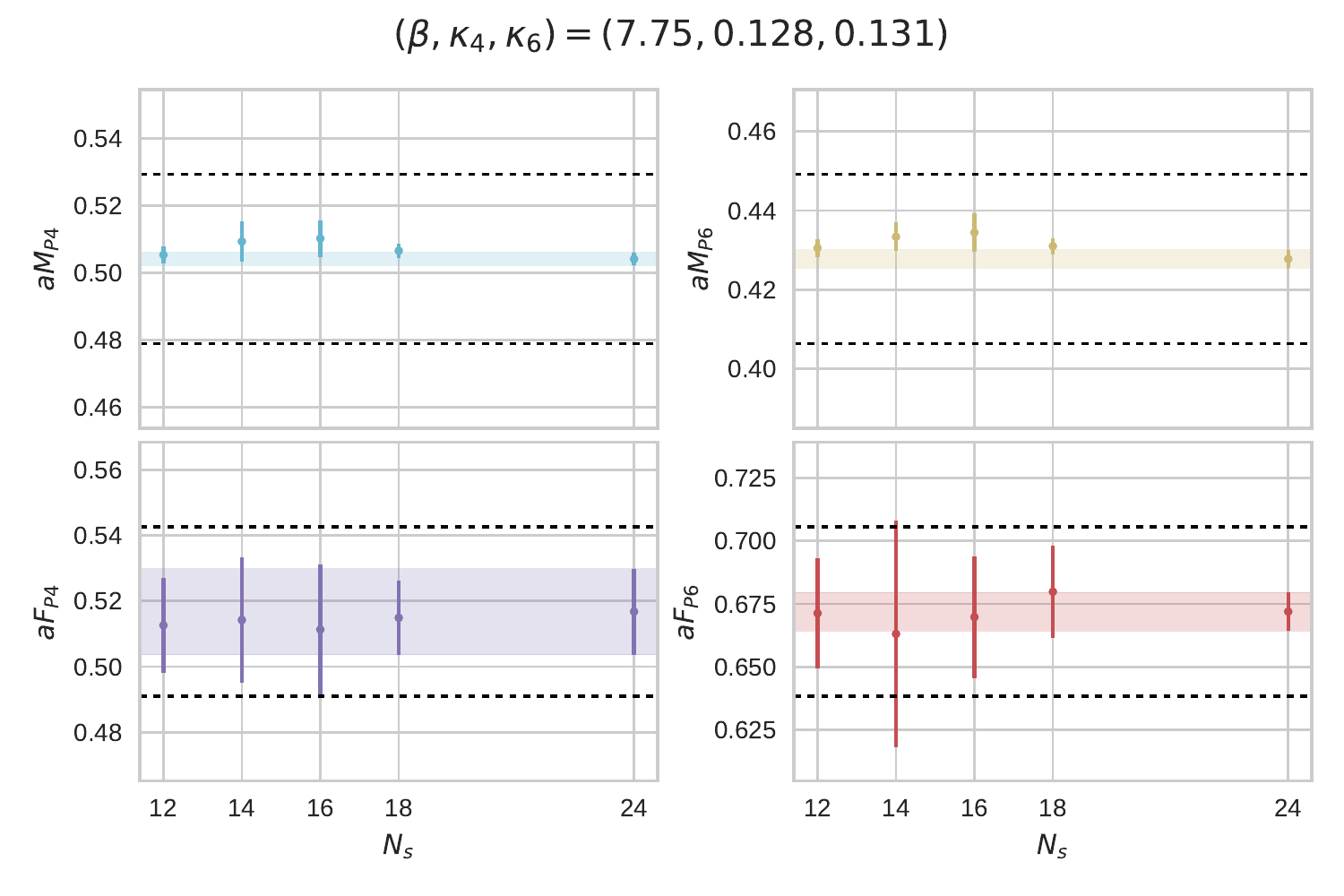}
\caption{
	Explicit test of the dependence of the pseudoscalar masses and decay constants on spatial volume at bare parameters
	$(\beta, \kappa_4, \kappa_6) = (7.75, 0.128, 0.131)$.  The dashed lines indicate variations of $\pm 5\%$ with respect to 
	the mean value of the rightmost $N_s = 24$ point.
	\label{fig:fv_test_1}
	}
\end{figure}

\begin{figure}[htb]
\includegraphics[width=0.8\textwidth]{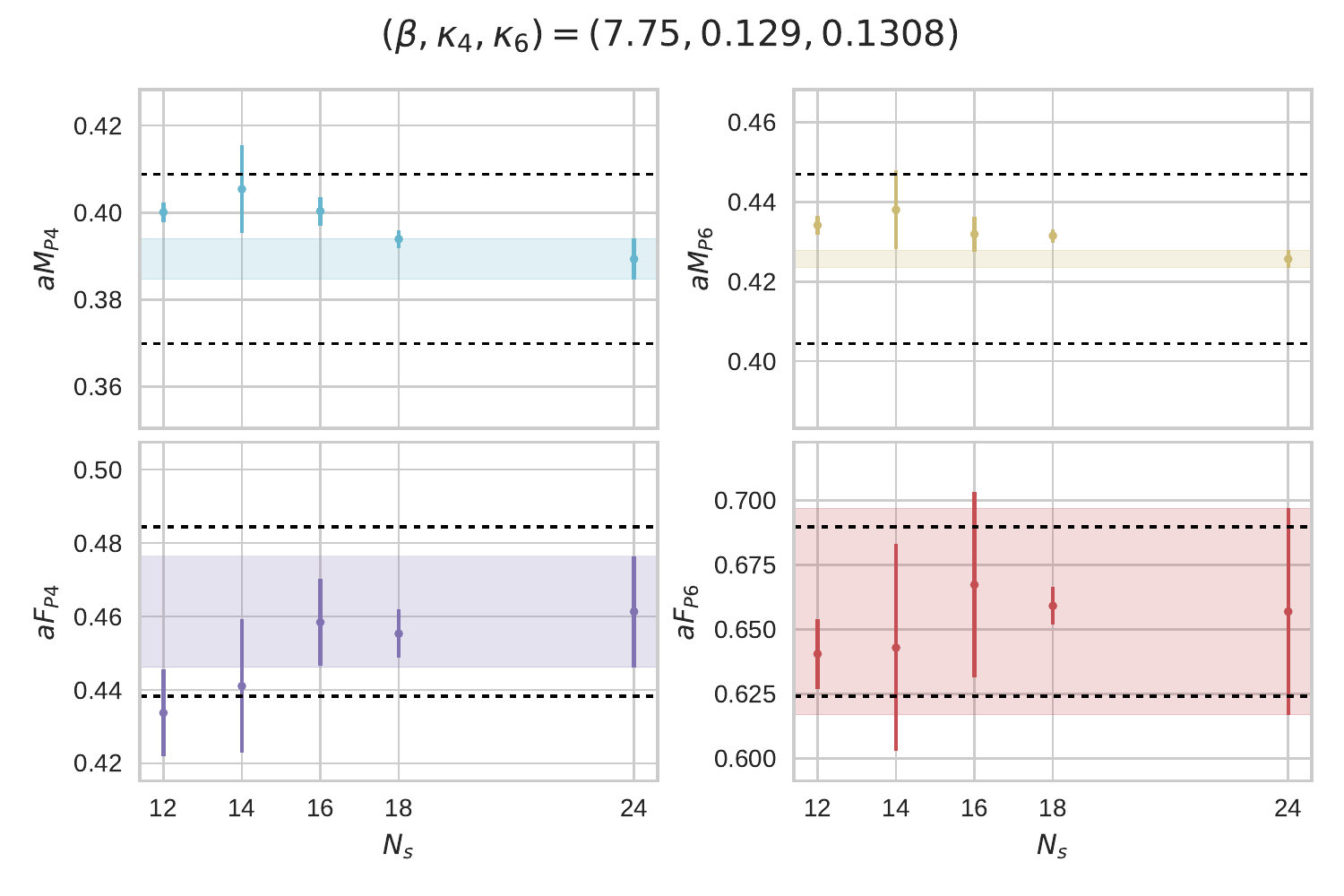}
\caption{
	Explicit test of the dependence of the pseudoscalar masses and decay constants on spatial volume at bare parameters
	$(\beta, \kappa_4, \kappa_6) = (7.75, 0.129, 0.1308)$.  The dashed lines indicate variations of $\pm 5\%$ with respect to 
	the mean value of the rightmost $N_s = 24$ point.
	\label{fig:fv_test_2}
	}
\end{figure}

\begin{figure}[htb]
\includegraphics[width=0.8\textwidth]{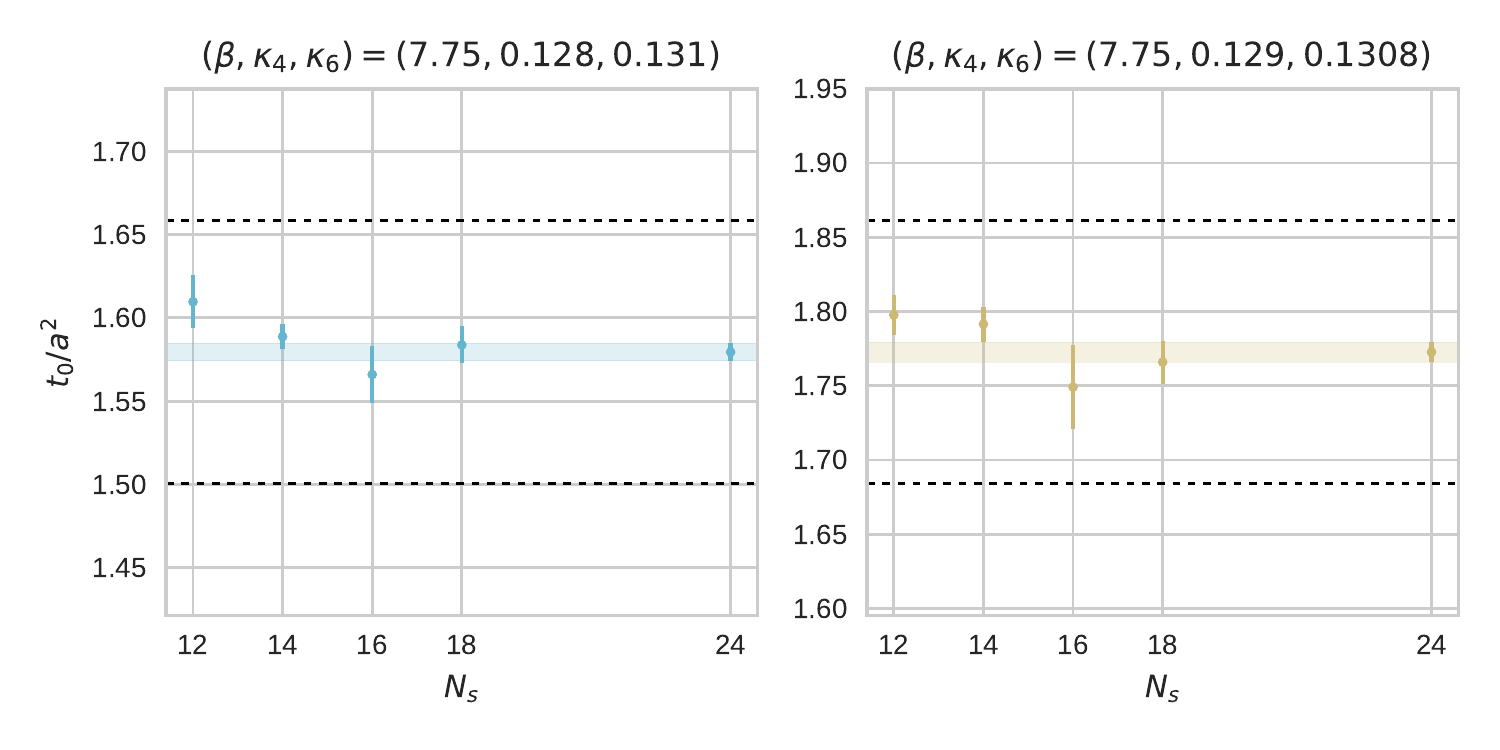}
\caption{
	Explicit test of the dependence of the Wilson flow scale $t_0/a^2$ on spatial volume at two sets of bare parameters,
	as described in the text.  The dashed lines indicate variations of $\pm 5\%$ with respect to 
	the mean value of the rightmost $N_s = 24$ point.
	\label{fig:fv_test_3}
	}
\end{figure}

\section{Technical matters---the axial current and W\cpt \label{app:tech-other}}

\subsection{\label{app:conserved_axial_current}Conserved $\U1$ Axial Current}

The conserved axial current is given by
\be
\label{J5}
  J_{5\mu} = \sum_{r=\textbf{4},\textbf{6}} q_r J_{5\mu}^{(r)} \ .
\ee
As usual, the normalization of U(1) currents is arbitrary.
We normalize the individual axial currents $J_{5\mu}^{(r)}$
such that all right-handed fields have unit charge.
The ratio $q_4/q_6$ is then fixed by the group traces
of the two representations.  For the normalization of $J_{5\mu}$ we adopt
the same prescription as in Ref.~\cite{DeGrand:2016pgq}.
The resulting charges are
\be
\label{Uiq}
  q_4 = \frac{2}{\sqrt{5}} \ , \qquad q_6 = -\frac{1}{\sqrt{5}} \ .
\ee
These charges were used in the \cpt\ formulae of Sec.~\ref{ssec:chifit}.

Tracing \Eqs{eq:chipt_mp4sq}--(\ref{eq:chipt_fp6}) back to the
general NLO expressions of Ref.~\cite{DeGrand:2016pgq},
one can check that they only depend on the ratios $q_r/F_\zeta$.
These ratios are independent of the choice of normalization for the
axial current, because a rescaling of $J_{5\mu}$ by a factor $\lambda$
implies a rescaling by the same factor of both the charges $q_4$ and $q_6$
and of the singlet's decay constant $F_\zeta$.  All other LECs
in \Eqs{eq:chipt_mp4sq}--(\ref{eq:chipt_fp6}) are independent of this
rescaling (for more details, see Ref.~\cite{DeGrand:2016pgq}).
Of course, once the normalization of the charges is fixed according to \Eq{Uiq},
the normalization of $F_\zeta$ is fixed as well.

\subsection{\label{ssec:appendix_wcpt}
  AWI mass and Wilson chiral perturbation theory}

In this appendix we review the proof of Ref.~\cite{Aoki:2009ri}
that the mass defined by imposing the axial Ward identity, $m_{\rm AWI}$,
is equal to the shifted mass $m_{\rm shifted}$, which is the mass parameter
occurring in the LO lagrangian of \wcpt, up to higher-order corrections.
For simplicity, we will consider
the GSM power counting used throughout this paper.

A nice feature of the GSM scheme is that the LO lagrangian
takes the same form as in the continuum.
The reason is that the only new operator at $O(a)$
is the Pauli term $a \bar\psi \sigma_{\mu\nu} F_{\mu\nu}\psi$,
which has the same chiral transformation properties as
the fermion mass term.  The Pauli term enters the Symanzik action
with a coefficient that depends linearly on the parameter $c_{SW}$
of the clover term in the Wilson action.  After the transition to the
chiral effective theory, the non-derivative terms in the LO lagrangian
take the form
\be
  {\cal L}_m = -\frac{F^2}{4} \tr(\chi^\dagger \Sigma + \Sigma^\dagger \chi)
  -\frac{F^2}{4} \tr(\hat{A}^\dagger \Sigma + \Sigma^\dagger \hat{A}) \ ,
\label{LOma}
\ee
where the mass term has been mapped to the first term on the right-hand side,
and the Pauli term to the second.
$\chi$ is the usual spurion of continuum \cpt, and
$\hat{A}$ is a new spurion with the same chiral transformation properties
as $\chi$.  The ``expectation values'' of the spurions are \cite{Sharpe:2004ny}
\be
  \chi = 2Bm_{\rm ctm} \ , \qquad \hat{A} = 2W_0 a,
\label{vevma}
\ee
where $B$ and $W_0$ are low-energy constants (LECs).
Substituting back into \Eq{LOma} gives
\be
\begin{split}
  {\cal L}_m &= -\frac{F^2}{4} (2Bm_{\rm ctm} +2W_0 a)
  \tr(\Sigma + \Sigma^\dagger)
\label{LOshift}\\
  &= -\frac{F^2}{2} Bm_{\rm shifted} \tr(\Sigma + \Sigma^\dagger) \ ,
\end{split}
\ee
where the shifted mass is defined by \Eq{mshift}.  For brevity, in the rest of this appendix we denote the shifted mass as $m$.

As explained in Sec.~\ref{sec:action}, in our numerical simulations
we define the fermion mass $m_{\rm AWI}$ by imposing the axial Ward identity,
\Eq{eq:pcac_continuum}.  Before we can use \wcpt,
we need to know the relation between $m_{\rm AWI}$ and the shifted mass $m$
of \Eq{mshift}, in terms of which the chiral expansion is done.

This relation was analyzed carefully in Ref.~\cite{Aoki:2009ri}
for the case of two-flavor QCD.
The pseudoscalar density that was considered there,
and which is also used in our work, is the usual local density,\footnote{%
  In this subsection we disregard renormalization factors.
}
\be
  P_{\rm loc}^a = \bar\psi \gamma_5 T^a \psi \ .
\label{Ploc}
\ee
For the axial current, Ref.~\cite{Aoki:2009ri} considered
\be
  A_\mu^a =
  A^a_{\mu,{\rm loc}} + a c_A \partial_\mu P_{\rm loc}^a \ ,
\label{Alocimp}
\ee
where $\partial_\mu$ stands for a lattice derivative;
the local axial current is
\be
  A^a_{\mu,{\rm loc}} = \bar\psi \gamma_5 \gamma_\mu T^a \psi \ .
\label{Aloc}
\ee
For the purpose of this discussion, we may consider $c_A$ in \Eq{Alocimp}
as a free parameter.  In our numerical simulations we use the
naive axial current, which corresponds to $c_A=0$.

To first order in the pion field,
the lattice operators are mapped to the effective theory according to
\begin{align}
  P_{\rm eff}^a &= \sqrt{2}FB \pi^a \bigl(1 + O(a)\bigr) \ ,
\label{Peff1st}\\
  A_{\mu,{\rm eff}}^a &= \sqrt{2}F \partial_\mu \pi^a \bigl(1 + O(a)\bigr) \ ,
\label{Amueff1st}
\end{align}
where $\pi^a$ is the effective pion field.  The precise form of the $O(a)$
corrections may be found in Ref.~\cite{Aoki:2009ri}.
In both equations, they depend linearly on the clover parameter $c_{SW}$.
In addition, the $O(a)$ correction in \Eq{Amueff1st} depends linearly on $c_A$.
Plugging this into \Eq{eq:pcac_continuum} and using the LO pion mass, given by
$M^2=2Bm$, we find
\be
  m_{\rm AWI} = m \bigl(1 + O(a)\bigr) + O(a^2) \ .
\label{mAWILOaa}
\ee
To the order we are working, in general one expects also
an $O(m^2)$ correction.  This correction vanishes, however,
because the continuum theory satisfies $m_{\rm AWI} = m$ identically.
While the derivation of Ref.~\cite{Aoki:2009ri} was given for
a complex representation, a similar argument applies to real (or pseudoreal)
representations.

Equation~(\ref{mAWILOaa}) is robust
in that changing the clover coefficient $c_{SW}$
or changing the parameter $c_A$ in \Eq{Alocimp} will change the $O(am)$
corrections, but will not affect the simultaneous vanishing
of $m_{\rm AWI}$ and the shifted mass $m$.  This feature is disrupted
only by $O(a^2)$ effects, which is as it should be.
Indeed, as shown in Ref.~\cite{Sharpe:1998xm},
depending on the sign of a particular $O(a^2)$ LEC,
in the region where $m\sim a^2$
one either encounters the Aoki phase, or a first-order discontinuity line
at which the pion mass reaches a non-zero minimum.

As a corollary of \Eq{mAWILOaa}, we may use the value of $m_{\rm AWI}$,
taken from the numerical simulations, for the shifted mass $m$.
At the order we are working, the differences between the two are absorbed
into a redefinition of NLO parameters of the chiral expansion.

\section{Gluon propagator and perturbative calculations for \lowercase{n}HYP links}
	\label{app:renormalization}

To perform one loop perturbation theory for the NDS action \cite{DeGrand:2014rwa}, we need
the gluon propagator. This appendix describes its construction
and gives perturbative
results for current renormalization factors.

Normalized hypercubic  links $V_{x,\mu}$ are constructed
from the dynamical gauge field $U_{x,\mu}$ via three successive
smearing steps \cite{Hasenfratz:2001hp,Hasenfratz:2007rf}.
Each step uses a weighted sum over staples,
which is then reunitarized.  Explicitly,
\begin{widetext}
\begin{subequations}
\begin{eqnarray}
  \Omega_{x,\rho;\xi} &=& (1-\alpha_3) U_{x,\rho}
  + \frac{\alpha_3}{2} \left(
    U_{x,\xi} U_{x+\hat\xi,\rho} U^\dagger_{x+\hat\rho,\xi}
    + U^\dagger_{x-\hat\xi,\xi} U_{x-\hat\xi,\rho} U_{x-\hat\xi+\hat\rho,\xi}
  \right)
\label{smeara}\\
  \rule{0ex}{3ex}
  \overline{V}_{x,\rho;\xi} &=& {\cal{P}}( \Omega_{\xi,\rho;\xi} )
\nonumber \\
  \rule{0ex}{3ex}
  \overline{\Omega}_{x,\mu;\nu} &=& (1-\alpha_2) U_{x,\mu}
  + \frac{\alpha_2}{4} \sum_{\stackrel{\scriptstyle \rho \ne \mu,\nu}{\xi \ne \mu,\nu,\rho}}
    \left(
    \overline{V}_{x,\rho;\xi} \overline{V}_{x+\hat\rho,\mu;\xi} \overline{V}^\dagger_{x+\hat\mu,\rho;\xi}
    + \overline{V}^\dagger_{x-\hat\rho,\rho;\xi} \overline{V}_{x-\hat\rho,\mu;\xi}
    \overline{V}_{x-\hat\rho+\hat\mu,\rho;\xi}
  \right)
  \hspace{5ex}
\label{smearb}\\
  \tilde{V}_{x,\mu;\nu} &=& {\cal{P}}\Big(\overline{\Omega}_{x,\mu;\nu}\Big)
\nonumber \\
  \tilde{\Omega}_{x,\mu} &=& (1-\alpha_1) U_{x,\mu}
  + \frac{\alpha_1}{6} \sum_{\nu \ne \mu} \left(
    \tilde{V}_{x,\nu;\mu} \tilde{V}_{x+\hat\nu,\mu;\nu} \tilde{V}^\dagger_{x+\hat\mu,\nu;\mu}
    + \tilde{V}^\dagger_{x-\hat\nu,\nu;\mu} \tilde{V}_{x-\hat\nu,\mu;\nu}
    \tilde{V}_{x-\hat\nu+\hat\mu,\nu;\mu}
  \right)
\label{smearc}\\
  V_{x,\mu} &=& {\cal{P}}\Big(\tilde{\Omega}_{x,\mu}\Big).
\nonumber
\end{eqnarray}
\end{subequations}
\end{widetext}
The reunitarization operator ${\cal{P}}$ is defined as
\begin{equation}
  V = {\cal{P}}(\Omega) \equiv \Omega Q^{-1/2} \ ,
\label{reu}
\end{equation}
where
\begin{equation}
  Q = \Omega^\dagger \Omega \ .
\label{Q}
\end{equation}

The best way to understand the smearing is to go in reverse order.
The staple sum extends into a different direction at each smearing step, such that
each fat link $V_{x,\mu}$ depends
on a particular thin link $U_{y,\nu}$, if and only if there exists a hypercube
to which both $V_{x,\mu}$ and $U_{y,\nu}$ belong.

The dislocation-suppressing action adds a new term to the pure-gauge action $S_g$
\begin{equation}
  S_g = S_{\textrm{plaq}} + S_{\textrm{NDS}} \ ,
\label{Sg}
\end{equation}
where the new term is
\begin{eqnarray}
  S_{\textrm{NDS}} &=& \frac{1}{2N_c} \sum_x \tr\!\left(
                  \gamma_1 \sum_{\mu} \tilde{Q}_{x,\mu}^{-1}
                + \gamma_2 \sum_{\mu\ne\nu} \overline{Q}_{x,\mu;\nu}^{-1} + \gamma_3 \sum_{\rho\ne\xi} Q_{x,\rho;\xi}^{-1}
      \right).
\label{dsa}
\end{eqnarray}
In practice we fix the $\alpha_i$'s and
take $\gamma_1=\gamma_2=\gamma_3=\gamma = z \beta$ where $z$ is held constant.
The weak-coupling expansion to be sketched below gives the bare coupling $g_0^2$ as
\begin{equation}
  \frac{1}{g_0^2} = \frac{\beta}{2N_c} +
  \frac{1}{N_c} \left( \frac{\gamma_1 \alpha_1}{3} + \gamma_2 \alpha_2 + \gamma_3 \alpha_3
  \right) \ .
\label{gbare}
\end{equation}

To construct the gluon propagator
we need to compute the gauge action in quadratic order.
This is pretty standard;
a good reference is Ref. \cite{Weisz:1982zw}.
We take the lattice action,
\be
S = a^4 \sum_x {\cal L}(\overline \psi, \psi, U),
\ee
and replace the link field by an expansion in terms of gauge fields
\be
U_\mu(x) = \exp[iga A_\mu(x)] = 1 + igaA_\mu(x) -
     \frac{1}{2} g^2a^2 A_\mu(x)^2 +\cdots,
\ee
where $A_\mu(x) = \sum_a (\lambda^a/2) A_\mu^a$ gives the
decomposition into color components.  There is an identical expansion
for the fat link, which we write as $V_\mu(x)=\exp[iga B_\mu(x)]$.

The  action has an expansion in powers of $A$.
In terms of the integral over the
four-dimensional Brillouin zone,
\index{Brillouin zone}
\be
\int_q \equiv \int_{-\pi/a}^{\pi/a} \frac{d^4q}{(2\pi)^4},
\ee
and the vector potential,
\be
A_\mu(x) = \int_q A_\mu(q)e^{iq(x + a \hat\mu/2)},
\ee
 the free gauge boson action is
\be
S_0^G = -\frac{1}{2}\int_{p p^\prime}    (2\pi)^4\delta^4(p+p^\prime)
   [  A^a_\mu(p^\prime)D^{ab}_{\mu\nu}(p) A^b_\nu(p)].
\ee
For the gauge boson,
$D^{ab}_{\mu\nu}=\delta^{ab}D_{\mu\nu}$. Just as in a continuum
theory, the gauge boson action cannot be inverted to give the
propagator without fixing the gauge.  A conventional choice for a
gauge fixing term is [introducing $\hat k^\mu = 2/a \, \sin
(ak_\mu/2)$]
\be
S_{\rm gf} = -\frac{1}{2} \sum_{\mu\nu}\int_k \Tr \frac{1}{\xi} \hat k_\mu \hat k_\nu  A_\mu^a(-k)A_\nu^a(k).
\ee
Then the gauge boson propagator is found by solving the field equation
\be
\sum_\nu \left[ \frac{1}{\xi} \hat k_\mu \hat k_\nu + D_{\mu\nu}(k)\right ]
   G_{\nu\tau}(k) = \delta_{\mu\tau}  .
\label{eq:gprop}
\ee
We simply do this numerically, inverting the four by four matrix for each $k$ value.

To perform the perturbative expansion of  the NDS action, we look at each term in turn. Consider
\be
Q_{x,\rho;\xi}= \Omega_{x,\rho;\xi}^\dagger \Omega_{x,\rho;\xi}\ .
\ee
Multiplying this out, we find that $Q$ is a sum of loops of perimeter four
and perimeter six, labeled $P$ and $E$ respectively.
The planar $E$ loops, $E_{\mu\nu}$, are $1\times 2$ loops extending in the $\pm \nu$ direction
from site $x$.
Hence,
\be
Q =1+ \alpha_3(1-\alpha_3) \sum _jP_j + \frac{\alpha_3}{4} \sum_k E_k.
\ee
Expanding each term out in terms of $A$'s, we discover that $Q^{-1} = 1 - S_Q$
where now the expressions are quadratic functions of the gauge fields. Slightly abusing notation,
we write $P$ and $E$ for the objects made of $A_\mu$'s, so that
\be
S_Q= \alpha_3(1-\alpha_3) \sum_j P_j  + \frac{\alpha_3}{4} \sum_k E_k
\ee

Nearly identical results obtain for the other two $Q$'s, with several small exceptions.
First, the analogs of the plaquettes are built of one thin link $A_\mu(x)$
while the other three links are fattened. Second, the perimeter-6 links are built entirely of fat links.
Finally, in addition to the $1\times 2$ $E_{\mu\nu}$ plaquettes, there are ``chair'' plaquettes
$C_{\mu\nu\rho}$ which are ``folded'' about the $\mu$ axis. They extend in four directions
($\pm \nu$, $\pm \rho$).

The analog of the fat to thin relation for the links ($U\rightarrow V$)
 is a fat to thin gauge field relation,
\be
B_\lambda(q) = \sum_\mu A_\lambda(q) h_{\mu\lambda}(q)
\label{eq:form}
\ee
This means that all the perimeter-six contributions can be easily computed beginning with
the fat links, whose gauge-unfixed action is $D^0_{\mu\nu}(q)$. The thin link action is
\be
D_{\mu\nu}(k) = \sum_{\lambda_1 \lambda_2}
 h_{\mu\lambda}(-q) D^0_{\lambda_1 \lambda_2}(q) h_{\nu \lambda}(q).
\ee

Finally, by convention the action $D_{\mu\nu}(q) \sim \delta_{\mu\nu}k^2 - k_\mu k_\nu $
so we have to rescale the action to remove explicit factors of the coupling constants, giving
\be
S = \frac{1}{N} \left(P + z\tilde{Q}^{-1} + z  \overline{Q}^{-1} + z Q^{-1} \right),
\ee
where $N=2N_c/(\beta g^2)= [1+ 2z(\alpha_1/3 + \alpha_2 + \alpha_3)]$, basically \Eq{gbare}.

The fattened terms in the action are awkward to compute. The plaquette is an example.
It is probably best to record
\be
F_{\mu\rho}(q) =  i\hat q_\mu B_\rho(q) + A_\mu(q) \exp(-iq_\rho/2) - B_\mu(q)  ,\exp(iq_\rho/2)
\ee
where $B_\mu$ is the fat link gauge potential, and
then to substitute in \Eq{eq:form}; to
write (schematically)
\be
F_{\mu\rho}(q) = \sum_\lambda A_\lambda(q)C_{\lambda \mu \nu}(q) \ ;
\ee
and then
\be
D_{\lambda_1 \lambda_2}(q)= \sum_{\mu\nu}C_{\lambda_1 \mu \nu}(-q)C_{\lambda_2 \mu \nu}(q).
\ee

We also need perturbative expressions for the partially fattened links. They are
\be
\overline{A}_{\mu,\rho\eta}(q) = A_\mu(q) + \frac{\alpha_3}{2}\left[ \hat q_\mu \hat q_\omega A_\omega(q) - \hat q_\omega^2 A_\mu(q)\right],
\ee
where $\omega \ne \mu,\nu,\rho$,
and
\be
\tilde{A}_{\mu,\rho}(q) = A_\mu(q) +
 \frac{\alpha_2}{2}\sum_{\eta \ne \mu,\rho} \hat q_\eta
\left(1 + \alpha_3 - \hat q_\omega^2 \frac{\alpha_3}{2}\right) %
\left[A_\eta(q)\hat q_\mu - A_\mu(q) \hat q_\eta\right],
\ee
where $\omega \ne \mu,\eta,\rho$.

Our simulations are done with $z=1/125$. At that value of $z$, the perturbative
properties of the NDS gauge action are almost indistinguishable from those of a pure Wilson action.
Here are three examples.

First, the plaquette has an expansion
${\rm Tr}\,U_{\rm plaq}/N_c= 1 - g^2 pC_F $ where $C_F$ is the quadratic Casimir for fundamentals
and $p$ is a constant. For the Wilson action,  $p=0.5$. This expression is
often replaced by
\be
-\ln \left\langle{\frac{1}{N_c} {\rm Tr}\,U_{\rm plaq}}\right\rangle = g^2 pC_F .
\ee
This defines a coupling $g^2$ in the so-called $\alpha_V$ scheme, because the potential is written as
\be
V(q) = - C_F \frac{4\pi \alpha_V(q)}{q^2}
\ee
The scale of the coupling is set by the Lepage--Mackenzie \cite{Lepage:1992xa} prescription,
\be
\label{eq:LM}
\log q^* =
 \frac{\int d^4 q \, \log q I(q)}{ \int d^4 q\, I(q)}\ .
\ee
Results for $p$ and $q^*$ for several values of $z$ are given in Table \ref{tab:plaqtable}.
\begin{table}[hbt]
\begin{ruledtabular}
\begin{tabular}{ddddddddddd}
z & p & q^*  &  z_V & q^*  &  z_A & q^*  & z_P & q^*   & z_S & q^*   \\
\hline
0     & 0.5   &   3.41  & -1.38 & 1.68  & -1.30 & 1.65  & -0.12 & 2.28    & 0.04 & 2.31  \\
0.008 & 0.504 & 3.41    & -1.37 & 1.70  & -1.30 & 1.66  &-0.11 & 2.42  &  0.05 & 2.48   \\
0.02 & 0.510  & 3.41   & -1.38 & 1.71  & -1.29 & 1.69  & -0.12 & 2.48   & 0.04 & 2.54   \\
0.05 & 0.525  & 3.41   & -1.40 & 1.70  &-1.31 & 1.69  &-0.15 & 2.40  &  0.02 & 2.45   \\
0.10 & 0.548  & 3.43   & -1.42 & 1.74  &-1.34  & 1.72 & -0.19 & 2.70   & -0.02 & 2.79 \\
\end{tabular}
\end{ruledtabular}
\caption{ Results of one loop lattice perturbation theory for selected observables, for
the NDS action for several values of $z$. Uncertainties are $\pm 2$ in the rightmost digit.
Each one is interleaved with its momentum scale from \Eq{eq:LM}.
The plaquette expectation is $p$. The quantities $z_V$, $z_A$, $z_P$ and $z_S$
are the renormalization factors for the local vector, axial vector, pseudoscalar, and scalar currents.
\label{tab:plaqtable}}
\end{table}

\begin{figure}[hbt]
\begin{center}
\includegraphics[width=0.8\columnwidth,clip]{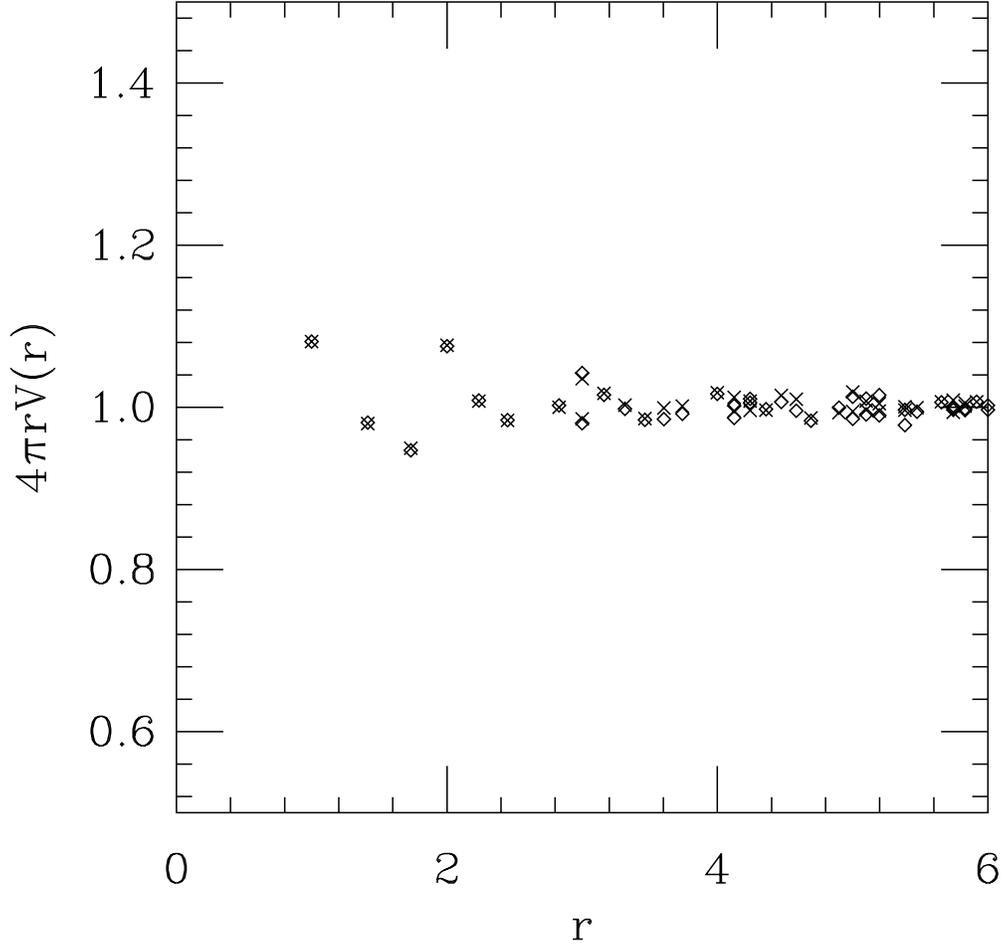}
\end{center}
\caption{
Comparison of the potential for the NDS action, with $\gamma/\beta=1/125$ (diamonds),
 with that of the usual Wilson action (crosses).}
\label{fig:vr}
\end{figure}

With the gluon propagator in hand we can immediately compute the static
Coulomb  potential.
With our conventions, the continuum potential is \( V(r)=1/(4\pi r) \),
and so plotting the rescaled lattice potential \( 4\pi rV(r) \) immediately
exposes the lattice artifacts of a particular action. We show results
for this quantity in Fig.~\ref{fig:vr} for $z=1/125$, plotted together with the Wilson action result.

Last, we have the renormalization factors for currents.
Calculations of matrix elements require a conversion to continuum regularization.
We adopt the old tadpole-improved procedure of Lepage and Mackenzie
\cite{Lepage:1992xa}.
In this scheme a continuum-regulated fermionic bilinear quantity $\bar{Q}$ with engineering
dimension $D$ [we have in mind finding the
$\overline{MS}$ (modified minimal subtraction) value at scale $\mu$] is related to the lattice value by
\be
\label{eq:Qmua}
\bar{Q}(\mu=1/a) = Q(a) \left(1 - \frac{3 \kappa}{4 \kappa_c}\right) Z_Q,
\ee
with 
\be
Z_Q= 1 + \alpha \frac{C_F}{4\pi} z_Q\ ,
\label{eq:zfactor}
\ee
where $\alpha=g^2/(4\pi)$, $C_F$ is the quadratic Casimir for the fermion,
 and $z_Q$ is
 a scheme matching number. Results for nHYP clover fermions without the NDS term are tabulated in  Ref.~\cite{DeGrand:2002vu}
and allow us to check the $z=0$ limit.  For more discussion of the calculation of the $z_Q$'s, see Ref.~\cite{DeGrand:2002va}.
Table \ref{tab:plaqtable} gives selected values of $z_Q$ for the
vector, axial vector, pseudoscalar, and
scalar currents.

To evaluate the final $Z$-factors, we run $\alpha_V(q)$ obtained at $aq^\star = 3.43$ using the Lepage--Mackenzie prescription to the appropriate $aq^\star$ for each $z_Q$ enumerated in the table.  Running of $\alpha_V$ is carried out by numerical integration of the two-loop $\beta$-function,
\be
\beta(\alpha_V) = q^2 \frac{d\alpha_V}{dq^2} = -\beta_0 \alpha_V^2 - \beta_1 \alpha_V^3 - \cdots,
\ee
where the required coefficients for a theory with multiple fermion representations are \cite{Shi:2015baa}
\bea
\beta_0 &=& \frac{1}{3(4\pi)} \left( 11 C_2(G) - 2 \sum_r N_r C_2(r) \right) \\
\beta_1 &=& \frac{1}{3(4\pi)^2} \left( 34 C_2(G)^2 - 2 \sum_r N_r T(r) \left[ 5 C_2(G) + 3 C_2(r) \right] \right).
\eea
Here $N_r$ denotes the number of Dirac flavors in representation $r$, while $T(r)$ and $C_2(r)$ are the standard trace and Casimir invariant for each representation.  For our SU(4) multirep theory, we obtain $\beta_0 = 53/(24\pi)$ and $\beta_1 = 1531/(192\pi^2)$.

\bibliography{su4_spectro}

\end{document}